\PassOptionsToPackage{table,xcdraw}{xcolor}
\documentclass[11pt,a4paper]{article}
\pdfoutput=1

\makeatletter
\def\@fpheader{}
\makeatother

\usepackage{jheppub}
\usepackage{gensymb}
\DeclareSymbolFont{matha}{OML}{txmi}{m}{it}
\DeclareMathSymbol{\varv}{\mathord}{matha}{118}
\usepackage{amsmath}
\usepackage{amssymb}
\usepackage{mathrsfs}
\usepackage{soul}
\usepackage[normalem]{ulem}
\usepackage{graphicx}
\usepackage{subfigure}
\usepackage{pstricks}
\usepackage{ulem}
\usepackage{bm}
\usepackage{pbox}
\usepackage{placeins}
\usepackage{booktabs}
\usepackage[T1]{fontenc}
\usepackage{footnote}
\usepackage{pdfpages}
\usepackage{hhline}
\usepackage{multirow}
\usepackage{multicol}
\usepackage{enumitem}
\usepackage[toc,page]{appendix}
\allowdisplaybreaks
\usepackage{comment}
\usepackage{float}
\usepackage{slashed}
\usepackage{xcolor}
\usepackage{textcomp}
\usepackage{multirow}
\usepackage[numbers]{natbib}
\usepackage{notoccite}
\usepackage{indentfirst}
\usepackage{gensymb}
\DeclareSymbolFont{matha}{OML}{txmi}{m}{it}
\DeclareMathSymbol{\varv}{\mathord}{matha}{118}
\usepackage{amsmath}
\usepackage{array}
\usepackage{amssymb}
\usepackage{graphicx}
\usepackage{subfigure}
\usepackage{pstricks}
\usepackage{bbm}
\usepackage{pbox}
\usepackage{placeins}
\usepackage{booktabs}
\usepackage[T1]{fontenc}
\usepackage{footnote}
\usepackage{pdfpages}
\usepackage{hhline}
\usepackage{multirow}
\usepackage{multicol}
\usepackage{enumitem}
\usepackage[toc,page]{appendix}
\allowdisplaybreaks
\usepackage{comment}
\usepackage{float}
\usepackage{slashed}
\usepackage{xcolor}
\usepackage{textcomp}
\usepackage{multirow}
\usepackage[numbers]{natbib}
\usepackage{notoccite}
\usepackage{soul}
\usepackage{extarrows}
\usepackage{makecell}
\usepackage{longtable}
\usepackage{physics}
\usepackage{comment}
\usepackage{blindtext}
\usepackage{csquotes}
\usepackage{soul}
\usepackage{indentfirst}
\DeclareUnicodeCharacter{0304}{ }
\usepackage{rotating}
\usepackage{tabularx}
\usepackage{extarrows}
\usepackage{physics}
\usepackage{comment}
\usepackage{cancel}
\usepackage{blindtext}
\usepackage{csquotes}
\usepackage[none]{hyphenat}
\DeclareUnicodeCharacter{0304}{ }
\usepackage{rotating}
\usepackage{tabularx}
\usepackage{cleveref}

\usepackage[many]{tcolorbox}
\definecolor{fg}{RGB}{34,139,34}

\renewcommand{\thetable}{\Roman{table}}

\def\figureautorefname~#1\null{Fig.\,#1\null}
\def\equationautorefname~#1\null{Eq.\,(#1)\null}
\def\tableautorefname~#1\null{Tab.\,#1\null}

\newcommand{\be}{\begin{equation}}
\newcommand{\ee}{\end{equation}}
\newcommand{\ba}{\begin{eqnarray}}
\newcommand{\ea}{\end{eqnarray}}
\newcommand{\bmat}{\begin{pmatrix}}
\newcommand{\emat}{\end{pmatrix}}

\def\ket{\rangle}

\def\ketbra{\rangle \langle}
\def\nn{\nonumber}

\def\la{\lambda}
\def\e{\epsilon}

\def\o{\omega}

\definecolor{MyDarkBlue}{rgb}{0.1, 0.1, 0.8} 
\definecolor{MyLightBlue}{rgb}{0.22,0.51,0.9}
\definecolor{MyGreen}{rgb}{0.0, 0.5, 0.0}
\definecolor{BrickRed}{rgb}{0.8, 0.25, 0.33}

\RequirePackage{hyperref}
\hypersetup{colorlinks=true, citecolor=MyGreen,linkcolor=BrickRed, urlcolor=MyLightBlue}
\usepackage{cleveref}
\crefname{equation}{Eq.}{Eqs.} 
\crefrangelabelformat{equation}{(#3#1#4--#5#2#6)}

\title{\boldmath Tripartite entanglement in $e^+ e^- \to t \bar{t} Z$}
\author[a]{Dorival Gon\c{c}alves,}
\author[b]{Alberto Navarro,}
\author[c]{Kazuki Sakurai}

\affiliation[a]{Department of Physics, Oklahoma State University, Stillwater, OK 74078, USA}
\affiliation[b]{Institute for Convergence of Basic Studies, Seoul National University of Science and Technology, Seoul 01811, Republic of Korea}
\affiliation[c]{Institute of Theoretical Physics, Faculty of Physics, University of Warsaw, Pasteura 5, 02-093, Warsaw, Poland}
\emailAdd{dorival@okstate.edu}\emailAdd{alberto.navarro1001@seoultech.ac.kr}\emailAdd{kazuki.sakurai@fuw.edu.pl}

\abstract{Multipartite entanglement is a uniquely quantum form of correlation that captures collective properties of a composite quantum state beyond those encoded in its bipartite subsystems. We investigate this phenomenon in the process $e^+e^-\to t\bar tZ$ at a future lepton collider, where the final state spins span the tripartite Hilbert space $\mathcal{H}=\mathbb{C}^{2}\otimes\mathbb{C}^{2}\otimes\mathbb{C}^{3}$.
Starting from the Standard Model helicity amplitudes, we reconstruct the full $12\times 12$ spin density matrix and characterise its entanglement structure through one-to-one negativities, one-to-other negativities, and the genuine multipartite negativity, evaluated at three increasingly inclusive levels of phase space integration.
Pairwise entanglement is generally suppressed relative to the collective (one-to-other) and the genuine multipartite entanglement, and all measures decrease as more kinematic information is integrated out.
Assuming quantum tomography in the fully leptonic decay channel at $\sqrt{s}=1$ TeV, we find that collective entanglement should be accessible at a realistic high-luminosity polarised lepton collider. By contrast, certifying genuine multipartite entanglement is more challenging, with only limited sensitivity projected for a specific polarisation benchmark within the expected ILC luminosity.
The study establishes $e^+e^-\to t\bar tZ$ as an attractive laboratory for probing multipartite entanglement in high-energy collisions and provides a general mixed state framework that applies to any tripartite spin system.}

\begin{document}
\maketitle

\begin{sloppypar}

\section{Introduction}
\label{sec:intro}

Polarisation and spin correlations are powerful probes of the dynamics of high-energy collisions, playing a central role in both precision measurements and searches for physics beyond the Standard Model. The interpretation of these observables through the lens of quantum information science has recently opened a new perspective on these studies, for instance, probing whether a system is entangled. Most analyses to date have focused on bipartite systems, including $t \bar t$~\cite{Afik:2020onf,Fabbrichesi:2021npl,Severi:2021cnj,Afik:2022dgh,Aoude:2022imd,Afik:2022kwm,Aguilar-Saavedra:2022uye,Fabbrichesi:2022ovb,Severi:2022qjy,Dong:2023xiw,Aguilar-Saavedra:2023hss,Han:2023fci,ATLAS:2023fsd,Cheng:2023qmz,Maltoni:2024tul,Barr:2024djo,Maltoni:2024csn,CMS:2024pts,White:2024nuc,Dong:2024xsg,Cheng:2024btk,Dong:2024xsb,CMS:2024zkc,Han:2024ugl,Altomonte:2024upf,Cheng:2025cuv,Afik:2025ejh,Afik:2025grr,Aoude:2025ovu,Nason:2025hix,Fabbrichesi:2025psr,Lin:2025eci,Low:2025aqq,Durupt:2025wuk,Gu:2025ijz,Gabrielli:2026tnl,Guo:2026yhz,Fang:2026ddi,Choi:2026omc,Altakach:2026fpl,Aoude:2026eeg}, $\tau^+ \tau^-$~\cite{Altakach:2022ywa,Ma:2023yvd,Ehataht:2023zzt,Fabbrichesi:2024wcd,Fabbrichesi:2025ywl,Han:2025ewp,Zhang:2025mmm,Jeans:2026eys,Lee:2026wlk} 
and electroweak gauge boson final states~\cite{Barr:2021zcp,Aguilar-Saavedra:2022wam,Aguilar-Saavedra:2022mpg,Ashby-Pickering:2022umy,Fabbrichesi:2023cev,Fabbrichesi:2023jep,Morales:2023gow,Aoude:2023hxv,Fabbri:2023ncz,Bernal:2023ruk,Bi:2023uop,Bernal:2024xhm,Ruzi:2024cbt,Grossi:2024jae,Wu:2024ovc,Sullivan:2024wzl,Aguilar-Saavedra:2024jkj,Ding:2025mzj,DelGratta:2025qyp,Goncalves:2025mvl,Aguilar-Saavedra:2025byk,Goncalves:2025xer,Ruzi:2025jql,Aguilar-Saavedra:2025njw,De:2025dpo,CMS-PAS-HIG-25-011,Pelliccioli:2026ltl,ATLAS:2026hye,Goncalves:2026njf}, where the quantum state is classified simply as separable or entangled. 
The theoretical framework for studying tripartite entanglement and nonlocal correlations in particle physics processes was introduced in Refs.~\cite{Sakurai:2023nsc,Horodecki:2025tpn}. 
Tripartite quantum correlations have subsequently been explored in a few collider studies~\cite{Aguilar-Saavedra:2024whi,Morales:2024jhj,Subba:2024mnl,Fabbrichesi:2025zpw}.
Nevertheless, a comprehensive characterisation of experimentally accessible tripartite spin systems, particularly in the presence of realistic mixed quantum states, remains limited.

Extending the study of quantum entanglement from bipartite to multipartite systems opens a qualitatively new frontier. Unlike bipartite states, multipartite quantum states admit a hierarchy of inequivalent separability classes. A tripartite state may be fully separable, bi-separable, or genuinely multipartite entangled (GME), and its quantum correlations cannot, in general, be reconstructed from one-body observables or pairwise correlations alone~\cite{Horodecki:2009zz,Vidal:2002zz,Jungnitsch:2011izf,Sakurai:2023nsc,Bernal:2023xzp,Horodecki:2025tpn}. Remarkably, genuine multipartite entanglement may persist even when all two-body entanglement measures vanish.  Understanding how such collective quantum correlations arise in collider processes, and whether they survive realistic phase space averaging and experimental reconstruction, constitutes an important open question at the interface of quantum information science and high-energy physics.

From the perspective of collider phenomenology, multipartite correlations represent a natural extension of the study of spin observables. While polarisation and pairwise spin correlations have become standard tools in new physics searches~\cite{ATLAS:2016rnf,ATLAS:2018gqq,Goncalves:2018fvn,Goncalves:2018ptp,ATLAS:2019zrq,CMS:2019nrx,Goncalves:2020vyn,MammenAbraham:2022yxp,Bhardwaj:2023ufl,Ackerschott:2023nax,Denner:2023ehn,CMS:2024ony,Denner:2024tlu,Carrivale:2025mjy}, the corresponding multipartite quantum correlations remain largely unexplored. Quantum tomography provides access to the complete spin density matrix, making it possible to move beyond one-body and two-body observables, probing the full structure of multipartite quantum states. This opens the possibility of identifying novel observables that are inaccessible to traditional analyses and may provide complementary information on the dynamics of particle interactions.

In this work, we perform a systematic study of tripartite spin entanglement in $e^+e^- \to t\bar t Z$ at future lepton colliders. The final state contains three spin-carrying particles and spans the Hilbert space
$\mathcal H = \mathbb C^2 \otimes \mathbb C^2 \otimes \mathbb C^3$, allowing direct access to pairwise, collective, and genuinely multipartite quantum correlations. We construct the full spin density matrix and characterise its separability structure through a hierarchy of entanglement measures applicable to mixed states: (1) one-to-one negativities, which probe pairwise entanglement between subsystems~\cite{Vidal:2002zz}; (2) one-to-other negativities, which quantify the entanglement between a single particle and the remaining pair; and (3) the genuine multipartite negativity (GMN)~\cite{Jungnitsch:2011izf}, which detects correlations that cannot be reduced to any bi-separable description. Unlike entanglement diagnostics that rely on pure state constructions, the  measures employed here remain valid for arbitrary mixed states, providing a natural framework for realistic collider measurements.
These observables resolve the full hierarchy of separability classes accessible to the $t\bar{t}Z$ spin state. We find that the dominant quantum correlations are collective rather than pairwise. Pairwise entanglement is already suppressed prior to phase space integration, while one-to-other entanglement and genuine multipartite entanglement remain sizeable across broad regions of phase space. We identify fiducial kinematic regions that maximise the retained multipartite entanglement and estimate the integrated luminosity required to probe these effects at future lepton colliders.

The prospects for studying these effects are particularly promising, since $e^+e^- \to t\bar t Z$ will be accessible at future high-energy lepton colliders. The process naturally produces a tripartite spin system and can be reconstructed with high precision at facilities operating above the $t\bar t Z$ threshold, including the high-energy stages of the ILC and CLIC~\cite{ILC:2013jhg,CLICdp:2018cto}. Their clean experimental environment, polarised beams, and excellent event reconstruction capabilities provide the ingredients required for detailed studies of multipartite spin states. Similar opportunities arise at the proposed multi-TeV muon collider through the analogous process $\mu^+\mu^- \to t\bar t Z$~\cite{InternationalMuonCollider:2025sys}. In this context, multipartite correlations complement the established top quark precision programme by providing a window into the collective quantum structure of the full spin density matrix.

This paper is organised as follows. In \autoref{sec:amp}, we set up the kinematics and construct the spin density matrix for $e^+e^- \to t\bar{t}Z$ from helicity amplitudes. In \autoref{sec:measures}, we introduce the entanglement measures employed in this work and review the separability structure of tripartite quantum systems. \autoref{sec:ttZ} presents the entanglement properties of the differential and fully inclusive spin states. In \autoref{sec:prospects}, we analyse partially inclusive states and assess the prospects for observing multipartite entanglement at a future lepton collider. Our conclusions are presented in \autoref{sec:concl}. The details of the quantum tomography reconstruction for the $t\bar{t}Z$ process are provided in~\autoref{app:tomography}.

\section{Helicity Amplitudes and Spin Quantum States}
\label{sec:amp}

In this section we set up the kinematics of $e^{+}e^{-}\to t\bar t Z$, compute the tree-level helicity amplitudes, and construct the spin density matrix of the final state in the Hilbert space ${\cal H} = {\mathbb C}^{2}\otimes{\mathbb C}^{2}\otimes{\mathbb C}^{3}$ (dimension $2\times 2\times 3 = 12$). The resulting density matrix is the central object on which all of the entanglement measures of \autoref{sec:measures} will be evaluated.

\subsection{Kinematics}
\label{sec:kinematics}

We consider the process
\begin{equation}
  e^{-}(k_{1},\la_1) \, + \, e^{+}(k_{2},\la_2) \;\longrightarrow\; t(p_{1},h_1) \, + \, \bar t(p_{2},h_2) \, + \, Z(p_{3},h_3)
\label{eq:process}
\end{equation}
at a future lepton collider with centre-of-mass energy $\sqrt{s}$.
The four-momenta $k_1$, $k_2$, $p_1$, $p_2$ and $p_3$ are assigned to the electron, positron, top, antitop and $Z$-boson, respectively, in the lab frame.
Similarly, $\la_1$, $\la_2$, $h_1$, $h_2$ and $h_3$ are the helicities of these particles. 
The spatial $z$-axis is chosen along the incoming electron beam, $\hat z \equiv \hat{\vec k}_{1}$. 
Neglecting the electron mass, one can write $k_1^\mu = \tfrac{\sqrt{s}}{2}(1,0,0,1)$ and $k_2^\mu = \tfrac{\sqrt{s}}{2}(1,0,0, - 1)$. We consider the idealised case with no initial state radiation.  In this limit, the lab-frame coincides with the centre-of-mass frame.

\begin{figure}[t]
\centering
\includegraphics[width=0.7\textwidth]{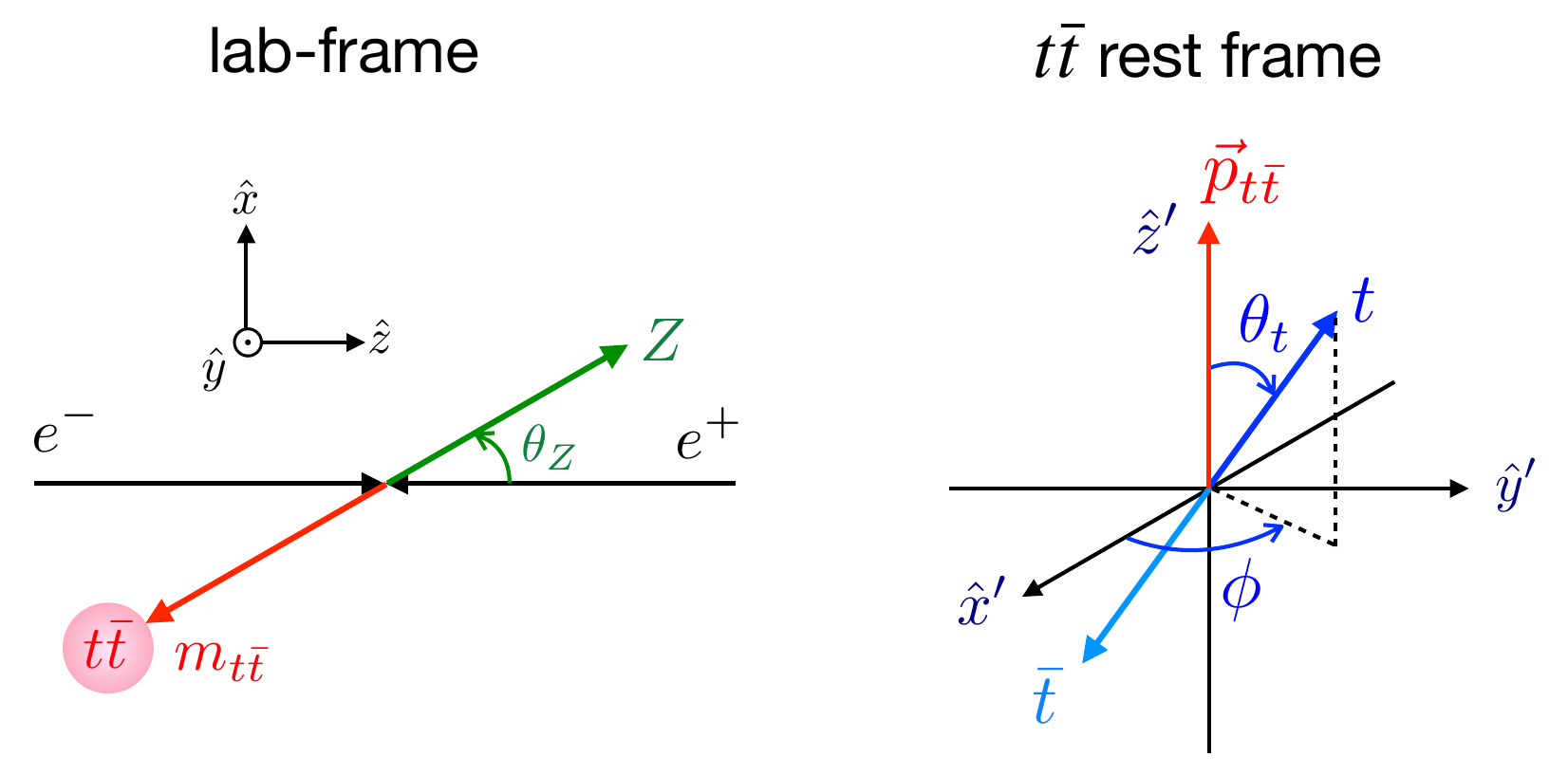}
\caption{\small Definition of the kinematical variables. \emph{Left:} lab-frame view defining $m_{t\bar t}$ and $\theta_{Z}$. \emph{Right:} $t\bar t$-rest-frame view defining $\theta_{t}$ and $\phi$.}
\label{fig:kinematics}
\end{figure}

Using the rotational symmetry around the beam axis, we fix the $x$-axis such that the $\hat x$-$\hat z$ plane is aligned with the production plane of $Z$ and the $t \bar t$ subsystem and the $x$-component of the $Z$'s momentum is positive (see the left panel of Fig.\ \ref{fig:kinematics}). 
Introducing the polar angle $\theta_Z$ for the $Z$'s momentum, one can write $p_3^\mu = (E_3, q \sin \theta_Z, 0, q \cos \theta_Z)$ with $E_3 = \sqrt{m_Z^2 + q^2}$\,.
In the lab frame, the $t \bar t$ subsystem is back-to-back with the $Z$-boson.  The momentum of $t \bar t$ subsystem is $p_{t \bar t}^\mu = (p_1 + p_2)^\mu = (E_{t \bar t}, -q \sin \theta_Z, 0, -q\cos\theta_Z)$ with
$E_{t \bar t} = \sqrt{m_{t \bar t}^2 + q^2}$, where $m_{t \bar t}^2 = (p_1 + p_2)^2$ is the $t \bar t$ invariant mass. 
The conservation of the energy, $\sqrt{s} = E_3 + E_{t \bar t}$, leads to $q = \frac{1}{2} \sqrt{s - 2(m_{t \bar t}^2 + m_Z^2) + \tfrac{1}{s}(m_{t \bar t}^2 - m_Z^2)^2 }$\,.

The kinematics of $t$ and $\bar t$ are determined as follows.
We first describe the $t$ and $\bar t$ momenta at the $t \bar t$ rest frame as $p'_1 = [\tfrac{m_{t\bar t}}{2}, p'(\sin \theta_t \cos \phi, \sin \theta_t \sin \phi, \cos \theta_t) ]$
and
$p'_2 = [\tfrac{m_{t\bar t}}{2}, -p'(\sin \theta_t \cos \phi, \sin \theta_t \sin \phi, \cos \theta_t) ]$ with $p' = \sqrt{ \tfrac{m_{t \bar t}^2}{4} - m_t^2  }$\,.
Here, $\theta_t$ and $\phi$ are the polar and azimuthal angles of the top momentum, where the polar axis is taken to be the boost direction of the $t \bar t$ subsystem (see the right panel of Fig.\ \ref{fig:kinematics}).
The lab-frame momenta, $p_1$ and $p_2$, are obtained by Lorentz-transforming $p_1'$ and $p_2'$ to the lab-frame.

With the above parametrisation, the lab-frame momenta $p_{1}$, $p_{2}$ and $p_{3}$ are completely fixed (up to the cylindrical symmetry of the collider) by the following four parameters:
\begin{itemize}
\item $m_{t\bar t}$: $t\bar t$ invariant mass, $m_{t\bar t}\in [\,2 m_{t},\;\sqrt{s}-m_{Z}\,]$;
\item $\theta_{Z}$: polar angle of $Z$ in the lab frame, $\cos\theta_{Z}\in[-1,1]$;
\item $\theta_{t}$: polar angle of $t$ in the $t\bar t$ rest frame, $\cos\theta_{t}\in[-1,1]$;
\item $\phi$: azimuthal angle of $t$ in the $t\bar t$ rest frame, $\phi\in[0,2\pi)$.
\end{itemize}

In this parametrisation, CP acts on the final state as $t\leftrightarrow\bar t$ together with $\vec p_i\to-\vec p_i$ for every final-state momentum. This reverses $\hat{\vec p}_Z$ in the lab frame, which sends $\phi_Z\to\phi_Z+\pi$; re-using the cylindrical symmetry of the collider to reset $\phi_Z=0$ (a rotation by $\pi$ about $\hat z$), CP acts on the four variables as
\be
{\rm CP}:\quad (m_{t\bar t},\,\cos\theta_Z,\,\cos\theta_t,\,\phi)\;\longrightarrow\;(m_{t\bar t},\,-\cos\theta_Z,\,-\cos\theta_t,\,-\phi)\,.
\label{eq:CP-rule}
\ee
The sign flips of $\cos\theta_t$ and $\phi$ are most easily checked by decomposing CP into its two factors. We treat CP as an active transformation acting only on the final state: the unpolarised initial state is itself CP-symmetric, so the lab axes $(\hat x,\hat y,\hat z)$ are kept fixed throughout. Under C, $\vec p_t$ is replaced by $\vec p_{\bar t}=-\vec p_t$ in the $t\bar t$ rest frame, contributing $\cos\theta_t\to-\cos\theta_t$ and $\phi\to\phi+\pi$. Under P (with the accompanying rotation), only the final-state $\hat p_Z$ flips in the lab (not $\hat z$); hence $\hat z'=-\hat p_Z$ and $\hat y'\propto\hat z\times\hat p_Z$ each change sign, while $\hat x'=\hat y'\times\hat z'$ picks up two such minus signs and is unchanged. Projecting onto the new triad then contributes an extra $\phi\to\pi-\phi$. Composing the two gives \eqref{eq:CP-rule}. For unpolarised beams the tree-level process is CP-invariant, so any CP-even observable is left invariant by \eqref{eq:CP-rule} and any CP-conjugate pair, e.g.\ $(N_{tZ},N_{\bar tZ})$, satisfies ${\cal O}_{tZ}(\cos\theta_Z,\cos\theta_t,\phi)={\cal O}_{\bar tZ}(-\cos\theta_Z,-\cos\theta_t,-\phi)$.

\subsection{Helicity Amplitudes}
\label{sec:helamp}

\begin{figure}[t]
\centering
\includegraphics[width=\textwidth]{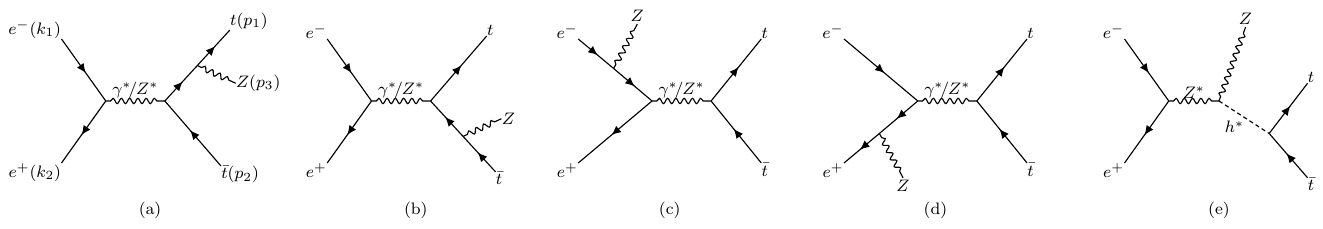}
\caption{\small Tree-level Feynman diagrams contributing to $e^{+}e^{-}\to t\bar t Z$ in the Standard Model. Diagrams~(a) and~(b): $Z$ radiation from $t$ and $\bar t$. Diagrams~(c) and~(d): $Z$ radiation from $e^{-}$ and $e^{+}$. Diagram~(e): Higgsstrahlung $e^{+}e^{-}\to Z h^{*}\to Z t\bar t$.}
\label{fig:diagrams}
\end{figure}

To obtain the spin quantum state of the final state, we compute the helicity amplitudes of $e^+ e^- \to t \bar t Z$ at the leading order. 
There are five types of tree-level diagrams, labelled by $I=a,\ldots,e$, shown in Fig.\ \ref{fig:diagrams}.
The amplitudes for fixed momenta and helicities are given as the sum of five diagrams:
\be
{\cal M}_{h_1,h_2,h_3}^{\la_1, \la_2}
\,=\,
\sum_{I = {a,\ldots,e}}
{\cal M}^{(I)}[ \bar u(k_1,\la_1), v(k_2,\la_2), u(p_1,h_1), \bar v(p_2,h_2), \e(p_3,h_3) ],
\label{eq:helamp}
\ee
where ${\cal M}^{(I)}$ is the contribution from the diagram $I$, which is a function of four Dirac spinors corresponding to $e^-$, $e^+$, $t$ and $\bar t$ and the polarisation vector $\epsilon^\mu$ of the $Z$-boson.
The fermion and anti-fermion spinors $u(p, \la)$ and $v(p, \la)$ with the helicity $\la = \pm$ and the momentum $p^\mu = (E, \vec{p})$ with $E=\sqrt{m^2 + |\vec{p}|^2}$ and $\vec{p} = |\vec{p}|(\sin\theta \cos\phi, \sin\theta \sin\phi, \cos \theta)$ are given by \cite{Hagiwara:1985yu,Hagiwara:1988pp,Murayama:1992gi}\footnote{Here $(\theta,\phi)$ denote the polar and azimuthal angles of $\vec p$ in the lab frame, and should not be confused with $\theta_{Z}$, $\theta_{t}$ and $\phi$ introduced in \autoref{sec:kinematics}.}
\ba
&& 
u(p,\la) \,=\, 
\bmat
\o_{- \la}(p) \chi_\la(p) \\
\o_{\la}(p) \chi_\la(p) \\
\emat,
\quad
v(p,\la) \,=\, 
\bmat
- \la \o_{\la}(p) \chi_{-\la}(p) \\
\la \o_{-\la}(p) \chi_{-\la}(p) \\
\emat,
\quad
\o_{\la}(p) = \sqrt{E + \la |\vec{p}|}\,,
\nn \\
&&
\qquad
\qquad
\qquad
\chi_+ \,=\,
\bmat
e^{-i \phi/2} \cos \tfrac{\theta}{2} \\
e^{i \phi/2} \sin \tfrac{\theta}{2} \\
\emat,
\qquad
\chi_- \,=\,
\bmat
-e^{-i \phi/2} \sin \tfrac{\theta}{2} \\
e^{i \phi/2} \cos \tfrac{\theta}{2} \\
\emat\,.
\ea
When $\theta = 0$ and $\pi$, the phase ambiguity in $\chi_\pm$ is fixed by setting $\phi=0$.
For the $Z$-boson momentum $p_{3}^{\mu} = (E_{3}, q \sin \theta_{Z}, 0, q \cos \theta_{Z})$, the polarisation vectors corresponding to helicities $h_{3}\in\{+1,0,-1\}$ are given by
\ba
\epsilon^\mu(p_3, \pm) 
\,=\,
\pm \frac{1}{\sqrt{2}}\left(0, -\cos \theta_Z, \pm i, \sin \theta_Z \right)\,,
\quad
\epsilon^\mu(p_3, 0) 
\,=\,
\left(
\frac{q}{m_Z},
\frac{E_3}{m_Z} (\sin \theta_Z, 0, \cos \theta_Z) 
\right)\,.
\nn \\
\ea
These polarisation vectors satisfy $p_3\cdot\epsilon(p_3,h_3)=0$ and the standard orthonormality relations. We explicitly evaluate the helicity amplitudes in Eq.~\eqref{eq:helamp} with these spinors and polarisation vectors. Since the electron mass is neglected, chirality conservation at the $e^+e^-\gamma^*/Z^*$ vertex implies helicity conservation along the fermion line, so the amplitudes vanish unless $\lambda_1=-\lambda_2$:
\be
{\cal M}_{h_1,h_2,h_3}^{+ +}
\,=\,
{\cal M}_{h_1,h_2,h_3}^{- -}
\,=\,
0\,.
\ee

\subsection{Spin Quantum States}
\label{sec:states}

For fixed external momenta and initial polarisation $(\lambda_1,\lambda_2)$, the unnormalised final spin state in the Hilbert space $H=\mathbb C^2\otimes\mathbb C^2\otimes\mathbb C^3$ is given by
\be
| \widetilde \psi_{\la_1,\la_2} \ket \,=\,
\sum_{h_1, h_2, h_3}
{\cal M}_{h_1,h_2,h_3}^{\la_1, \la_2}
| h_1,h_2,h_3 \ket\,.
\ee
The corresponding unnormalised $12 \times 12$ density matrix is 
\be
R^{\la_1,\la_2}_{(h_1, h_2, h_3),(h_1', h_2', h_3')}
\,=\, 
{\cal M}_{h_1,h_2,h_3}^{\la_1, \la_2}\,
[{\cal M}_{h_1',h_2',h_3'}^{\la_1, \la_2}]^*\,.
\label{eq:R}
\ee
The normalised density matrix is then obtained as
$\rho^{\la_1,\la_2} = R^{\la_1,\la_2}/{\rm Tr}[R^{\la_1,\la_2}]$.

In a realistic experimental situation, the incoming beams are not 100\% polarised.
We introduce the polarisation parameters ${\cal P}_{e^\pm} \in [-1, 1]$ to describe the $e^-$ and $e^+$ beams as \cite{Altomonte:2024upf,Altakach:2026fpl}
\be
\rho_{e^\pm} \,=\,\frac{1}{2}(1 + {\cal P}_{e^\pm}) | + \ketbra + |_{e^\pm}
\,+\,
\frac{1}{2}(1 - {\cal P}_{e^\pm}) | - \ketbra - |_{e^\pm}\,.
\ee
For imperfectly polarised beams with ${\cal P}_{e^-}$ and ${\cal P}_{e^+}$, the final state density matrix is in general mixed and is obtained as the cross-section-weighted average over initial helicities.
The unnormalised density matrix in this case is given by
\be
R\,[{\cal P}_{e^-},{\cal P}_{e^+}]
\,=\,
\sum_{\la_1,\la_2} P_{\la_1,\la_2}\, R^{\la_1,\la_2} 
\,,
\quad
P_{\la_1,\la_2} \,=\, \frac{1}{4}(1 + \la_1 {\cal P}_{e^-})(1 + \la_2 {\cal P}_{e^+})\,,
\label{eq:rho_polarised}
\ee
where $P_{\la_1,\la_2}$ is the joint probability of having initial helicities $(\la_1,\la_2)$.
Since the SM amplitudes vanish for $\la_1=\la_2$, only the $(+,-)$ and $(-,+)$ helicity configurations contribute. Throughout \autoref{sec:ttZ} we use the unpolarised configuration ${\cal P}_{e^-}={\cal P}_{e^+}=0$, for which
\be
\rho^{\rm unpol} \,=\, \frac{R[0,0]}{{\rm Tr} \,R[0,0]},
\qquad
R[0,0]
\,=\,
\frac{1}{4} \left[ R^{+-} +\, R^{-+} \right]\,.
\ee
In \autoref{sec:prospects}, where we estimate the prospects for an entanglement observation at a future $e^+ e^-$ collider, we additionally consider two polarised benchmarks, Pol$+$ with $({\cal P}_{e^-},{\cal P}_{e^+})=(+0.8,-0.3)$ and Pol$-$ with $(-0.8,+0.3)$; the corresponding $\rho[{\cal P}_{e^-},{\cal P}_{e^+}]$ is obtained directly from \autoref{eq:rho_polarised}.

The diagonal elements of the $R$-matrix correspond to the differential cross-section
\be
\frac{d \sigma^{\la_1,\la_2}_{h_1,h_2,h_3}}{d \Phi} \,=\, 
\frac{1}{2s} R^{\la_1,\la_2}_{(h_1, h_2, h_3),(h_1, h_2, h_3)}\,.
\ee
In an experimental setup, the density matrix $\rho$ must be averaged over the events contained in some selected phase space region $\Sigma$, which leads to
\be
\rho^{\la_1,\la_2}_\Sigma
\,=\,
\frac{ R^{\la_1,\la_2}_\Sigma}{{\rm Tr}[ R^{\la_1,\la_2}_\Sigma]}
,
\qquad
 R^{\la_1,\la_2}_\Sigma
\,=\,
\int_\Sigma d \sigma^{\la_1,\la_2} \, \rho^{\la_1,\la_2}
\,=\,
\int_\Sigma 
\frac{d\Phi}{2s} 
 \,
R^{\la_1,\la_2}\,,
\label{eq:fictitious-state}
\ee
where $d \sigma^{\la_1,\la_2} = \sum_{h1,h2,h3} d \sigma^{\la_1,\la_2}_{h1,h2,h3}$ and the full phase space integration is
\be
\int_{\rm full} d \Phi 
\,=\,
\frac{1}{1024 \, \pi^4}
\int_{(m_{t \bar t}^{\rm min})^2}^{(m_{t \bar t}^{\rm max})^2} d m^2_{t \bar t}
\, \kappa_{ttZ}^{\frac{1}{2}} \, \kappa_{tt}^{\frac{1}{2}} 
\int_{-1}^1 d \cos \theta_Z 
\int_{-1}^1 d \cos \theta_t 
\int_0^{2 \pi} d \phi
\ee
with $m_{t \bar t}^{\rm min} = 2 m_t$,
$m_{t \bar t}^{\rm max} = \sqrt{s} - m_Z$,
$\kappa_{ttZ} = \lambda(s,m_{t\bar{t}}^2,m_Z^2)/s^2$,  
and
$\kappa_{tt} = 1 - 4 m_t^2 / m^2_{t \bar t}$, where
$\lambda(x,y,z)=x^2+y^2+z^2-2xy-2xz-2yz$
is the K\"all\'en function.

The averaging procedure generally makes the state mixed due to the entanglement between the spin and the momentum degrees of freedom.
Note that for each event the spin quantisation axes are taken along the directions of the momentum of each particle (\emph{i.e.}, the helicity basis). This means that in the phase space integration, we sum density matrices defined with different spin-quantisation axes.
In the literature, the result of such an average is called the ``fictitious quantum state'' \cite{Afik:2022kwm,Cheng:2023qmz}. It is this object on which all the entanglement measures introduced in \autoref{sec:measures} will be evaluated.
For unpolarised beams the fictitious final state is therefore given by
\be
\rho_{\Sigma}^{\rm unpol}
\,=\,
\frac{R_\Sigma[0,0]}{{\rm Tr}\, R_\Sigma[0,0] },
\qquad
R_\Sigma[0,0]
\,=\,
\frac{1}{4} \left[ R_{\Sigma}^{+-} +\, R_{\Sigma}^{-+} \right] \,;
\label{eq:rho-unpol}
\ee
the polarised analogue $\rho_\Sigma\,[{\cal P}_{e^-},{\cal P}_{e^+}]$ used in \autoref{sec:prospects} is obtained by normalising $R_\Sigma\,[{\cal P}_{e^-},{\cal P}_{e^+}]\,=\,\sum_{\lambda_1,\lambda_2} P_{\lambda_1,\lambda_2}\,R_\Sigma^{\lambda_1,\lambda_2}$, with the joint probabilities $P_{\lambda_1,\lambda_2}$ given in \autoref{eq:rho_polarised}.

\section{Entanglement Measures}
\label{sec:measures}

The spin density matrix $\rho$ (and $\rho_\Sigma$) constructed in the previous sections describes a tripartite spin state of the $t\bar t Z$ system in the Hilbert space ${\cal H} = {\mathbb C}^{2}_{t}\otimes{\mathbb C}^{2}_{\bar t}\otimes{\mathbb C}^{3}_{Z}$, which has dimension twelve.
In this section, we introduce several entanglement measures to characterise the multipartite entanglement structure of the state. 

\subsection{Separability and LOCC Monotonicity}
\label{sec:sep}

One of the defining properties of quantum entanglement is non-separability. 
For a bipartite mixed state $\rho_{A B}$ acting on the Hilbert space
${\cal H}_{A}\otimes{\cal H}_{B}$, the state is called \emph{separable} if it admits a convex decomposition,
\be
\rho^{AB}_{\rm sep} \,=\, \sum_{i} q_{i}\,
\rho^A_i \otimes \rho^B_i,
\qquad 
\rho^A_i \in {\cal S}({\cal H}_A),
\quad
\rho^B_i \in {\cal S}({\cal H}_B),
\label{eq:sep}
\ee
with $q_i \geq 0$ and $\sum_i q_i = 1$, where ${\cal S}({\cal H})$ is the set of all density matrices of ${\cal H}$. The set of entangled states is defined as the complement of the set of separable states. Based on this notion of separability, we consider an entanglement measure, a non-negative function $E(\rho) \geq 0$, which vanishes for all separable states. If $E(\rho) > 0$ for every entangled state, \emph{i.e.}, every non-separable state, the measure $E$ is called {\it faithful}.  

For a tripartite system ${\cal H} = {\cal H}_A \otimes {\cal H}_B \otimes {\cal H}_C$, the structure of separability becomes richer, since several inequivalent notions of separability may arise. First, the state $\rho$ is said to be \emph{fully separable} if it admits a convex decomposition
\be
\rho^{\rm fs} \,=\, \sum_{i} q_{i}\,
\rho^A_i \otimes \rho^B_i \otimes \rho^C_i,
\label{eq:fs}
\ee
with $q_i \ge 0$, $\sum_i q_i = 1$
and $\rho^I_i \in {\cal S}({\cal H}_I)$ for
$I = A,B,C$.
Fully separable states contain only classical correlations among the three subsystems, and  all entanglement measures vanish for such states. 

Beyond fully separable states, there is a set of states that are separable for a given bipartition $M | \overline M$. A state is called \emph{$M$-separable} if it admits a decomposition
\be
\rho^{M|\overline M} \,=\, \sum_{i} q_{i}\,
\rho^M_i \otimes \rho^{\overline M}_i\,,
\label{eq:M-sep}
\ee
with $q_i \ge 0$ and $\sum_i q_i = 1$.
In the tripartite system, there are three types of such states, $\rho^{A|BC}$, $\rho^{B|AC}$ and $\rho^{C|AB}$.
Below we will define entanglement measures that can probe $M|\overline M$ entanglement. 

\begin{figure}[t]
\centering
\includegraphics[width=0.52\textwidth]{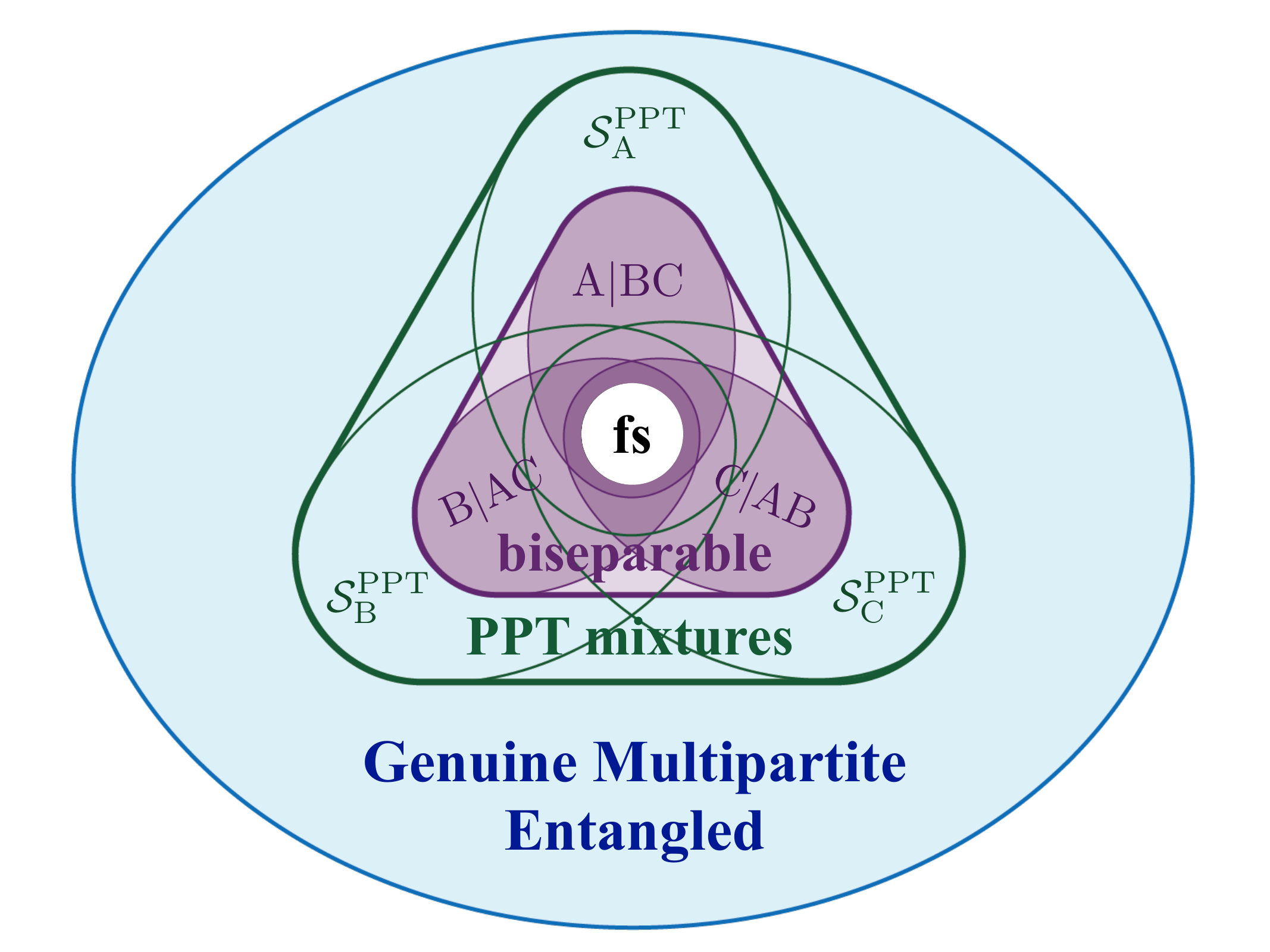}
\caption{\small Separability classification of tripartite quantum states (cf.\ \cite{Guhne:2008qic}).
The outer ellipse represents the set of all tripartite states.
The small disk in the centre, labelled `fs', is the set of fully separable states.
The purple rounded triangle is the set of bi-separable states, given by the convex hull of the three $M$-separable subsets ${\cal S}^{M|\overline M}$ ($M=A,B,C$), indicated by $A|BC$, $B|AC$ and $C|AB$.
States outside the purple region are {\it genuine multipartite entangled} (GME).
The green rounded triangle is the set of PPT mixtures, the convex hull of the three single-cut PPT subsets ${\cal S}^{\rm PPT}_M$.
Every bi-separable state is a PPT mixture, but the converse may fail: the region between the purple and green triangles consists of bound-entangled PPT mixtures, which exist whenever the local Hilbert-space dimensions are larger than $2\times 3$ for at least one cut.}
\label{fig:GME_diagram}
\end{figure}

A third notion of separability is bi-separability. 
A tripartite state $\rho$ is called \emph{bi-separable} if it can be expressed as a convex mixture of states that are separable across different bipartitions:
\be
\rho^{\rm bs} \,=\,
 q_1\, \rho^{A|BC} \,+\, q_2\, \rho^{B|AC} \,+\,
 q_3\, \rho^{C|AB}\,,
\label{eq:bisep}
\ee
with $q_i \geq 0$ and $q_1 + q_2 + q_3 = 1$.
A general mixed state may be bi-separable even if it is non-$M$-separable for all cuts, $M=A$, $B$ and $C$. 
A state that is not bi-separable is called {\it genuine multipartite entangled} (GME). 
\autoref{fig:GME_diagram} illustrates the different separability classes in the tripartite quantum system.

\medskip

Another fundamental property of entanglement is the monotonicity under the local operations and classical communication (LOCC). 
Under the LOCC protocol, distant observers can perform unitary transformations and measurements only on their own subsystem and exchange only classical information. Since LOCC cannot generate entanglement, no non-classical correlations can be created through such operations.
An entanglement measure $E$ is called an entanglement monotone if it is non-increasing under LOCC,
\be
E( \Lambda_{\rm LOCC}(\rho) )  \;\le\;  E(\rho)\,,
\label{eq:LOCC-monotone}
\ee
where $\Lambda_{\rm LOCC}(\rho)$ denotes the post-LOCC state of $\rho$.
This property gives the magnitude of $E$ an operational meaning: it imposes a partial ordering on quantum states according to their entanglement content and identifies maximally entangled states as those achieving the largest possible value of $E$ within a given Hilbert space.

Finally, an entanglement measure $E$ is said to be \emph{convex} if it does not increase under probabilistic mixing:
\be
E\left( \sum_i q_i \rho_i \right) \,\leq\, \sum_i q_i E(\rho_i)\,.
\label{eq:convex}
\ee
Convexity reflects the intuitive fact that classical uncertainty cannot increase quantum entanglement on average.

\medskip

\subsection{Partial Transpose and Peres--Horodecki Criterion}
\label{sec:PT}

While the definitions of separability introduced in \autoref{sec:sep} are conceptually fundamental, they are generally not practical for determining whether a given mixed state is entangled. In particular, checking separability requires establishing the existence of a convex decomposition of the form of Eqs.~\eqref{eq:sep}--\eqref{eq:bisep}, which is in general a highly non-trivial problem. For this reason, operational entanglement criteria that can be efficiently evaluated are of central importance. One of the most widely used criteria is the Peres--Horodecki criterion~\cite{Peres:1996dw,Horodecki:1997vt}, based on the partial transpose operation.

For a state $\rho$ acting on $\mathbb{C}^{d_{M}}\otimes\mathbb{C}^{d_{\overline M}}$, the partial transpose with respect to subsystem $M$ is defined by
\be
\left(
\rho^{T_M}
\right)_{i_M i_{\overline M},\, j_M j_{\overline M}}
\,=\,
\rho_{j_M i_{\overline M},\, i_M j_{\overline M}}\,,
\label{eq:partial-transpose}
\ee
If a state $\rho$ is $M$-separable, its partial transpose takes the form
\be
\rho^{T_M} \,=\, \sum_i q_i \, (\rho^M_i)^{T} \otimes \rho^{\overline M}_i\,.
\ee
with $q_i \geq 0$ and $\sum_i q_i = 1$.
Since $(\rho^M_i)^{T}$ is a positive semi-definite Hermitian matrix with a unit trace, the same holds for $\rho^{T_M}$. Hence,  the presence of at least one negative eigenvalue in $\rho^{T_M}$ immediately rules out $M$-separability and certifies entanglement across the bipartition $M|\overline M$~\cite{Peres:1996dw,Horodecki:1997vt}. A state whose partial transpose is positive semi-definite is called \emph{positive-partial-transpose} (PPT), while a state with at least one negative eigenvalue in $\rho^{T_M}$ is called \emph{negative-partial-transpose} (NPT).

Whether the converse statement holds depends on the dimensions of the bipartite system.
For $2\times2$ and $2\times3$ systems, the Peres--Horodecki criterion is both necessary and sufficient for separability, implying that every entangled state is necessarily NPT. For larger systems, however, there exist entangled states whose partial transpose remains positive semi-definite. Such states are known as PPT-entangled or bound-entangled states.

\subsection{Negativity}
\label{sec:neg}

Our primary variable to characterise the entanglement structure of the $t \bar t Z$ state is the {\it negativity}. 
For a state $\rho$ acting on $\mathbb{C}^{d_{M}}\otimes\mathbb{C}^{d_{\overline M}}$, 
the negativity is defined by~\cite{Vidal:2002zz}
\be
N_M(\rho) \,=\, \frac{|| \rho^{T_M} ||_1 - 1}{2},
\ee
where $|| {\cal O} ||_1 = {\rm Tr} \sqrt{ {\cal O}^\dagger {\cal O} }$ is the trace norm or the sum of the singular values of ${\cal O}$.
Since ${\rm Tr}\rho^{T_M} = 1$, $N_{M}$ can also be computed in terms of the eigenvalues $\lambda_i$ of $\rho^{T_M}$ as
\be
N_M(\rho) \,=\, 
\sum_i \frac{|\lambda_i| - \lambda_i}{2}
\,=\,
\sum_{\lambda_i < 0} |\lambda_i|\,.
\ee

By construction, it is clear that $N_M$ is non-negative, invariant under local basis changes, and positive iff $\rho$ is NPT with respect to $M$.
Since all $M$-separable states are PPT with respect to $M$, $N_M(\rho) = 0$ for all $M$-separable states.  
For $2 \times 2$ and $2 \times 3$ systems,
every entangled state is necessarily NPT,
and the negativity is therefore a faithful entanglement measure. For larger quantum systems, however, positivity of the partial transpose is no longer sufficient for separability, since PPT-entangled states may exist. Thus, the negativity is not a faithful entanglement measure in general.

The negativity has several other desired properties: 
it is convex, non-increasing under LOCC, and symmetric under $M \leftrightarrow \overline M$.
The general upper bound on $N_M$ for a $d_{M}\!\times\! d_{\overline M}$ system is
\be
N_M(\rho) \,\le\, \frac{d_{\min} - 1}{2}\,,
\qquad
d_{\min} \,\equiv\, \min(d_{M},d_{\overline M})\,.
\label{eq:neg-upper-bound}
\ee
The upper bound is saturated by a maximally entangled state of the form
 $\tfrac{1}{\sqrt{d_{\min}}}\sum_{i=1}^{d_{\min}}|i\rangle_{M}|i\rangle_{\overline M}$. 

Using the negativity, we measure two types of bipartite entanglement in the $t \bar t Z$ system: one-to-one entanglement and one-to-other entanglement.

\subsection{One-to-One Entanglement}
\label{sec:one-to-one}

For a tripartite quantum state $\rho$ describing a system of $ABC$, it is natural to ask how strongly two individual subsystems are entangled with each other. We call this type of entanglement {\it one-to-one entanglement}. 
To measure the entanglement between $A$ and $B$, one first traces out subsystem $C$ and constructs the {\it reduced density matrix} to describe the $AB$ subsystem 
\be
\rho_{AB} \,\equiv\, {\rm Tr}_C [ \rho ]\,.
\ee
The entanglement between $A$ and $B$ can then be quantified by the negativity $N_A(\rho_{AB})=N_B(\rho_{AB})$.

For a state $\rho$ of the $t \bar t Z$ spin system, one can define three reduced density matrices describing different pairs:
\be
\rho_{t \bar t} \,=\, {\rm Tr}_{Z}[\rho] \quad(2 \times 2),
\qquad
\rho_{t Z} \,=\, {\rm Tr}_{\bar t}[\rho] \quad(2 \times 3),
\qquad
\rho_{\bar t Z} \,=\, {\rm Tr}_{t}[\rho] \quad(2 \times 3),
\label{eq:reduced-states}
\ee
The parenthetical labels denote the Hilbert-space dimensions of the retained subsystems after the partial trace. Correspondingly, we define the {\it one-to-one negativities} as 
\be
N_{t \bar t} \,\equiv\, N_t(\rho_{t \bar t}),
\qquad
N_{t Z} \,\equiv\, N_Z(\rho_{t Z}),
\qquad
N_{\bar t Z} \,\equiv\, N_Z(\rho_{\bar t Z})\,.
\ee
From \autoref{eq:neg-upper-bound}, these quantities have the same range 
\be
({\rm separable})\quad 0 \,\leq\, N_{t \bar t},\,
N_{t Z},\,
N_{\bar t Z}
\,\leq\, \frac{1}{2} \quad ({\rm maximally~entangled})\,.
\label{eq:neg-bounds1}
\ee
Since the Peres--Horodecki criterion is a necessary and sufficient condition for entanglement in the $2 \times 2$ and $2 \times 3$ systems, 
these one-to-one negativities are faithful entanglement measures: they vanish if and only if the corresponding reduced state is separable.

\subsection{One-to-Other Entanglement}
\label{sec:one-to-other}

Another natural question for the $t \bar t Z$ state is whether each particle is entangled with the rest of the system. This type of entanglement is called {\it one-to-other entanglement}.
Taking the partial transpose with respect to the single subsystem $Z$, $t$ and $\bar t$, we define the following \emph{one-to-other negativities}
\ba
N(Z|t \bar t) \equiv N_Z(\rho) \;\; (3\times 4),
~~~
N(t|\bar t Z) \equiv N_t(\rho) \;\; (2\times 6),
~~~
N(\bar t|t Z) \equiv N_{\bar t}(\rho) \;\; (2\times 6)\,.
\nn \\
\label{eq:one-to-other}
\ea
The parentheses denote the dimensions of the two subsystems defined by the partition.
Hence, the corresponding negativities are not faithful entanglement measures.
The upper bound \autoref{eq:neg-upper-bound} gives the following ranges:
\begin{equation}
\text{(PPT)} \qquad
\begin{aligned}
&~~~~~~0\, \leq\, N(Z|t\bar{t}) \,\leq \,1& 
\\
&0\, \leq\, N(t|\bar{t}Z),\, N(\bar{t}|tZ) \leq \,\frac{1}{2}& 
\end{aligned}
\qquad \text{(maximally entangled)}
\label{eq:neg-bounds2}
\end{equation}
Since all $M$-separable states are PPT with respect to the bipartition $M|\overline M$, a non-vanishing negativity, $N_M(\rho)>0$, guarantees non-$M$-separability, certifying one-to-other entanglement for the corresponding bipartition.

\subsection{Genuine Multipartite Negativity}
\label{sec:gmn}

One of the central questions for the $t\bar t Z$ spin state is whether it exhibits quantum correlations that are genuinely tripartite in nature.
The genuine multipartite entanglement (GME) is defined as non-bi-separability (see \autoref{eq:bisep}).
Even if all one-to-other negativities are nonzero, the state can be bi-separable possessing no GME.  

In Ref.~\cite{Jungnitsch:2011izf} Jungnitsch, Moroder and G{\"u}hne (JMG) introduced a GME measure called the {\it genuine multipartite negativity} (GMN), which is applicable and calculable for arbitrary mixed states. 
To understand this measure, we first define {\it PPT mixtures}, which are the states that admit the decomposition  
\be
\rho^{\rm PPT\text{-}mix} \,=\,
p_A\, \rho^{\rm PPT}_A \, + \,
p_B\, \rho^{\rm PPT}_B \, + \,
p_C\, \rho^{\rm PPT}_C \, .
\ee
with $p_M \ge 0$ and $\sum_M p_M = 1$ ($M = A,B,C$).
In this decomposition, $\rho^{\rm PPT}_M \in {\cal S}^{\rm PPT}_M$, where ${\cal S}^{\rm PPT}_M$ is the set of all states that are PPT with respect to $M$. Since $M$-separable states are necessarily PPT with respect to $M$, 
$\rho^{M | \overline M} \in {\cal S}^{\rm PPT}_M$.
Thus, all bi-separable states in \autoref{eq:bisep}  belong to the set of all PPT mixtures ${\cal S}^{\rm PPT\text{-}mix}$.
The idea of JMG is to construct a tight witness that can detect non-PPT-mixtures, $\rho \notin {\cal S}^{\rm PPT\text{-}mix}$.
Such a state cannot be bi-separable and is therefore GME (see \autoref{fig:GME_diagram}).

The construction of Ref.~\cite{Jungnitsch:2011izf} is based on a particular class of entanglement witnesses called {\it fully decomposable witnesses}.
A Hermitian operator ${\cal W}$ is called fully decomposable if, for every bipartition $M = A, B, C$, it admits a decomposition
\be
{\cal W} \,=\, P_M \,+\, Q_M^{T_M}\,,
\qquad
0 \,\leq\, P_M\,,
\qquad
0 \,\leq\, Q_M \,\leq\, 1\,,
\label{eq:full-decomp}
\ee
for some Hermitian operators $P_M$ and $Q_M$.
The pairs $(P_M, Q_M)$ are allowed to be different for different $M$ but the operator ${\cal W}$ on the left-hand side is common to all $M$.

For any state $\rho^{\rm PPT}_M \in {\cal S}^{\rm PPT}_M$ and any ${\cal W}$ admitting the $M$-decomposition \eqref{eq:full-decomp}, we have
\be
{\rm Tr}[\rho^{\rm PPT}_M \,{\cal W}] \,=\, {\rm Tr}[\rho^{\rm PPT}_M \,P_M] \,+\, {\rm Tr}[(\rho^{\rm PPT}_M)^{T_M} \,Q_M] \,\geq\, 0\,,
\label{eq:Wpsd}
\ee
because the operators in both traces are positive semi-definite.
A fully decomposable ${\cal W}$ satisfies \eqref{eq:Wpsd} for {\it every} bipartition $M$, and therefore its expectation value on an arbitrary PPT mixture is also non-negative,
\be
{\rm Tr}[\rho^{\rm PPT\text{-}mix} \,{\cal W}] \,=\, p_A \,{\rm Tr}[\rho^{\rm PPT}_A \,{\cal W}] 
\,+\,
p_B \,{\rm Tr}[\rho^{\rm PPT}_B \,{\cal W}]
\,+\,
p_C \,{\rm Tr}[\rho^{\rm PPT}_C \,{\cal W}]
\,\geq\, 0\,,
\ee
where each term is bounded by applying \eqref{eq:Wpsd} with the $M$-decomposition of ${\cal W}$ matching that term.
Hence, finding a fully decomposable ${\cal W}$ with ${\rm Tr}[\rho \,{\cal W}] < 0$ certifies $\rho \notin {\cal S}^{\rm PPT\text{-}mix}$, which in turn implies that $\rho$ is GME.

To quantify how strongly a state violates the PPT mixture condition, we minimise ${\rm Tr}[\rho \,{\cal W}]$ over the set of all fully decomposable witnesses:
\be
\overline {\cal N}_G(\rho) \,\equiv\, \min_{\{P_M, Q_M\}} \,{\rm Tr}[\rho \,{\cal W}]\,,
\qquad
{\rm subject~to~\eqref{eq:full-decomp}~for~all}~M\,.
\label{eq:GMN-def}
\ee
The optimisation variables are the three Hermitian operator pairs $(P_M, Q_M)$ for $M = A, B, C$.
Note that ${\cal W} = 0$ (with $P_M = Q_M = 0$) is always feasible, so $\overline {\cal N}_G(\rho) \leq 0$ for any $\rho$.
The {\it genuine multipartite negativity} (GMN) is then defined as
\be
{\cal N}_G(\rho) \,\equiv\, -\,\overline {\cal N}_G(\rho) \,\geq\, 0\,.
\label{eq:GMN}
\ee

By construction, ${\cal N}_G(\rho) > 0$ certifies $\rho \notin {\cal S}^{\rm PPT\text{-}mix}$.
Remarkably, the converse statement also holds:
\be
\rho \notin {\cal S}^{\rm PPT\text{-}mix}
\quad \Longleftrightarrow \quad
{\cal N}_G(\rho) \,>\, 0\,.
\label{eq:GMN-iff}
\ee
Indeed, for any $\rho \notin {\cal S}^{\rm PPT\text{-}mix}$, the hyperplane separation theorem applied to the closed convex set ${\cal S}^{\rm PPT\text{-}mix}$ guarantees the existence of a Hermitian operator ${\cal W}$ satisfying ${\rm Tr}[\rho \,{\cal W}] < 0$ and ${\rm Tr}[\sigma \,{\cal W}] \geq 0$ for all $\sigma \in {\cal S}^{\rm PPT\text{-}mix}$.
For each $M$, any state PPT with respect to that single cut belongs trivially to ${\cal S}^{\rm PPT\text{-}mix}$, so ${\cal S}^{\rm PPT}_M \subseteq {\cal S}^{\rm PPT\text{-}mix}$ and the same operator ${\cal W}$ satisfies ${\rm Tr}[\sigma\,{\cal W}]\geq 0$ for all $\sigma \in {\cal S}^{\rm PPT}_M$.
The Horodecki characterisation of decomposable witnesses~\cite{Horodecki:1997vt} states that any Hermitian operator non-negative on ${\cal S}^{\rm PPT}_M$ admits a decomposition $P_M + Q_M^{T_M}$ with $P_M, Q_M \geq 0$.
Applying this characterisation separately for each $M=A,B,C$ then provides three (in general different) decomposition pairs $(P_M, Q_M)$ of the same ${\cal W}$, one per cut, showing that ${\cal W}$ is fully decomposable.
By a positive rescaling of ${\cal W}$, one can always make the largest eigenvalue of $Q_M$ less than or equal to $1$ for every $M$ simultaneously, ensuring $0 \le Q_M \le 1$, while preserving the sign of ${\rm Tr}[\rho\,{\cal W}]$.

\medskip

A few remarks are in order.
First, the set of PPT mixtures contains the set of bi-separable states, but the converse may not hold due to the existence of PPT-entangled states (in dimensions $d_M \times d_{\overline M}$ larger than $2 \times 3$).
A state with ${\cal N}_G(\rho) = 0$ is therefore a PPT mixture but is not guaranteed to be bi-separable, and the GMN provides only a sufficient (but not necessary) condition for GME.
Second, the optimisation in \eqref{eq:GMN-def} is a semi-definite program and can be solved efficiently with standard numerical routines \cite{Jungnitsch:2011izf,Hofmann:2014ywl}.
Third, the GMN inherits several desirable properties from the underlying witness construction \cite{Jungnitsch:2011izf,Hofmann:2014ywl}:
it is non-negative, convex, non-increasing under LOCC and invariant under local unitary transformations.

Finally, the GMN is bounded from above by the smallest one-to-other negativity,
\be
{\cal N}_G(\rho) \,\leq\, \min_M \, N(M | \overline M)\,.
\label{eq:GMN-NM-bound}
\ee
The argument is immediate from the definition: for any fully decomposable witness ${\cal W} = P_M + Q_M^{T_M}$, the constraints $P_M\geq 0$ and $0\leq Q_M\leq 1$ imply
\be
{\rm Tr}[\rho\,{\cal W}] \,=\, {\rm Tr}[\rho\,P_M] \,+\, {\rm Tr}[\rho^{T_M}\,Q_M]\,,
\ee
where the first term is non-negative because $P_M\geq 0$ and $\rho\geq 0$, and the second is bounded below by $-N_M(\rho)$ because $0\leq Q_M\leq 1$ and the most negative value of ${\rm Tr}[\rho^{T_M}\,Q_M]$ is $\sum_{\lambda_i<0}\lambda_i=-N_M(\rho)$, attained when $Q_M$ projects onto the negative eigenspace of $\rho^{T_M}$.
Minimising the left-hand side over all fully decomposable ${\cal W}$ gives ${\cal N}_G(\rho)\leq N_M(\rho)$ for each $M$ separately, and therefore \eqref{eq:GMN-NM-bound}.
The bound \eqref{eq:GMN-NM-bound} is generally strict for mixed states but is saturated on pure states, ${\cal N}_G(|\psi\rangle\langle\psi|) = \min_{M} N_M(|\psi\rangle\langle\psi|)$\,\cite{Jungnitsch:2011izf}.
Combining \eqref{eq:GMN-NM-bound} with the universal bound on the negativity \eqref{eq:neg-upper-bound}, we obtain the state-independent upper bound
\be
{\cal N}_G(\rho) \,\leq\, \frac{d_{\min} - 1}{2}\,,
\qquad
d_{\min} \,\equiv\, \min(d_A, d_B, d_C)\,.
\label{eq:GMN-bound}
\ee
This universal bound is saturated for GHZ-type maximally entangled states.

Applying these state-dependent and -independent bounds to the $t \bar t Z$ system, we have
\be
{\cal N}_G(\rho) \,\leq\, \min \Big[\,
N(Z|t \bar t),\, N(t|\bar t Z),\, N(\bar t|t Z)\,
\Big]\,,
\label{eq:GMN-bound-ttz}
\ee
and
\be
({\rm PPT\text{-}mixture})~~~0 \,\leq\, {\cal N}_G(\rho) \,\leq\, \frac{1}{2}
~~~({\rm maximally~ GME})
\,,
\ee
respectively.

\section{Entanglement in \texorpdfstring{$t \bar t Z$}{t t̄ Z}}
\label{sec:ttZ}

We now evaluate the entanglement measures of \autoref{sec:measures} on the spin quantum states $\rho$ and $\rho_{\Sigma}$ defined in \autoref{sec:states} for $e^{+}e^{-}\to t\bar t Z$ in the Standard Model.
Throughout this section we focus on unpolarised $e^+ e^-$ beams. The impact of beam polarisation is discussed in \autoref{sec:prospects}.
The numerical results are organised along two complementary levels of phase space (non)integration, which differ in how much of the kinematic information has been traced out:
\begin{itemize}
\item
{\em Differential} (a phase space point).
For a fixed choice of $(m_{t\bar t}, \,\cos\theta_{Z}, \,\cos\theta_{t}, \,\phi)$ the state $\rho$ is a cross-section-weighted sum of two pure states, $\rho \propto R^{+-} + R^{-+}$, originating from the $+-$ and $-+$ initial helicity configurations.
At this level, $\rho$ is a {\em genuine} quantum state and the entanglement measures are invariant under local unitary transformations and therefore independent of the choice of spin-quantisation axes.
We use this representation to explore the rich structure of the $t\bar t Z$ spin state across the four-dimensional phase space.

\item {\em Fully inclusive}. All four phase space variables are integrated against the squared matrix element, leaving the spin density matrix $\rho_{\Sigma}$ as a function of $\sqrt{s}$ only (for fixed beam polarisation). This is the most inclusive measurement, corresponding to a measurement that sums over all $t\bar t Z$ kinematic configurations. The averaging mixes density matrices defined with different spin-quantisation axes, so $\rho_{\Sigma}$ is a {\em fictitious} quantum state in the sense of \autoref{eq:fictitious-state}. Hence, the resulting entanglement measures depend on the choice of basis. Throughout this work we adopt the helicity basis.
\end{itemize}

\subsection{Entanglement in Differential States}
\label{sec:ttZ-diff}

We start by discussing the entanglement at a fixed phase space point.
The differential state $\rho$ depends on four kinematic variables, $(m_{t\bar t},\,\cos\theta_{Z},\,\cos\theta_{t},\,\phi)$.
Throughout this subsection we set $\sqrt{s}=1$~TeV and examine two representative parameter planes:
\begin{itemize}
\item[(i)]
The ($m_{t\bar t}$, $\cos\theta_{Z}$) plane, with the remaining variables fixed to $\cos\theta_{t} = \phi = 0$. 
\item[(ii)]
The ($\cos\theta_{Z}$, $\phi$) plane, with $m_{t\bar t}=600$~GeV and $\cos\theta_{t}=0$ fixed. 
\end{itemize}

We first discuss the purity of the full and the reduced spin states, which provides a simple measure of their mixedness. For a state $\rho$ of a $d$-dimensional quantum system, the purity is defined as $\gamma(\rho) \,\equiv\, {\rm Tr}(\rho^2)$ with
\be
({\rm maximally~mixed})~~\frac{1}{d} \,\leq\, \gamma \,\leq\, 1~~({\rm pure}).  
\ee
The lower bound $\frac{1}{d}$ is saturated only by the maximally mixed state, $\rho = \tfrac{1}{d} \,\mathbbm{1}$, for which all entanglement measures vanish. While purity does not, in general, quantify entanglement, it provides useful insight into the correlation structure of the state: a strong reduction in the purity of a reduced density matrix signals substantial correlations between the traced and retained subsystems. As we will see below, the regions with the lowest reduced-state purities also exhibit the largest multipartite entanglement.

\begin{figure}[t]
\centering
\includegraphics[width=0.33\textwidth]{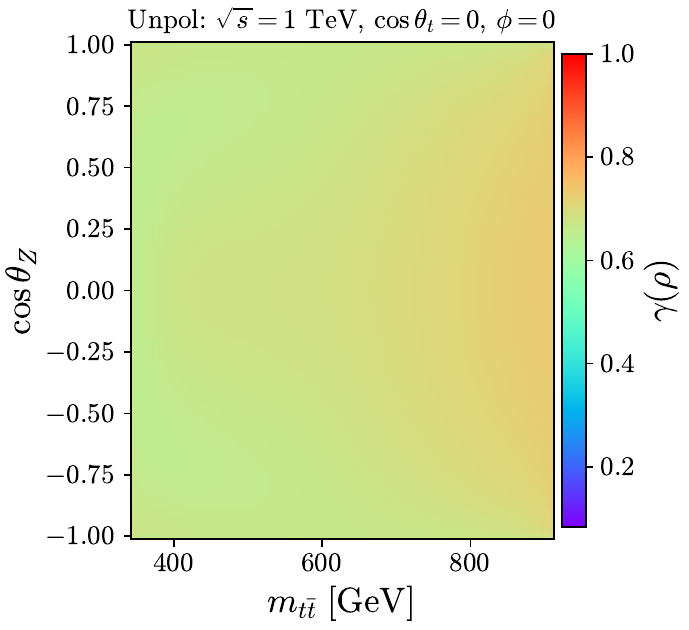}
\includegraphics[width=0.33\textwidth]{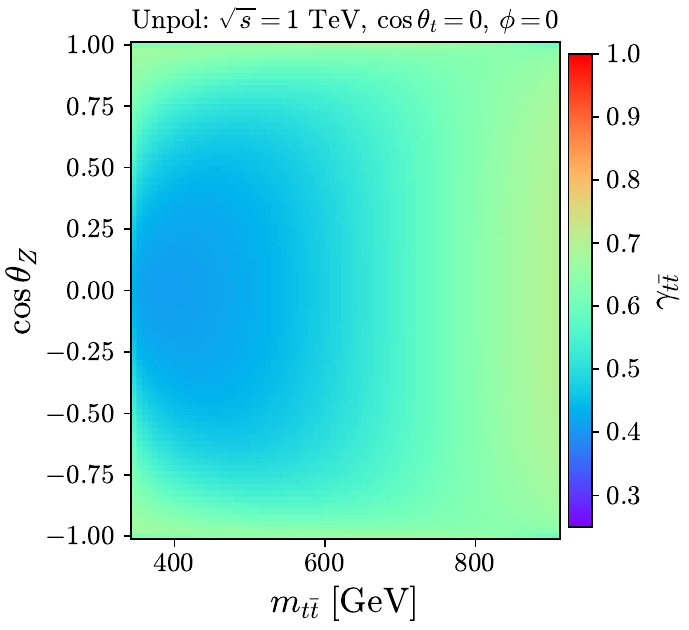}
\includegraphics[width=0.33\textwidth]{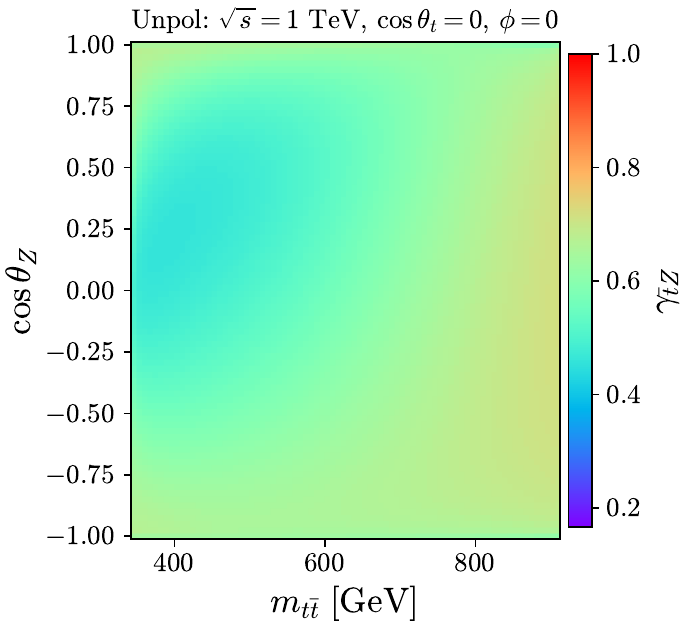}
\includegraphics[width=0.33\textwidth]{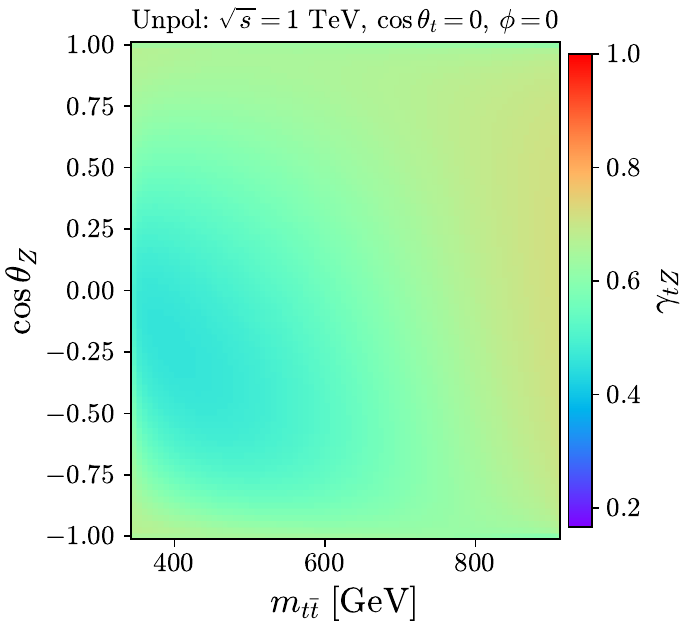}
\caption{\small Purity of the differential $t\bar t Z$ spin state at $\sqrt{s}=1~\mathrm{TeV}$ for $\cos\theta_t=\phi=0$, shown in the $(m_{t\bar t},\cos\theta_Z)$ plane for unpolarised beams. The panels display the purity of the full state, $\gamma(\rho)={\rm Tr}(\rho^2)$ (top left), and the purities of the reduced states $\gamma_{t\bar t}$ (top right), $\gamma_{tZ}$ (bottom left), and $\gamma_{\bar t Z}$ (bottom right), obtained by tracing out the $Z$, $\bar t$, and $t$ degrees of freedom, respectively. The lower limits of the colour scales correspond to the maximally mixed values $\frac{1}{12}$, $\frac{1}{4}$, $\frac{1}{6}$ and $\frac{1}{6}$ for the respective states.}
\label{fig:pure_slices}
\end{figure}

\autoref{fig:pure_slices} shows the purity of the differential $t\bar t Z$ spin state, $\gamma(\rho)$, and the three reduced purities $\gamma_{t\bar t} = {\rm Tr}(\rho_{t \bar t}^2)$, $\gamma_{tZ} = {\rm Tr}(\rho_{tZ}^2)$ and $\gamma_{\bar tZ} = {\rm Tr}(\rho_{\bar t Z}^2)$ in the $(m_{t\bar t},\cos\theta_{Z})$ plane.
The lower colour-bar limits correspond to the maximally mixed values $\frac{1}{12}$, $\frac{1}{4}$, $\frac{1}{6}$ and $\frac{1}{6}$.
The full state purity $\gamma(\rho)$ is uniformly high, $\sim 0.65\text{--}0.70$.
The only source of mixing here is the cross-section-weighted sum over the two initial helicities $+-$ and $-+$, whose amplitudes have comparable magnitudes across the plane, leaving the mixture close to an even superposition rather than dominated by one helicity. The reduced purities are significantly smaller, reflecting the correlations between the traced and retained subsystems. The most pronounced reduction occurs in $\gamma_{t\bar t}$, which reaches $\simeq 0.30\text{--}0.35$ around $\cos\theta_Z\simeq 0$ and $m_{t\bar t}\simeq 400\text{--}500$ GeV. This indicates strong correlations between the $t\bar t$ pair and the $Z$ boson, such that tracing out the latter leaves a highly mixed $t\bar t$ state. The  $\gamma_{tZ}$ and $\gamma_{\bar t Z}$ purities display shallower minima of approximately $0.4$ in different regions of phase space. The two are CP-conjugate, $\gamma_{tZ}(\cos\theta_Z)=\gamma_{\bar tZ}(-\cos\theta_Z)$, while $\gamma_{t\bar t}$ is CP-even and is therefore symmetric under $\cos\theta_Z\to-\cos\theta_Z$, in line with~\eqref{eq:CP-rule}.

\begin{figure}[t]
\centering
\includegraphics[width=0.3\textwidth]{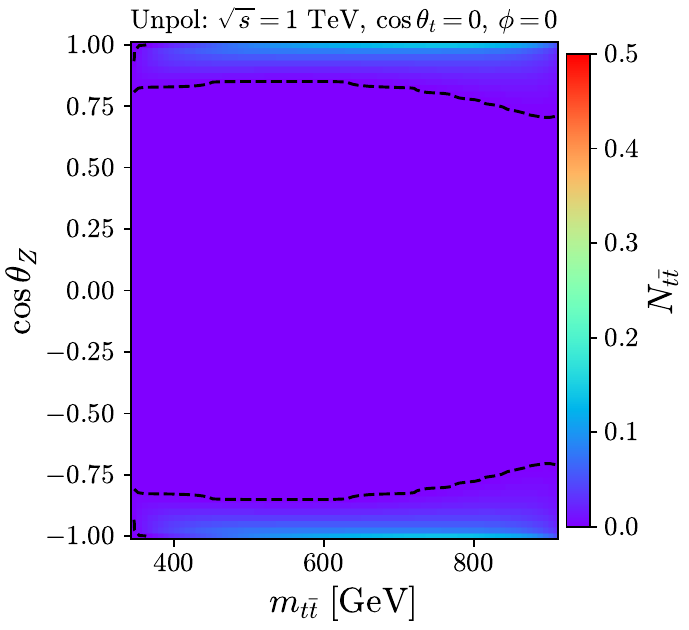}
\includegraphics[width=0.3\textwidth]{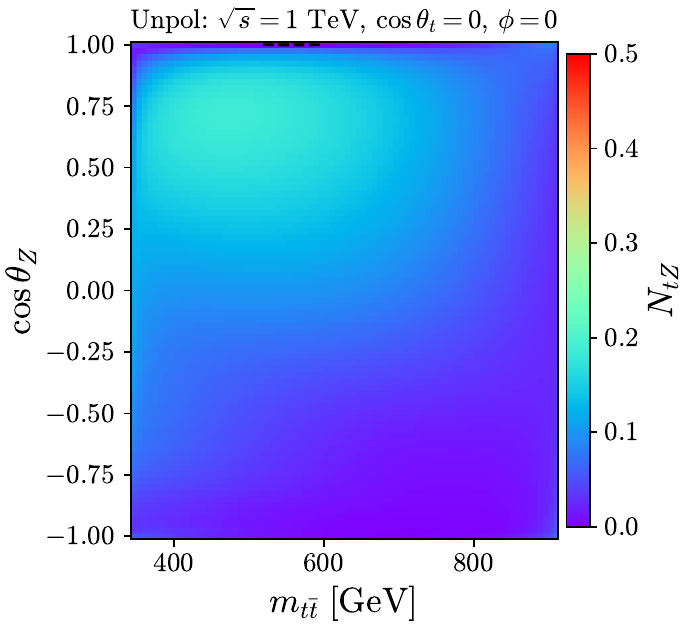}
\includegraphics[width=0.3\textwidth]{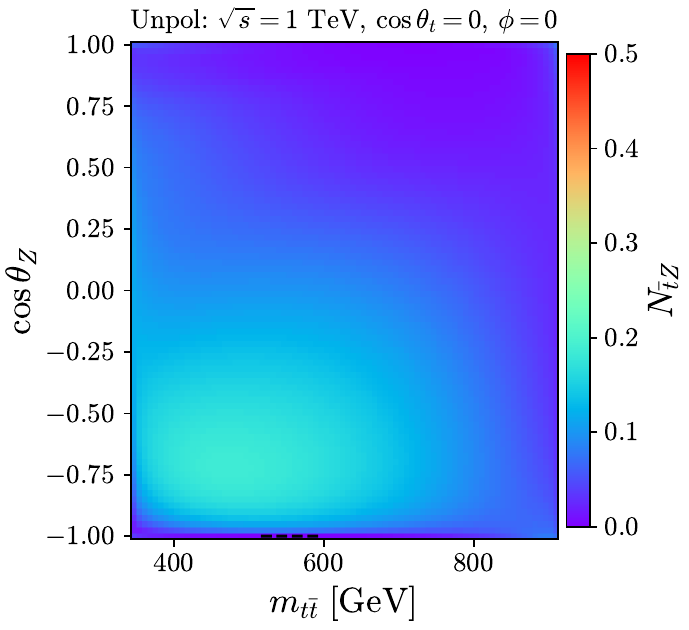}
\includegraphics[width=0.3\textwidth]{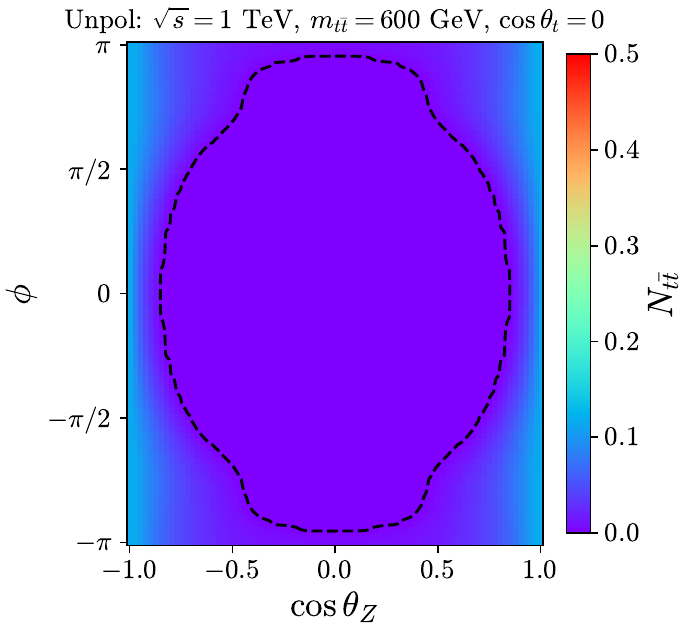}
\includegraphics[width=0.3\textwidth]{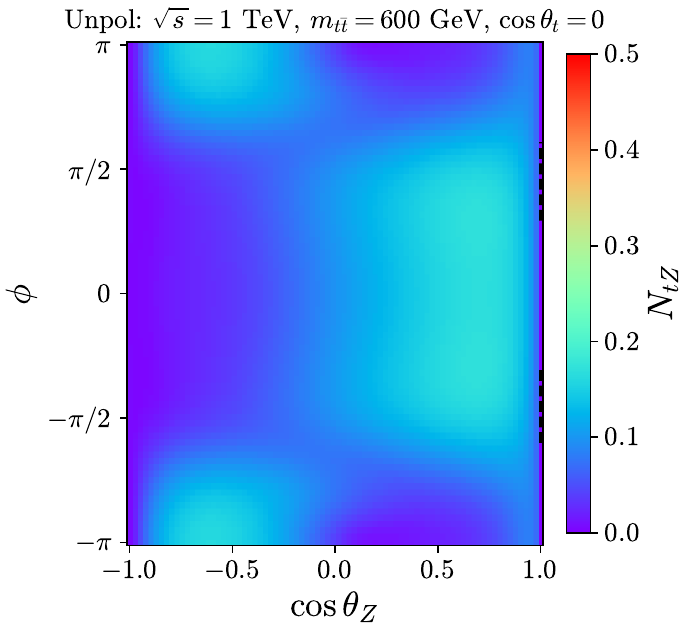}
\includegraphics[width=0.3\textwidth]{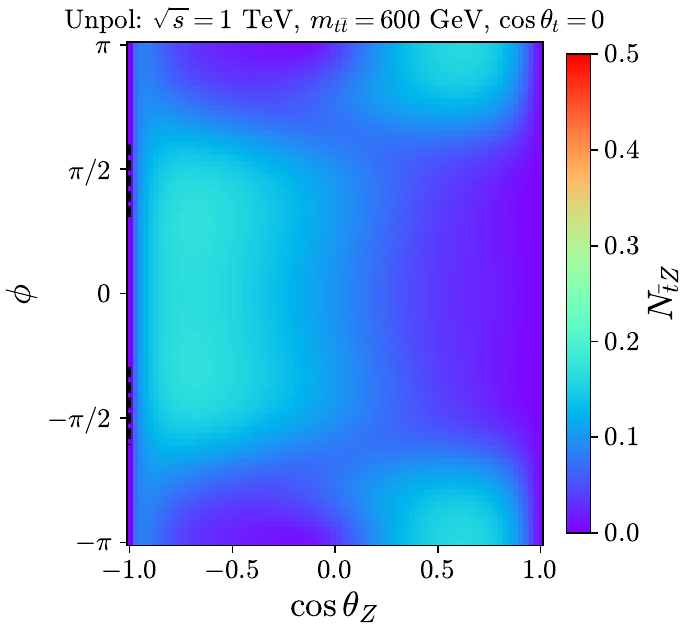}
\caption{\small One-to-one negativities of the differential $t\bar t Z$ spin state at $\sqrt{s}=1$~TeV for unpolarised beams: $N_{t\bar t}$ (left), $N_{tZ}$ (middle), and $N_{\bar tZ}$ (right).
\emph{Top row:} $(m_{t\bar t},\cos\theta_{Z})$ plane at fixed $\cos\theta_{t}=\phi=0$.
\emph{Bottom row:} $(\cos\theta_{Z},\phi)$ plane at fixed $m_{t\bar t}=600$~GeV and $\cos\theta_{t}=0$.
The dashed black contours mark the boundary at $N_{t\bar t}=10^{-3}$, below which $N_{t\bar t}$ is effectively vanishing.
For all panels, the colour bar saturates at the algebraic upper bound $N=1/2$.}
\label{fig:1-1_negative_slices}
\end{figure}

\autoref{fig:1-1_negative_slices} shows the three one-to-one negativities $N_{t\bar t}$, $N_{tZ}$, $N_{\bar tZ}$. The dashed contour in the $N_{t\bar t}$ panel marks $N_{t\bar t}=10^{-3}$, the threshold below which $N_{t\bar t}$ is treated as effectively vanishing. 
$N_{t\bar t}$ is nearly zero throughout most of phase space, becoming non-zero only near the soft-$Z$ boundary, $m_{t\bar t}\to\sqrt{s}-m_{Z}\simeq 909$~GeV, and for $|\cos\theta_Z|\to 1$. 
Both are limits in which the $Z$ effectively decouples from the $t\bar t$ pair: at the upper $m_{t\bar t}$ boundary the $Z$ momentum vanishes, and in the collinear limit the initial-state-radiation diagrams \autoref{fig:diagrams} (c,d) dominate, with the matrix element approximately factorising into a universal $e\to eZ$ splitting times the on-shell $e^{+}e^{-}\to t\bar t$ amplitude in the high energy limit. In both cases $\rho\to\rho_{t\bar t}\otimes\rho_{Z}$ and the $t\bar t$ reduced state roughly recovers the standard $e^{+}e^{-}\to t\bar t$ entanglement \cite{Altakach:2026fpl}.
This behaviour is closely related to the strong $t\bar t$--$Z$ correlations in the purity analysis: over most of phase space, tracing out the $Z$ leaves a highly mixed $t\bar t$ state, suppressing the entanglement in $\rho_{t\bar t}$. 
Only in regions where the $t\bar t$--$Z$ correlations weaken does the reduced $t\bar t$ state retain enough coherence to exhibit non-zero $t\bar t$ entanglement.
In contrast, $N_{tZ}$ and $N_{\bar tZ}$ are non-zero over large regions of phase space, reaching comparable maxima of approximately $0.25$, and are exact CP mirror images of each other under \eqref{eq:CP-rule}.

\begin{figure}[t]
\centering
\includegraphics[width=0.3\textwidth]{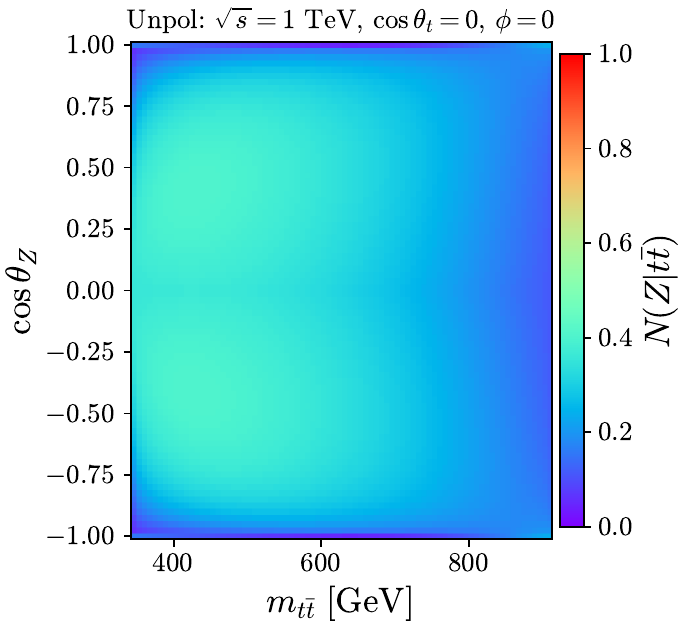}
\includegraphics[width=0.3\textwidth]{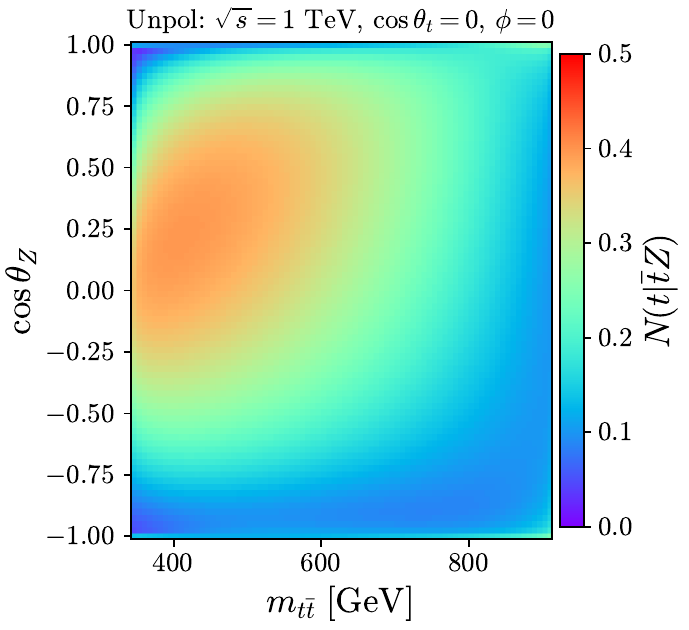}
\includegraphics[width=0.3\textwidth]{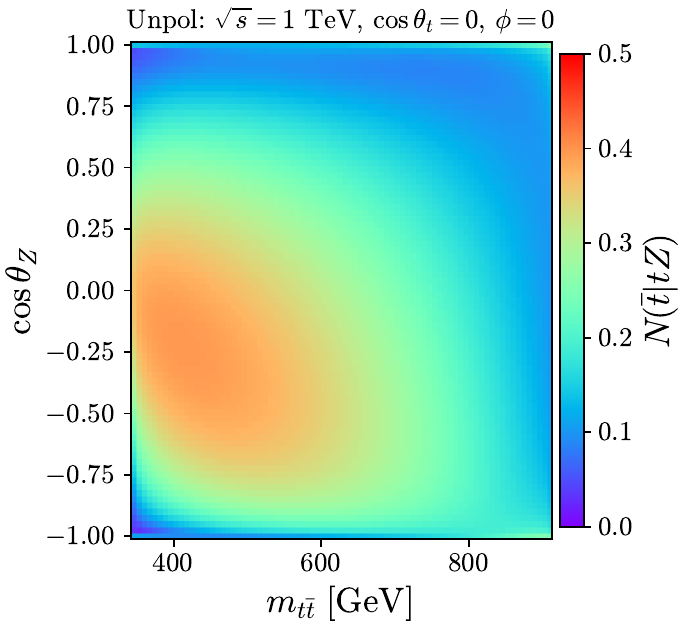}
\includegraphics[width=0.3\textwidth]{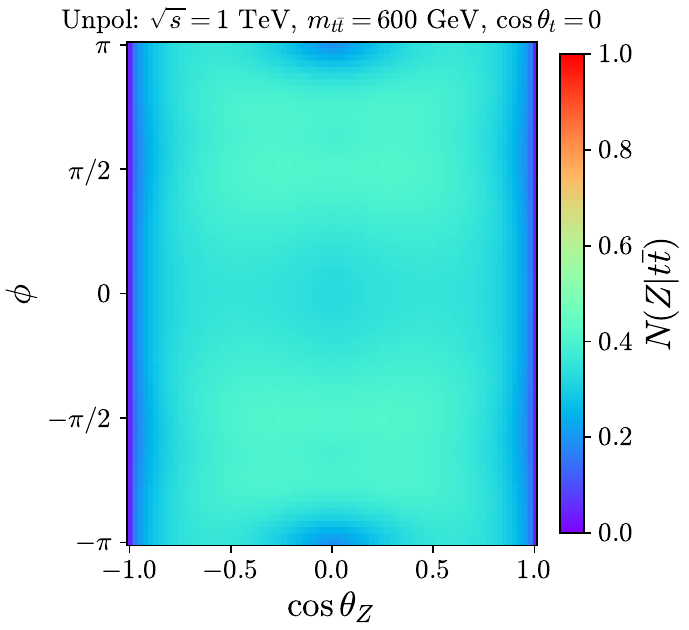}
\includegraphics[width=0.3\textwidth]{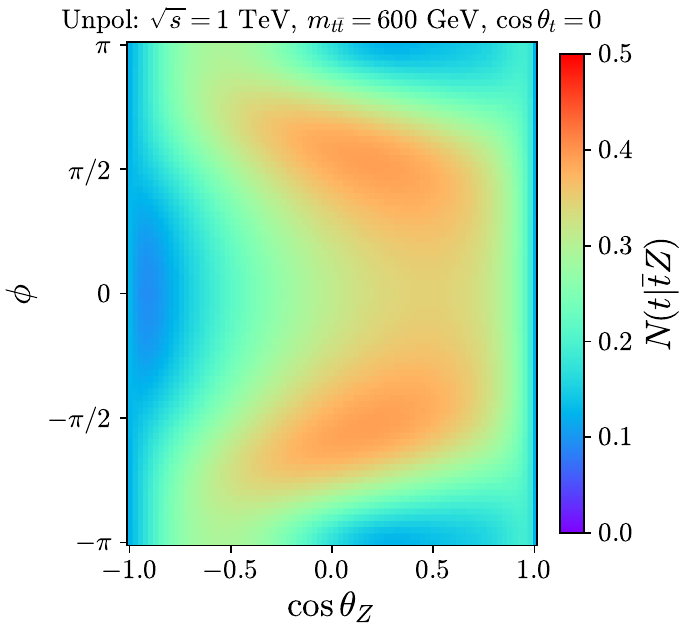}
\includegraphics[width=0.3\textwidth]{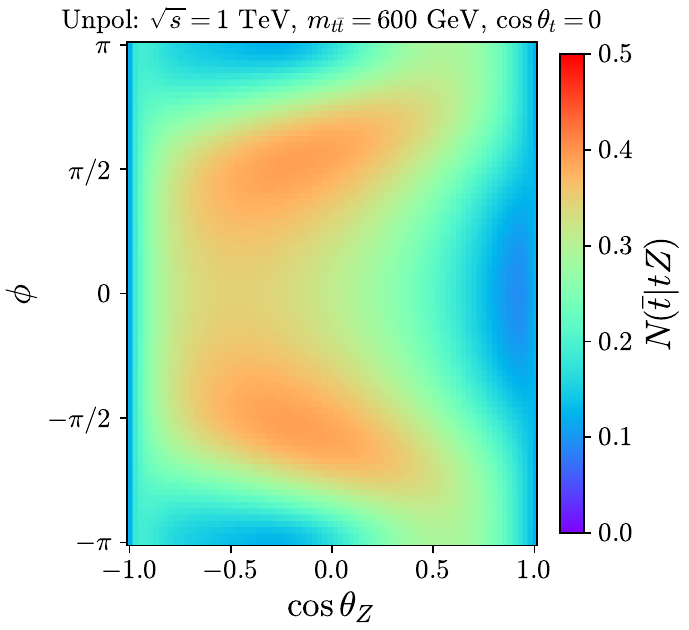}
\caption{\small One-to-other negativities of the differential $t\bar t Z$ spin state at $\sqrt{s}=1$~TeV for unpolarised beams: $N(Z|t\bar t)$ (left), $N(t|\bar tZ)$ (middle), and $N(\bar t|tZ)$ (right).
\emph{Top row:} $(m_{t\bar t},\cos\theta_{Z})$ plane at fixed $\cos\theta_{t}=\phi=0$.
\emph{Bottom row:} $(\cos\theta_{Z},\phi)$ plane at fixed $m_{t\bar t}=600$~GeV and $\cos\theta_{t}=0$.
The colour bar saturates at the algebraic upper bound, $1$ for the $3\times 4$ partition $N(Z|t\bar t)$ and $\frac{1}{2}$ for the $2\times 6$ partitions $N(t|\bar tZ)$ and $N(\bar t|tZ)$.}
\label{fig:1-2_negative_slices}
\end{figure}

\autoref{fig:1-2_negative_slices} shows the three one-to-other negativities $N(Z|t\bar t)$, $N(t|\bar tZ)$, $N(\bar t|tZ)$. 
All three are positive over essentially the entire $(m_{t\bar t},\cos\theta_{Z})$ plane, signalling robust entanglement across every bipartition. $N(Z|t\bar t)$ reaches $\sim 0.5\text{--}0.6$ in the central region $\cos\theta_{Z}\simeq 0$, $m_{t\bar t}\simeq 500\text{--}700$~GeV (about half of its algebraic maximum), while $N(t|\bar tZ)$ and $N(\bar t|tZ)$ reach $\simeq 0.40\text{--}0.45$ in broad central regions, approaching their algebraic bound $\tfrac{1}{2}$.\footnote{Recall that the algebraic upper bounds are $N(Z|t\bar t)\le 1$ for the $3\times4$ partition and $N(t|\bar t Z), N(\bar t|t Z)\le \frac{1}{2}$ for the two $2\times6$ partitions, as shown in \autoref{eq:neg-bounds2}.} As in \autoref{fig:1-1_negative_slices}, $N(Z|t\bar t)$ is CP-even and $N(t|\bar tZ)\leftrightarrow N(\bar t|tZ)$ under~\eqref{eq:CP-rule}. Combined with the $N_{t\bar t}\simeq 0$, this pattern points to a GHZ-like entanglement structure, in which each particle is strongly entangled with the remaining pair while the reduced $t\bar t$ state remains nearly separable. 

\begin{figure}[t]
\centering
\includegraphics[width=0.35\textwidth]{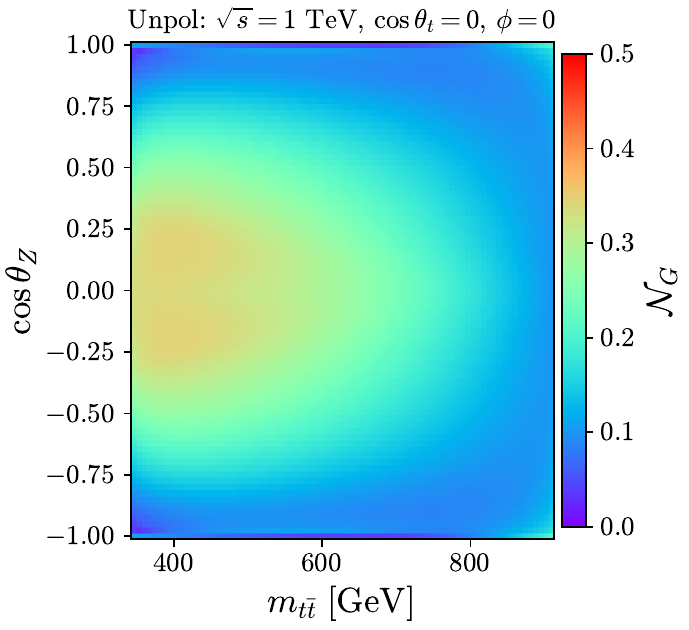}
\hspace{3mm}
\includegraphics[width=0.35\textwidth]{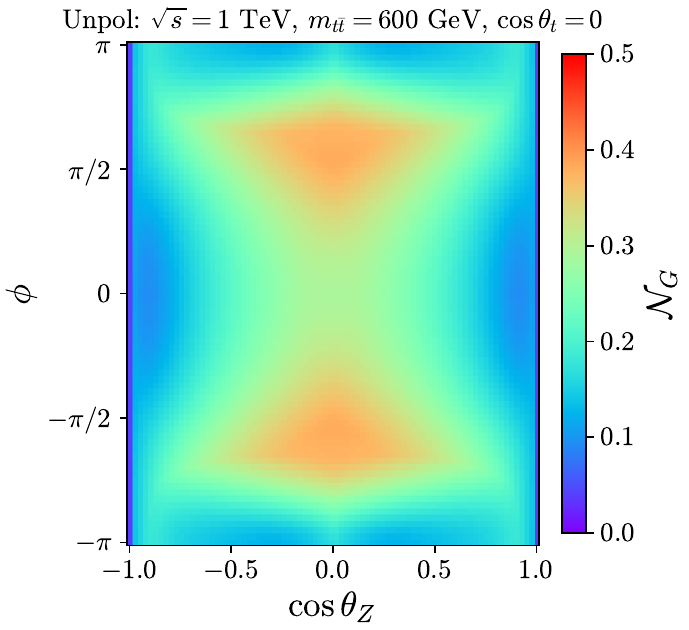}
\caption{\small Genuine multipartite negativity ${\cal N}_{G}$ of the differential $t\bar t Z$ spin state at $\sqrt{s}=1$~TeV for unpolarised beams.
\emph{Left:} $(m_{t\bar t},\cos\theta_{Z})$ plane at fixed $\cos\theta_{t}=\phi=0$.
\emph{Right:} $(\cos\theta_{Z},\phi)$ plane at fixed $m_{t\bar t}=600$~GeV and $\cos\theta_{t}=0$.
The colour bar saturates at the GHZ upper bound $(d_{\min}-1)/2=\frac{1}{2}$ for the $2\times 2\times 3$ Hilbert space.}
\label{fig:GMN_slices}
\end{figure}

\autoref{fig:GMN_slices} shows the genuine multipartite negativity ${\cal N}_G$. It is positive throughout most of the central region of phase space, reaching values of approximately $0.40$, around $80\%$ of the algebraic GHZ bound $(d_{\min}-1)/2=\frac{1}{2}$ for the $2\times 2\times 3$ Hilbert space. This unambiguously certifies genuine tripartite entanglement in the differential $t\bar t Z$ spin state over a large fraction of phase space -- a basis-independent statement. The largest values occur in the same kinematic region where the reduced-state purities are lowest and the one-to-other negativities are largest, confirming that the dominant correlations in the differential state are collective rather than pairwise. The state-dependent bound in \autoref{eq:GMN-NM-bound}, ${\cal N}_G(\rho)\leq\min_M N_M(\rho)$, is respected, but close to saturation. In the region where ${\cal N}_G$ reaches its maximum, the smallest one-to-other negativity is approximately $0.40\text{--}0.45$, only slightly larger than ${\cal N}_G$. ${\cal N}_G$ is CP-even, and the displayed slices are correspondingly symmetric under \eqref{eq:CP-rule}.

A coherent picture emerges from Figs.~\ref{fig:pure_slices}--\ref{fig:GMN_slices}. The differential $t\bar t Z$ spin state exhibits a clear hierarchy of
quantum correlations. Pairwise entanglement is generally suppressed,
especially within the $t\bar t$ subsystem, while one-to-other
entanglement remains large throughout most of phase space. The
positivity of ${\cal N}_G$ across broad kinematic regions further
demonstrates that these correlations are genuinely multipartite.
Overall, the dominant quantum structure of the differential state is
collective rather than pairwise.

\subsection{Entanglement in Fully Inclusive States}
\label{sec:ttZ-integrated}

\begin{figure}[t]
\centering
\includegraphics[width=0.45\textwidth]{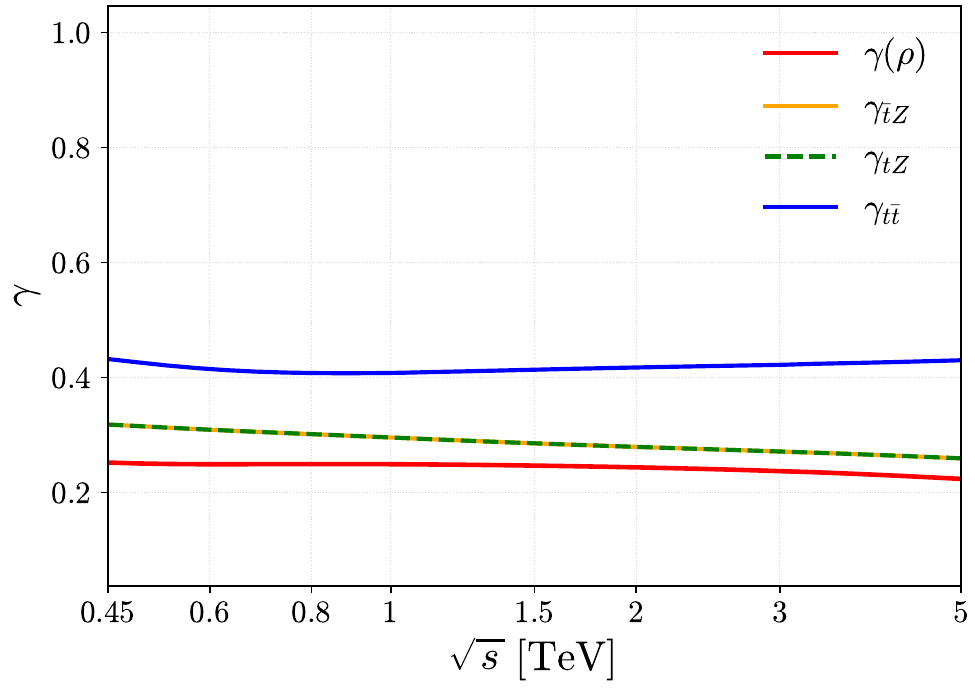}
\hspace{3mm}
\includegraphics[width=0.45\textwidth]{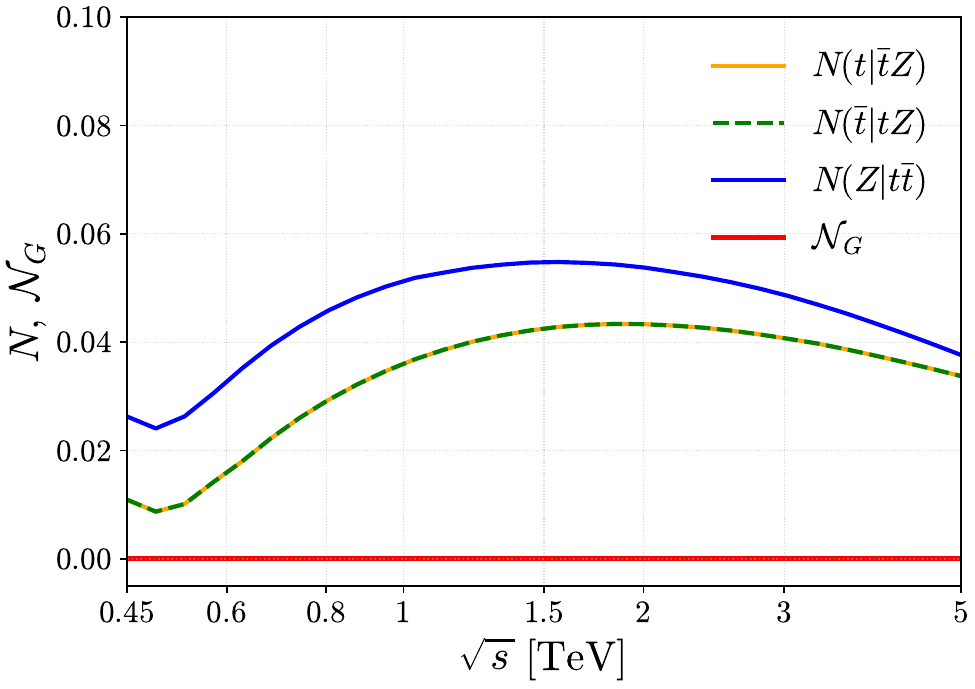}
\caption{\small Entanglement measures of the fully integrated fictitious $t\bar t Z$ spin state $\rho_{\Sigma}^{\rm unpol}$ in the helicity basis, as functions of the centre-of-mass energy $\sqrt{s}$ for unpolarised beams.
\emph{Left:} purity of the full state, $\gamma(\rho_{\Sigma})$ (red), and those of the three reduced states, $\gamma_{\bar tZ}$ (orange), $\gamma_{tZ}$ (green dashed) and $\gamma_{t\bar t}$ (blue).
The curves $\gamma_{tZ}$ and $\gamma_{\bar tZ}$ coincide as a consequence of CP invariance.
\emph{Right:} one-to-other negativities $N(t|\bar tZ)$ (orange), $N(\bar t|tZ)$ (green dashed), $N(Z|t\bar t)$ (blue), and the genuine multipartite negativity ${\cal N}_{G}$ (red).
The one-to-one negativities $N_{t\bar t}$, $N_{tZ}$ and $N_{\bar tZ}$ vanish identically across the whole range and are not shown.}
\label{fig:pure_GMN_full}
\end{figure}

We now turn to the fully inclusive state, obtained by integrating over the complete $t\bar t Z$ phase space. The resulting density matrix $\rho_{\Sigma}^{\rm unpol}$ is a fictitious quantum state (see~\autoref{sec:states}). The left panel of \autoref{fig:pure_GMN_full} shows the purity of the full state and of the three two-body reduced states. All purities depend only mildly on $\sqrt{s}$. The equality $\gamma_{tZ}=\gamma_{\bar tZ}$ follows from CP invariance after full phase space integration. The full state is the most mixed, with $\gamma(\rho_{\Sigma})\simeq0.22\text{--}0.25$,  while the reduced states
are less mixed, with
$\gamma_{t\bar t}\simeq0.41\text{--}0.43$
and
$\gamma_{tZ}=\gamma_{\bar tZ}\simeq0.26\text{--}0.32$.

For the one-to-one negativities, we find
\be
N_{t \bar t} \,=\, N_{t Z} \,=\, N_{\bar t Z} \,=\, 0 \qquad {\rm for~all}~\sqrt{s}\,.
\ee
Since the corresponding reduced states act on $2\times2$ and $2\times3$ Hilbert spaces, the Peres--Horodecki criterion is both necessary and sufficient for separability. Hence, the vanishing negativities imply that the $t\bar t$, $tZ$, and $\bar t Z$ systems are all separable once the full phase space is integrated over. 

The right panel of \autoref{fig:pure_GMN_full} shows the one-to-other negativities and the genuine multipartite negativity. The one-to-other negativities remain small but non-zero over the entire energy range. By CP invariance, $N(t|\bar tZ)=N(\bar t|tZ)$, reaching a maximum of approximately $0.043$ around $\sqrt{s}\sim2$~TeV, while $N(Z|t\bar t)$ peaks at approximately $0.055$ around $\sqrt{s}\sim1.5$~TeV. Thus, full phase space integration removes all pairwise entanglement but does not erase all collective entanglement across single-particle cuts. For the GMN, however, we obtain ${\cal N}_G = 0$ for all $\sqrt{s}$. The fully inclusive $\rho_{\Sigma}^{\rm unpol}$ is therefore a PPT mixture. Since PPT-entangled states may exist for all three bipartitions, ${\cal N}_G=0$ does not imply bi-separability. Rather, it shows that genuine multipartite entanglement is not detected by the PPT-mixture criterion.

\section{Prospects for Entanglement Observation at a Future $e^+ e^-$ Collider}
\label{sec:prospects}

In the previous section, we showed that the differential $t\bar t Z$ spin state exhibits a rich entanglement structure, including strong one-to-other entanglement and genuine multipartite entanglement over broad regions of phase space. These correlations are highly sensitive to phase space averaging. After full integration, the one-to-one negativities vanish, the surviving one-to-other negativities are strongly suppressed, and the GMN no longer certifies genuine multipartite entanglement. 
A realistic collider study cannot reconstruct the density matrix at a single phase space point.  The reconstruction of the spin density matrix from experimental data, known as \emph{quantum state tomography}, requires a sufficiently large sample of $t\bar{t}Z$ events from which the decay angular distributions of the $t$, $\bar t$ and $Z$ can be measured in their respective rest frames. The details of this procedure are presented in \autoref{app:tomography}. The central question is therefore whether one can identify kinematic regions that retain a substantial fraction of the differential entanglement while still providing enough events for its reconstruction and reliable certification.

In \autoref{sec:partial}, we analyse the partially inclusive spin state $\rho_{\Sigma}$ obtained by integrating over $\cos\theta_{t}$ and $\phi$, and use the resulting entanglement maps in the $(m_{t\bar t},\cos\theta_{Z})$ plane to define fiducial kinematic regions in which the most informative measures remain large.
In \autoref{sec:feesibility}, we then translate the fiducial entanglement values into the statistical significance and integrated luminosity needed to establish each measure as non-zero. Throughout the section we fix $\sqrt{s}=1$~TeV, near the maximum of the unpolarised $\sigma(e^{+}e^{-}\to t\bar t Z)$, and consider three benchmark beam polarisation configurations:
\be
\renewcommand{\arraystretch}{1.4}
\setlength{\arraycolsep}{12pt}
\begin{array}{c|ccc}
                                     & \text{Unpol} & \text{Pol}+   & \text{Pol}-    \\ \hline
(\mathcal{P}_{e_-},\mathcal{P}_{e_+}) & (0,\,0)      & (+0.8,\,-0.3) & (-0.8,\,+0.3)
\end{array}
\ee
We focus on measuring the three one-to-other negativities and
the genuine multipartite negativity, as they are the most informative quantities characterising the tripartite entanglement structure and the most promising to access experimentally, owing to their relatively large values in the differential state.

\subsection{Entanglement in Partially Inclusive States}
\label{sec:partial}

In this subsection we analyse the partially inclusive spin state $\rho_{\Sigma}$ obtained by integrating $\rho$ over $\cos\theta_{t}$ and $\phi$ with the matrix-element-squared weight.
This is the relevant state for an analysis in which the kinematic information used to reconstruct $\rho$ is the $t\bar t$ invariant mass and the polar angle of the $Z$ in the lab frame, while the orientation of $t$ in the $t\bar t$ rest frame is summed over.
We work in the helicity basis throughout.
We start with the unpolarised configuration. The results for the Pol$+$ and Pol$-$ benchmarks, which are qualitatively similar, are collected in \autoref{app:pol}.

\begin{figure}[t]
\centering
\includegraphics[width=0.32\textwidth]{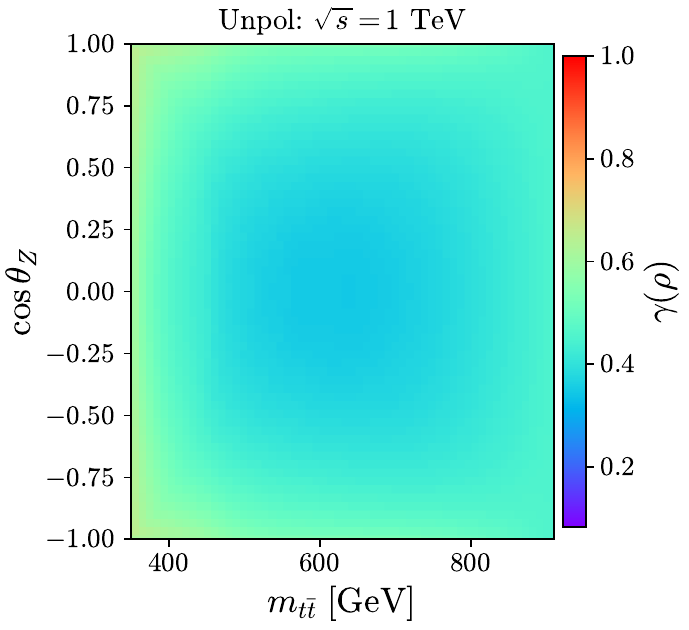}
\includegraphics[width=0.32\textwidth]{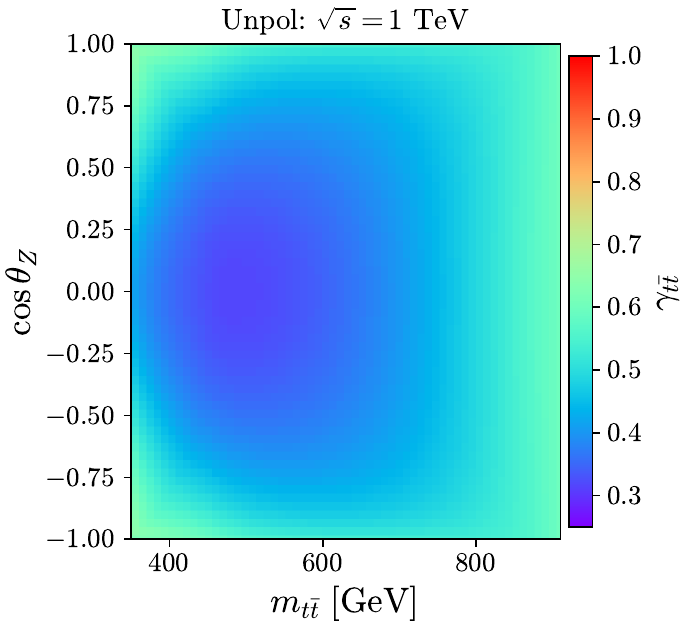}
\includegraphics[width=0.32\textwidth]{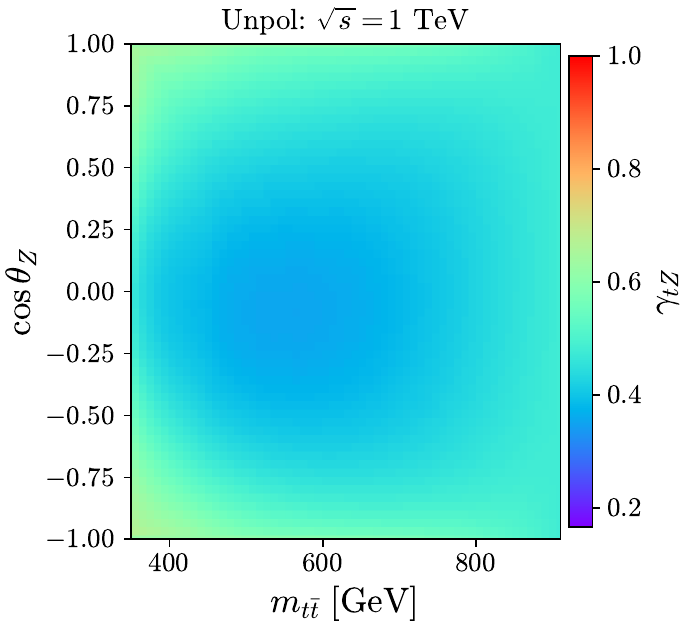}
\caption{\small Purity of the partially inclusive fictitious $t\bar t Z$ spin state $\rho_{\Sigma}$ at $\sqrt{s}=1$~TeV for unpolarised beams, in the $(m_{t\bar t},\cos\theta_{Z})$ plane after integration over $\cos\theta_{t}$ and $\phi$ in the helicity basis.
\emph{Left:} purity of the full state, $\gamma(\rho_{\Sigma})$.
\emph{Middle:} purity of the $t\bar t$ reduced state, $\gamma_{t\bar t}$.
\emph{Right:} purity of the $tZ$ reduced state, $\gamma_{tZ}$, which equals $\gamma_{\bar tZ}$ by CP invariance.}
\label{fig:pure_partial}
\end{figure}

\medskip

\autoref{fig:pure_partial} shows the purity of the full state $\gamma(\rho_{\Sigma})$ (left) and those of the reduced states $\gamma_{t\bar t}$ (middle) and $\gamma_{tZ}=\gamma_{\bar tZ}$ (right).
The exact equality $\gamma_{tZ}=\gamma_{\bar tZ}$ holds point by point across the plane due to the CP invariance of the neutral interaction. 
All three purities are markedly smaller than in the differential case, reflecting the dilution of the spin information by the phase space averaging.
$\gamma(\rho_{\Sigma})$ drops to $\sim 0.35$ in the central region $\cos\theta_{Z}\simeq 0$, $m_{t\bar t}\simeq 600$~GeV and rises towards $\sim 0.6$ near the corners of the plane, where the available solid angle in $(\cos\theta_{t},\phi)$ shrinks and the integration mixes a narrower set of orientations.
$\gamma_{t\bar t}$ follows the same overall shape with a minimum of $\sim 0.25$, close to the lower bound $\frac{1}{4}$, signalling that the $t\bar t$ reduced state is nearly maximally mixed in this region.
$\gamma_{tZ}=\gamma_{\bar tZ}$ takes intermediate values $\sim 0.35\text{--}0.55$.

\begin{figure}[t]
\centering
\includegraphics[width=0.32\textwidth]{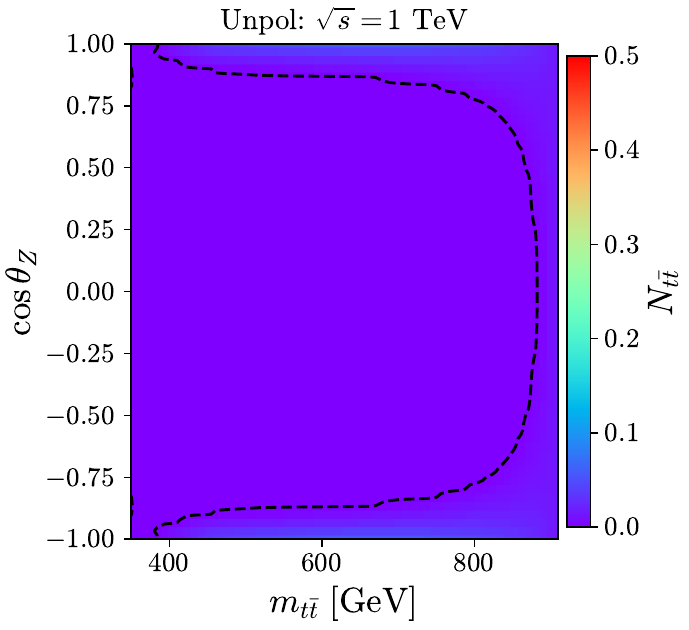}
\includegraphics[width=0.32\textwidth]{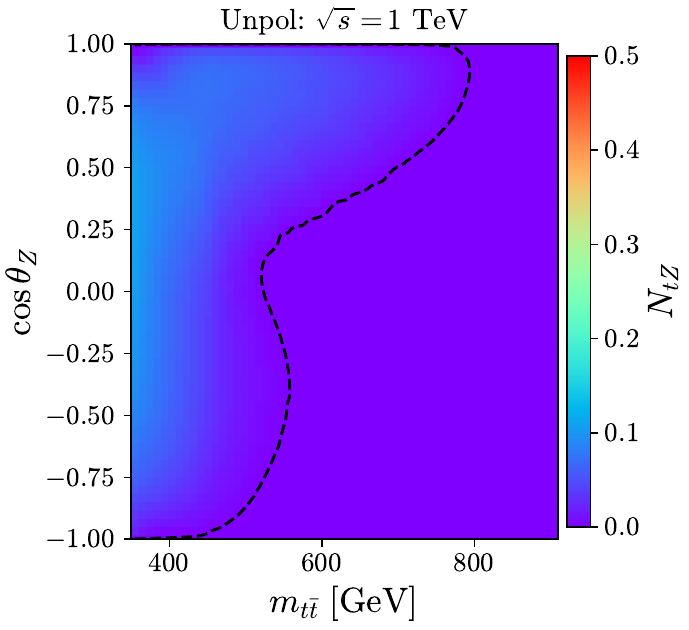}
\includegraphics[width=0.32\textwidth]{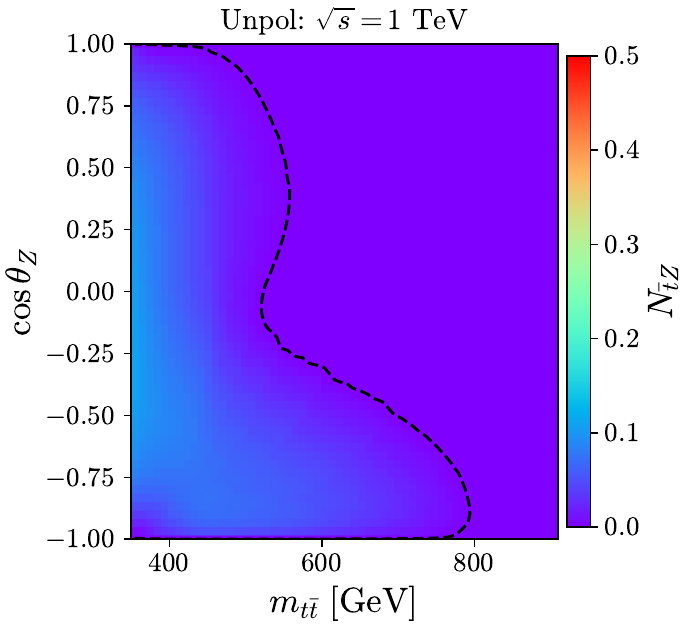}
\caption{\small One-to-one negativities of the partially inclusive $t\bar t Z$ spin state $\rho_{\Sigma}$ at $\sqrt{s}=1$~TeV for unpolarised beams, in the $(m_{t\bar t},\cos\theta_{Z})$ plane after integration over $\cos\theta_{t}$ and $\phi$ in the helicity basis: $N_{t\bar t}$ (left), $N_{tZ}$ (middle), and $N_{\bar tZ}$ (right).
The dashed black contours mark the threshold $N=10^{-3}$, below which each measure is treated as effectively vanishing.
The colour bar saturates at the algebraic upper bound $\frac{1}{2}$.}
\label{fig:1-1_negativity_partial}
\end{figure}

\medskip

\autoref{fig:1-1_negativity_partial} shows the three one-to-one negativities $N_{t\bar t}$ (left), $N_{tZ}$ (middle) and $N_{\bar tZ}$ (right) for unpolarised beams. The corresponding results for Pol$+$ and Pol$-$ are presented in \autoref{fig:1-1_negativity_partial_pol} and display a similar behaviour.
The dashed black contours mark the threshold $N=10^{-3}$, below which each measure is treated as effectively vanishing. 
We can see the relation $N_{tZ}(m_{t\bar t},\cos\theta_{Z}) = N_{\bar tZ}(m_{t\bar t},-\cos\theta_{Z})$, anticipated by the CP invariance of the interaction. Compared with the differential results of \autoref{sec:ttZ-diff}, all three one-to-one negativities are strongly depleted by phase space averaging. The $t\bar t$ negativity survives only in a narrow region near the soft-$Z$ and collinear boundaries, where the $t\bar t$ subsystem approximately decouples from the $Z$ boson and recovers the familiar entanglement structure of $e^{+}e^{-}\to t\bar t$. The negativities $N_{tZ}$ and $N_{\bar tZ}$ remain non-zero only at low invariant masses, $m_{t\bar t}\lesssim 550$~GeV, with peak values below $0.10$. Pairwise entanglement is therefore absent over a large fraction of the $(m_{t\bar t},\cos\theta_Z)$ plane, with all three two-body reduced states classified as separable by the negativity criterion.

\begin{figure}[t]
\centering
\includegraphics[width=0.32\textwidth]{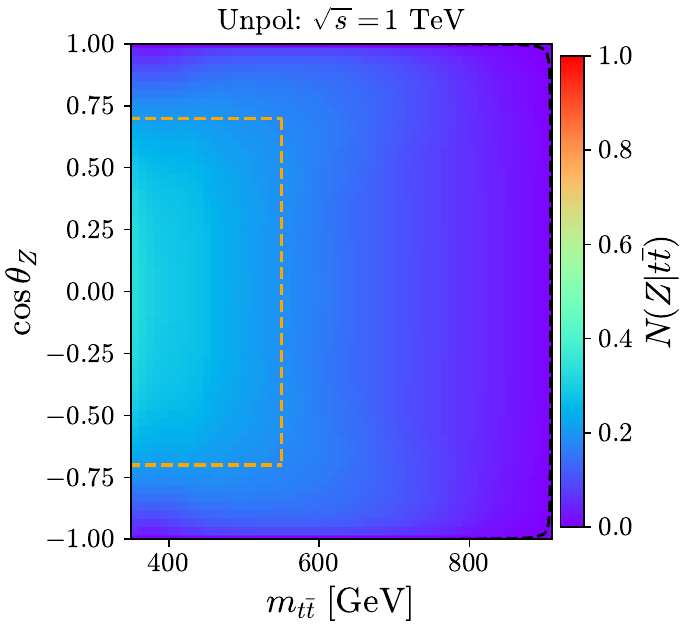}
\includegraphics[width=0.32\textwidth]{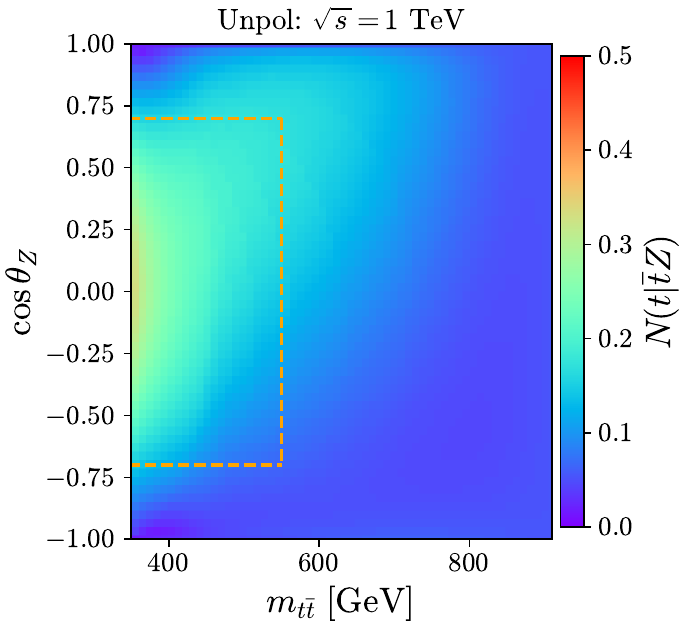}
\includegraphics[width=0.32\textwidth]{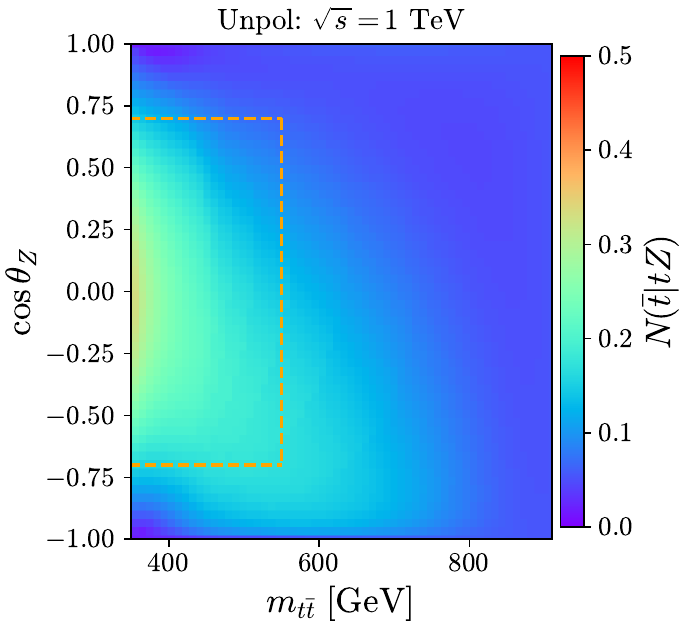}
\caption{\small One-to-other negativities of the partially inclusive $t\bar t Z$ spin state $\rho_{\Sigma}$ at $\sqrt{s}=1$~TeV for unpolarised beams, in the $(m_{t\bar t},\cos\theta_{Z})$ plane after integration over $\cos\theta_{t}$ and $\phi$ in the helicity basis: $N(Z|t\bar t)$ (left), $N(t|\bar tZ)$ (middle), and $N(\bar t|tZ)$ (right).
The orange dashed lines mark the fiducial region $\Sigma_{1|2}$ defined in \autoref{eq:sig_12}.
The colour bar saturates at the algebraic upper bound, $1$ for $N(Z|t\bar t)$ and $\frac{1}{2}$ for $N(t|\bar tZ)$ and $N(\bar t|tZ)$.}
\label{fig:1-2_negativity_partial}
\end{figure}

\medskip

\autoref{fig:1-2_negativity_partial} shows the three one-to-other negativities $N(Z|t\bar t)$ (left), $N(t|\bar tZ)$ (middle), and $N(\bar t|tZ)$ (right) for unpolarised beams. The corresponding results for Pol$+$ and Pol$-$ are shown in \autoref{fig:1-2_negativity_partial_pol} and display essentially the same shape and magnitude.
In contrast to the one-to-one negativities, all three measures are positive throughout an extended region of the plane.
$N(Z|t\bar t)$ remains of order $0.1\text{--}0.3$ over the entire central region of $(m_{t\bar t},\cos\theta_{Z})$, signalling that even after the angular integration the $Z$ helicity is non-trivially entangled with the $t\bar t$ pair taken as a whole.
$N(t|\bar tZ)$ and $N(\bar t|tZ)$ are mirror images of each other under $\cos\theta_{Z}\to -\cos\theta_{Z}$ (again by CP) and peak near the $t\bar t$ threshold, $m_{t\bar t}\simeq 2m_{t}$, where they reach ${\sim}0.3\text{--}0.4$, close to their algebraic maximum $\frac{1}{2}$.
For $m_{t\bar t}\gtrsim 700$~GeV both negativities fall rapidly and become very small.
The hierarchy among the three one-to-other measures suggests defining a fiducial region in which all three negativities are simultaneously large.
A simple rectangular choice is
\be
\Sigma_{1|2}\,:\quad |\cos\theta_{Z}| < 0.7
~~\text{and}~~ m_{t\bar t} < 550~{\rm GeV}\,,
\label{eq:sig_12}
\ee
indicated by the orange dashed contour in \autoref{fig:1-2_negativity_partial}. We will use $\Sigma_{1|2}$ in \autoref{sec:feesibility} as the fiducial region for the one-to-other entanglement observation.

\begin{figure}[t]
\centering
\includegraphics[width=0.32\textwidth]{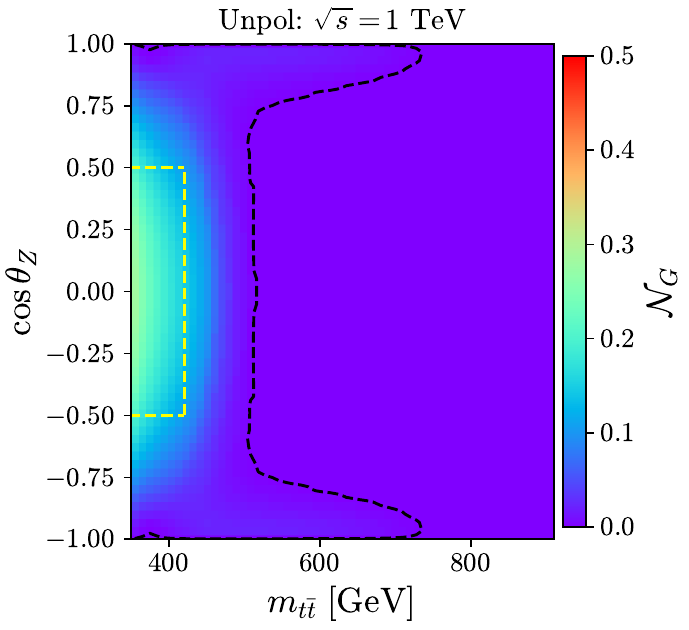}
\includegraphics[width=0.32\textwidth]{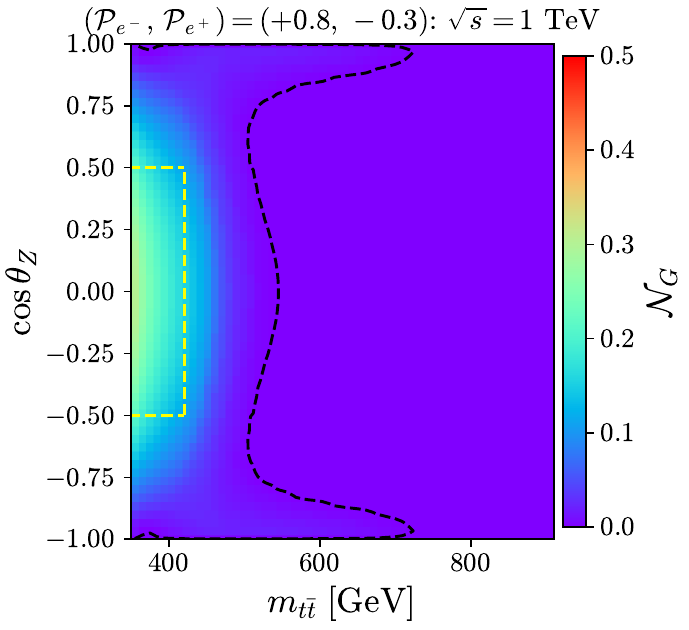}
\includegraphics[width=0.32\textwidth]{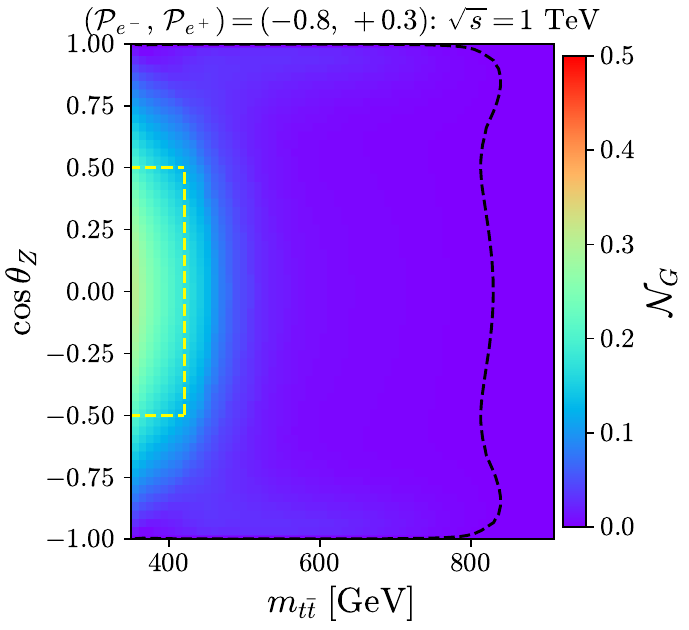}
\caption{\small Genuine multipartite negativity ${\cal N}_{G}$ of the partially inclusive $t\bar t Z$ spin state $\rho_{\Sigma}$ at $\sqrt{s}=1$~TeV, in the $(m_{t\bar t},\cos\theta_{Z})$ plane in the helicity basis, for the three beam polarisation configurations: Unpol (left), Pol$+$ (middle), and Pol$-$ (right).
The dashed black contours mark the threshold ${\cal N}_{G}=10^{-3}$, below which the GMN is treated as effectively vanishing; the yellow dashed contour marks the fiducial region $\Sigma_{\rm GME}$ defined in \autoref{eq:sig_GME}.}
\label{fig:GMN_beam}
\end{figure}

\medskip

\autoref{fig:GMN_beam} shows the genuine multipartite negativity ${\cal N}_{G}$ of $\rho_{\Sigma}$ for the three beam configurations.
The pattern is essentially the same in all three: ${\cal N}_{G}>0$ in a contiguous region near threshold, $m_{t\bar t}\lesssim 500$~GeV, at central $\cos\theta_{Z}$, where it reaches peak values $\sim 0.25$ --- halfway to the GHZ-like upper bound $\frac{1}{2}$ and certifying GME of $\rho_{\Sigma}$ in this region.
Narrow strips of ${\cal N}_{G}>0$ also appear near the collinear-$Z$ edges $|\cos\theta_{Z}|\to 1$, paralleling the $N_{t\bar t}>0$ regions of \autoref{fig:1-1_negativity_partial}. For Pol$-$, the positive ${\cal N}_{G}$ region extends to $m_{t\bar t}\simeq 700\text{--}850$~GeV.
Outside these regions ${\cal N}_{G}$ vanishes within numerical precision, so the corresponding $\rho_{\Sigma}$ is a PPT mixture. Because PPT mixtures in $2\times 6$ subsystems can still be entangled, this does not strictly exclude bi-separability, but it does mean that the GMN measure can no longer certify GME from the data. The GMN profile is more sharply localised in $(m_{t\bar t},\cos\theta_{Z})$ than the one-to-other negativities, so a smaller fiducial region is appropriate:
\be
\Sigma_{\rm GME}\,:\quad |\cos\theta_{Z}| < 0.5
~~\text{and}~~ m_{t\bar t} < 420~{\rm GeV}\,,
\label{eq:sig_GME}
\ee
shown as the yellow dashed contour in \autoref{fig:GMN_beam}.

\subsection{Feasibility for Observing Tripartite Entanglement}
\label{sec:feesibility}

The fiducial regions $\Sigma_{1|2}$ and $\Sigma_{\rm GME}$ defined in \autoref{sec:partial} were tailored to maximise the entanglement signal at the price of a restricted phase space volume.
In this subsection we quantify the resulting trade-off between signal strength and event rate, and translate it into the integrated luminosity required to observe each entanglement measure at a future $e^{+}e^{-}$ collider.

\begin{figure}[t]
\centering
\includegraphics[width=0.5\textwidth]{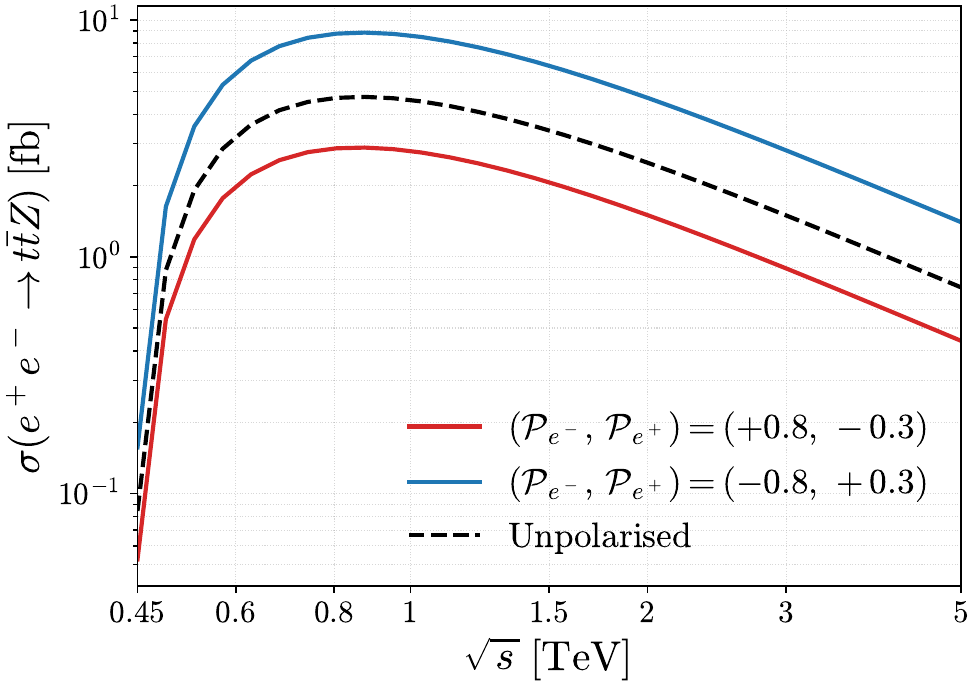}
\caption{\small Tree-level Standard Model cross section for $e^{+}e^{-}\to t\bar t Z$ as a function of the centre-of-mass energy $\sqrt{s}$, for the three beam polarisation configurations Pol$+$ with $({\cal P}_{e^{-}},{\cal P}_{e^{+}})=(+0.8,-0.3)$ (red), Pol$-$ with $(-0.8,+0.3)$ (blue), and unpolarised (black dashed).}
\label{fig:xsec}
\end{figure}

\autoref{fig:xsec} shows the leading-order Standard Model cross section $\sigma(e^{+}e^{-}\to t\bar t Z)$ as a function of the centre-of-mass energy, for the three beam polarisation configurations.
The cross section rises rapidly above the kinematic threshold $\sqrt{s}\gtrsim 2m_{t}+m_{Z}\simeq 0.44$~TeV, reaches a maximum around $\sqrt{s}\simeq 0.8\text{--}0.9$~TeV, with peak values of approximately $9$, $5$, and $3$~fb for Pol$-$, Unpol, and Pol$+$, respectively, and then decreases at higher energies.
The enhancement of the Pol$-$ configuration originates from the chiral electroweak couplings of the electron, which favour left-handed electrons and right-handed positrons in the dominant production amplitudes. As a result, Pol$-$ configuration offers the largest event sample for a given integrated luminosity and is the natural choice when statistics dominates the measurement. In the following, we fix $\sqrt{s}=1$~TeV for each beam-polarisation scenario. This energy lies close to the maximum of the corresponding $e^+e^- \to t\bar t Z$ production cross section and is a benchmark energy for future high-energy lepton colliders.

\begin{figure}[t]
\centering
\includegraphics[width=0.32\textwidth]{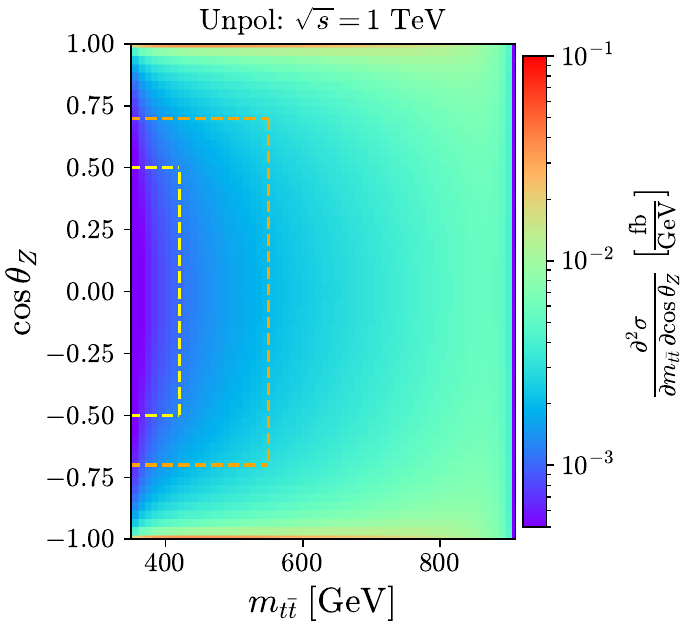}
\includegraphics[width=0.32\textwidth]{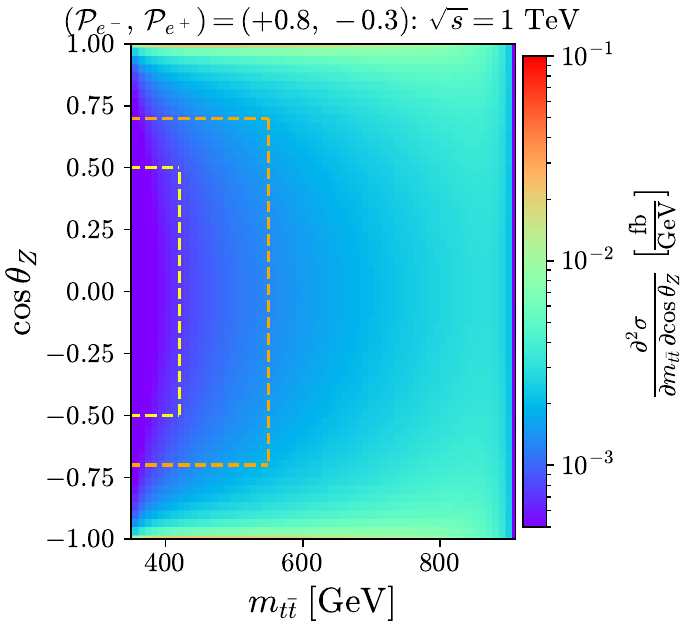}
\includegraphics[width=0.32\textwidth]{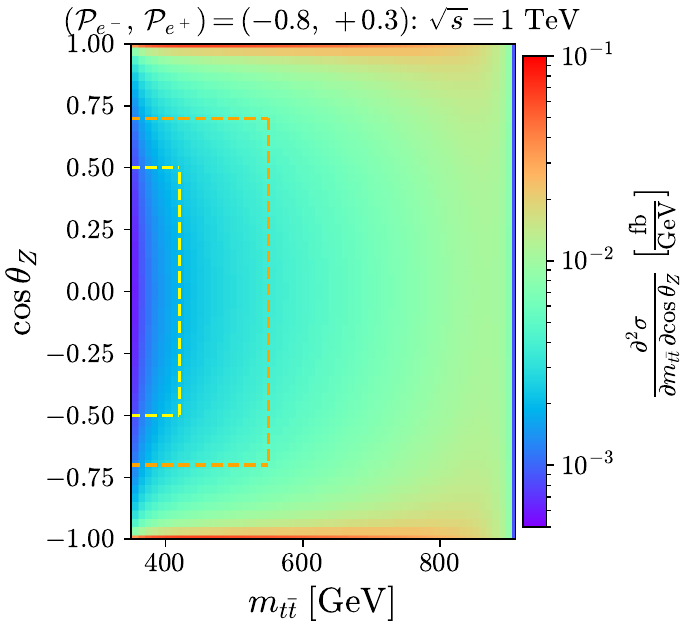}
\caption{\small Doubly differential cross section $\partial^{2}\sigma/(\partial m_{t\bar t}\,\partial\cos\theta_{Z})$ in fb/GeV at $\sqrt{s}=1$~TeV, in the $(m_{t\bar t},\cos\theta_{Z})$ plane for the three beam configurations: Unpol (left), Pol$+$ (middle), and Pol$-$ (right).
The orange and yellow dashed contours mark the fiducial regions $\Sigma_{1|2}$ and $\Sigma_{\rm GME}$ defined in \autoref{eq:sig_12} and \autoref{eq:sig_GME}.}
\label{fig:diff_xsec}
\end{figure}

\begin{table}
\begin{center}
\renewcommand{\arraystretch}{1.2} 
\setlength{\tabcolsep}{12pt}    
\begin{tabular}{ c||c|c||c|c|c } 
 & $\sigma_{1|2} \,[\rm fb]$ & $\sigma_{\rm GME} \,[\rm fb]$ & $N(Z| t \bar t)$ & $N(t | \bar t Z)$  & ${\cal N}_G$ \\ 
 \hline 
 Unpol     & 0.41 & 0.054 & 0.199 & 0.136 & 0.156 \\ 
 \hline
 Pol$+$  & 0.27 & 0.036 & 0.216 & 0.150 & 0.169  \\ 
 \hline
 Pol$-$ & 0.74 & 0.099 & 0.244 & 0.170 & 0.181
\end{tabular}
\end{center}
\caption{\small \label{tab:xsec}
\small
Fiducial cross sections $\sigma_{1|2}$ and $\sigma_{\rm GME}$ (in fb) for the regions $\Sigma_{1|2}$ and $\Sigma_{\rm GME}$, defined in \cref{eq:sig_12,eq:sig_GME}, at $\sqrt{s}=1$~TeV. We also report the corresponding  fiducial values of $N(Z|t\bar t)$ and $N(t|\bar tZ)$ in $\Sigma_{1|2}$, and of ${\cal N}_{G}$ in $\Sigma_{\rm GME}$, for the three beam configurations. By the CP symmetry of the neutral-current production, we have $N(\bar t|tZ)=N(t|\bar tZ)$ in $\Sigma_{1|2}$.
}
\end{table}

\autoref{fig:diff_xsec} shows the doubly differential cross section $\partial^{2}\sigma/(\partial m_{t\bar t}\,\partial\cos\theta_{Z})$ in fb/GeV at $\sqrt{s}=1$~TeV.
In all three polarisations the events are concentrated near $|\cos\theta_{Z}|\to 1$ (collinear $Z$ emission) and at intermediate $m_{t\bar t}$, in marked contrast to the entanglement maps of \autoref{sec:partial}, which favour the central region $|\cos\theta_{Z}|\lesssim 0.5$ and low $m_{t\bar t}$.
The fiducial regions $\Sigma_{1|2}$ and $\Sigma_{\rm GME}$ therefore sit in the kinematically suppressed corner of the cross section, and the price for retaining a large fiducial entanglement is a small fiducial event rate.

The relevant numbers are summarised in \autoref{tab:xsec}.
For the unpolarised beam, the wider fiducial region $\Sigma_{1|2}$ retains about $10\%$ of the total $\sigma(e^{+}e^{-}\to t\bar t Z)$ at $\sqrt{s}=1$~TeV ($\sigma_{1|2}\simeq 0.41$~fb), while the smaller $\Sigma_{\rm GME}$ retains only about $1\%$ ($\sigma_{\rm GME}\simeq 0.054$~fb).
Both fiducial cross sections scale by a factor of $\sim 2.7$ from Pol$+$ to Pol$-$, mirroring the rate hierarchy of \autoref{fig:xsec}.
The fiducial entanglement values, in contrast, depend only weakly on the polarisation: $N(Z|t\bar t)\simeq 0.20\text{--}0.24$, $N(t|\bar tZ)\simeq 0.14\text{--}0.17$, and ${\cal N}_{G}\simeq 0.16\text{--}0.18$, with the strongest signals consistently for Pol$-$ across all three measures.
Pol$-$ is therefore unambiguously the preferred running condition for the feasibility study below.

\medskip

To assess the feasibility of the entanglement observation, we consider quantum tomography in the fully leptonic decay channel ($t \to b\,\ell^{+}\nu$,  $\bar t \to \bar b \ell^{-} \bar\nu$, $Z \to \ell^{+}\ell^{-}$, where $\ell = e,\mu,\tau$),
which provides a clean final state with two charged leptons from the $Z$ decay and two additional charged leptons from the leptonic top decays.
The corresponding branching ratio is 
${\rm BR} \,\simeq\, (0.33)^{2}\cdot 0.10 \,\simeq\, 0.010$.
For a fiducial cross section $\sigma_{\Sigma}$ and an integrated luminosity ${\cal L}$, the expected number of fully leptonic events inside the fiducial region $\Sigma$ is
\be
n_{\Sigma} \,=\, \sigma_{\Sigma}\cdot{\rm BR}\cdot{\cal L}\,.
\ee
To obtain an order-of-magnitude estimate of the statistical requirements for entanglement certification, we adopt a simple scaling model for the uncertainty on an entanglement variable $v\in\{N(Z|t\bar t),\,N(t|\bar tZ),\,N(\bar t|tZ),\,{\cal N}_{G}\}$. For a sample of $n_{\Sigma}$ events, we estimate
\be
\delta_{v} \,=\, c_{\rm sys}\cdot \frac{v_{\rm max}}{\sqrt{n_{\Sigma}}}\,,
\label{eq:delta_v}
\ee
where $v_{\rm max}$ is the algebraic upper bound on $v$, namely $v_{\rm max}=1$ for $N(Z|t\bar t)$ (the $3\times 4$ partition) and $v_{\rm max}=\frac{1}{2}$ for $N(t|\bar tZ)$, $N(\bar t|tZ)$ and ${\cal N}_{G}$.
The coefficient $c_{\rm sys}\ge 1$ parametrises departures from the naive scaling $v_{\rm max}/\sqrt{n_{\Sigma}}$, including reconstruction inefficiencies, finite detector resolution, uncertainties in the quantum state reconstruction procedure, and any backgrounds.
Throughout this section we adopt the optimistic benchmark $c_{\rm sys}=1$ in order to isolate the impact of event statistics.
The resulting luminosity requirements should be interpreted as indicative statistical targets rather than as the outcome of a complete detector-level tomography analysis. A realistic experimental study would generally increase $\delta_v$ by a factor of a few, with the corresponding luminosity requirements scaling as $c_{\rm sys}^{2}$. The naive significance at which a non-zero value of $v$ can be established in the fiducial region is
\be
\sigma_{v} \,\equiv\, \frac{v_{\Sigma}}{\delta_{v}}
\,=\,
\frac{v_{\Sigma}}{v_{\rm max}}\,\sqrt{\sigma_{\Sigma}\cdot{\rm BR}\cdot{\cal L}}\,,
\ee
and the luminosity required to reach a target significance $\sigma_{\rm tgt}$ is
\be
{\cal L}_{\sigma_{\rm tgt}}
\,=\,
\left(\frac{\sigma_{\rm tgt}\,v_{\rm max}}{v_{\Sigma}}\right)^{2}
\frac{1}{\sigma_{\Sigma}\cdot{\rm BR}}\,.
\label{eq:L_target}
\ee
Because $v_{\rm max}/v_{\Sigma}\sim 3\text{--}5$ for the $t\bar tZ$ system, $\sigma_{\Sigma}$ is sub-fb, and ${\rm BR}\simeq 10^{-2}$, the required luminosities naturally fall in the multi-ab$^{-1}$ regime, making the observation of the most robust entanglement observables a plausible target for a future high-luminosity lepton collider.

\begin{table}[t!]
\centering
\renewcommand{\arraystretch}{1.6}
\begin{tabular}{|c||c|c|c|c|c|c|c|c|c|}
\hline
 & \multicolumn{3}{c|}{$N(Z|t\bar{t})$} & \multicolumn{3}{c|}{$N(t|\bar{t}Z)$} & \multicolumn{3}{c|}{$\mathcal{N}_G$} \\
\hline
\footnotesize\makecell{significance,\\[-2pt] luminosities}
 & $\sigma_v$ & $\mathcal{L}_{1\sigma}$ & $\mathcal{L}_{2\sigma}$
 & $\sigma_v$ & $\mathcal{L}_{1\sigma}$ & $\mathcal{L}_{2\sigma}$
 & $\sigma_v$ & $\mathcal{L}_{1\sigma}$ & $\mathcal{L}_{2\sigma}$ \\
\hline \hline
Unpol & 1.1 & 6.2 & 24.8 & 1.5 & 3.3 & 13.2 & 0.65 & 19.1 & 76.4 \\
\hline
Pol$+$ & 1.0 & 7.9 & 31.7 & 1.4 & 4.1 & 16.4 & 0.57 & 24.6 & 98.4 \\
\hline
Pol$-$ & 1.9 & 2.3 & 9.1 & 2.6 & 1.2 & 4.7 & 1.01 & 7.8 & 31.1 \\
\hline
\end{tabular}
\caption{\small \small
Naive significances $\sigma_{v}$ for the observation of each entanglement measure at ${\cal L}=8$~ab$^{-1}$, together with the integrated luminosities ${\cal L}_{1\sigma}$ and ${\cal L}_{2\sigma}$ (in ab$^{-1}$) required to reach $1\sigma$ and $2\sigma$ significance, for the three beam configurations.
The one-to-other negativities $N(Z|t\bar t)$ and $N(t|\bar tZ)$ are evaluated inside $\Sigma_{1|2}$; the GMN ${\cal N}_{G}$ is evaluated inside $\Sigma_{\rm GME}$.
All entries assume $c_{\rm sys}=1$.
}
\label{tab:significance}
\end{table}

\autoref{tab:significance} summarises the naive significances $\sigma_{v}$ expected at a benchmark integrated luminosity ${\cal L}=8$~ab$^{-1}$, together with the integrated luminosities ${\cal L}_{1\sigma}$ and ${\cal L}_{2\sigma}$ required to reach $1\sigma$ and $2\sigma$ significance through \autoref{eq:L_target}.
By construction ${\cal L}_{2\sigma}=4\,{\cal L}_{1\sigma}$, since $\sigma_{v}\propto\sqrt{{\cal L}}$.
Three clear messages emerge.

First, the easiest measure to certify is $N(t|\bar tZ)$ (equivalently $N(\bar t|tZ)$, identical by CP).
This is the result of a fortunate combination: the fiducial value $\sim 0.14\text{--}0.17$ is a sizeable fraction of the upper bound $v_{\rm max}=\frac{1}{2}$, and the fiducial region $\Sigma_{1|2}$ used to evaluate it has the largest cross section among the two fiducial choices.
Already with ${\cal L}\simeq 3.3$~ab$^{-1}$ (Unpol) or $1.2$~ab$^{-1}$ (Pol$-$) the measure can be established as non-zero at the $1\sigma$ level, and $2\sigma$ certification requires only $13.2$ and $4.7$~ab$^{-1}$ respectively.
These luminosities are comparable to those envisioned for the ILC at $\sqrt{s}=1$~TeV.

Second, $N(Z|t\bar t)$ is intrinsically harder to certify than $N(t|\bar tZ)$ and $N(\bar t|tZ)$ despite its larger fiducial value ($\sim 0.20\text{--}0.24$), because the relevant algebraic maximum is $v_{\rm max}=1$ rather than $\frac{1}{2}$.
The ratio $v_{\Sigma}/v_{\rm max}$ that drives $\sigma_{v}$ is therefore smaller, and the luminosity needed to reach a given significance is correspondingly larger: $24.8$~ab$^{-1}$ at $2\sigma$ for Unpol, dropping to $9.1$~ab$^{-1}$ for Pol$-$.

Third, certifying genuine multipartite entanglement through ${\cal N}_{G}$ remains the most demanding goal.
The fiducial value of ${\cal N}_{G}$ ($\sim 0.16\text{--}0.18$) is comparable to that of $N(t|\bar tZ)$, so the dominant penalty for the GMN comes from the fiducial cross section in $\Sigma_{\rm GME}$, which is $\sim 7\text{--}8$ times smaller than in $\Sigma_{1|2}$.
The resulting luminosities are $19$ and $8$~ab$^{-1}$ for $1\sigma$ in Unpol and Pol$-$, and $76$ and $31$~ab$^{-1}$ for $2\sigma$.
With the optimistic $c_{\rm sys}=1$ assumed here, a $1\sigma$ GME certification at $\sqrt{s}=1$~TeV for Pol$-$ is achievable within the projected ILC luminosity of $\sim 8$~ab$^{-1}$~\cite{LinearColliderVision:2025hlt}; the $2\sigma$ target for Pol$-$ requires a factor of $\sim 4$ more than this baseline.
More optimisation for the fiducial cuts beyond the ($m_{t \bar t}$, $\cos \theta_Z$) plane, 
optimisation for the quantisation axes (beyond the helicity basis) in reconstructing fictitious quantum states,
a more refined GME observable, or inclusion of hadronic decay channels would all be valuable directions for further study.

\section{Conclusions}
\label{sec:concl}

We have presented a systematic study of the tripartite entanglement structure of the spin state produced in $e^{+}e^{-}\to t\bar t Z$.
The final state spin Hilbert space ${\cal H}={\mathbb C}^{2}\otimes{\mathbb C}^{2}\otimes{\mathbb C}^{3}$ has dimension $12$.
Starting from the tree-level Standard Model helicity amplitudes for all five contributing diagrams, we constructed the $12\times 12$ density matrix as a function of the four independent phase space variables $(m_{t\bar t},\cos\theta_{Z},\cos\theta_{t},\phi)$ and evaluated on it a hierarchy of negativity-based entanglement measures that remain valid for mixed states: the three {\em one-to-one} negativities $N_{t\bar t}$, $N_{tZ}$, $N_{\bar tZ}$, the three {\em one-to-other} negativities $N(Z|t\bar t)$, $N(t|\bar tZ)$, $N(\bar t|tZ)$, and the {\em genuine multipartite negativity} (GMN) ${\cal N}_{G}$ of Refs.~\cite{Jungnitsch:2011izf,Hofmann:2014ywl}.
These quantities were studied at three increasingly inclusive levels: differential (a fixed phase space point), partially inclusive (integrated over $\cos\theta_{t}$ and $\phi$ in the helicity basis), and fully inclusive (integrated over all four kinematic variables).

\medskip

At the differential level the state $\rho$ is nearly pure ($\gamma(\rho)\simeq 0.65\text{--}0.70$) and genuinely tripartite entangled over a large fraction of phase space, with ${\cal N}_{G}\sim 0.40$ in the central region.
Its entanglement pattern is GHZ-like: the one-to-other negativities $N(t|\bar tZ)$ and $N(\bar t|tZ)$ nearly saturate their algebraic maxima of $\frac{1}{2}$, while the pairwise negativities remain small or vanish over most of phase space. Hence, each particle is largely entangled with the remaining two, but each pair carries little extractable entanglement.

At the fully inclusive level the state $\rho_{\Sigma}$ becomes nearly maximally mixed ($\gamma(\rho_{\Sigma})\simeq 0.22\text{--}0.25$, only slightly above the $\frac{1}{12}$ floor): all one-to-one negativities vanish identically, ${\cal N}_{G}=0$ for all $\sqrt{s}$, and only the one-to-other negativities survive with small but non-zero values ($N(Z|t\bar t)\lesssim 0.06$ and $N(t|\bar tZ)=N(\bar t|tZ)\lesssim 0.04$ around $\sqrt{s}\simeq 1.5\text{--}2$~TeV).
Thus, the genuine tripartite quantum correlations of $e^{+}e^{-}\to t\bar t Z$ live in the differential states rather than in the inclusive ones.

Partial integration over $(\cos\theta_{t},\phi)$ at fixed $(m_{t\bar t},\cos\theta_{Z})$ preserves much of the differential entanglement in the threshold region.
$N(t|\bar tZ)$ and $N(\bar t|tZ)$ peak at $\sim 0.3\text{--}0.4$ near $m_{t\bar t}\simeq 2m_{t}$, $N(Z|t\bar t)$ remains of order $0.1\text{--}0.3$ over a broad central region, and ${\cal N}_{G}$ reaches $\sim 0.25$ at low $m_{t\bar t}$ and central $\cos\theta_{Z}$.
The one-to-one negativities, by contrast, remain comparatively small in this region.
Motivated by these observations, we defined the fiducial regions $\Sigma_{1|2}$ (\autoref{eq:sig_12}) and $\Sigma_{\rm GME}$ (\autoref{eq:sig_GME}) for the one-to-other and GMN measurements, respectively.

\medskip
Assuming quantum tomography in the fully leptonic decay channel of $t\bar t Z$ at $\sqrt{s}=1$~TeV with the most optimistic statistical assumption $c_{\rm sys}=1$, our naive luminosity estimates for $2\sigma$ certification are
\begin{itemize}
\item $N(t|\bar tZ)$ (equivalently $N(\bar t|tZ)$ by CP): ${\cal L}\simeq 13$~ab$^{-1}$ (Unpol) or $5$~ab$^{-1}$ (Pol$-$),
\item $N(Z|t\bar t)$: ${\cal L}\simeq 25$~ab$^{-1}$ (Unpol) or $9$~ab$^{-1}$ (Pol$-$),
\item ${\cal N}_{G}$: ${\cal L}\simeq 76$~ab$^{-1}$ (Unpol) or $31$~ab$^{-1}$ (Pol$-$).
\end{itemize}
The first two targets are accessible within the projected high-luminosity polarised programme at the ILC~\cite{LinearColliderVision:2025hlt}. The GMN target at $1\sigma$ for Pol$-$ is comparable to the projected ILC luminosity at $\sqrt{s}=1$~TeV, while the $2\sigma$ target requires a factor of $\sim 4$ more than this baseline.
Beam polarisation gives a $\sim 2\text{--}3$ luminosity-saving factor across all three measures.
The GME prospects could be improved by developing refined GME observables (the GMN considered here is a sufficient but not necessary certificate of GME), optimising the fiducial kinematic regions beyond the ($m_{t \bar t}$, $\cos \theta_Z$) plane, 
optimising the quantisation axes in reconstructing fictitious quantum states, 
or including hadronic channels in the quantum state tomography~\cite{Dong:2023xiw,Dong:2024xsg,Dong:2024xsb}.
Another promising future direction is to extend the present quantum information programme of the $t\bar t Z$ final state beyond entanglement to other genuinely quantum correlations and resources, such as tripartite Bell nonlocality~\cite{Horodecki:2025tpn} and ``magic'' (non-stabiliserness) \cite{White:2024nuc}.

Beyond the Standard Model, the multipartite spin density matrix studied here is a natural probe of new physics in the $t\bar t Z$ interaction (anomalous neutral-current vertices, CP-violating contributions, or new heavy mediators), which can leave a distinctive imprint on the entanglement maps presented in \autoref{sec:ttZ} and \autoref{sec:partial}.
The framework developed in this paper, combining helicity amplitudes, multi-level phase space integration, and a hierarchy of negativity-based entanglement measures, applies to any tripartite spin system.
Other natural collider targets include $e^{+}e^{-}/pp \to\tau^{+}\tau^{-}Z$, $W^+ W^- Z$ and $h\to\tau^{+}\tau^{-}Z$. Furthermore, the present analysis can be naturally extended to
$pp\to t\bar tZ$ at the LHC and the FCC-hh.
The numerical results and methodology presented here lay out a quantitative quantum information programme for the study of multipartite quantum correlations at current and next-generation colliders.

\section*{Acknowledgments}
We thank Błażej Rozwoda for contributions in the early stages of this project.
The work of DG is supported in part by US Department of Energy Grant Number DE-SC 0016013. Some computing for this project was performed at the High Performance Computing Center at Oklahoma State University, supported in part through the National Science Foundation grant OAC-1531128. 
KS thanks Hai-Chau Nguyen, Ties-Albrecht Ohst and
Gilberto Tetlamatzi-Xolocotzi for helpful discussions on the
computation of genuine multipartite entanglement measures. 
KS thanks Yuki Nakasone for producing the schematic illustration in \autoref{fig:GME_diagram}.


\appendix

\section{Quantum State Tomography}
\label{app:tomography}

In \autoref{sec:amp} we described how the $t\bar t Z$ spin density matrix is constructed from the tree-level helicity amplitudes, and the results presented in \autoref{sec:ttZ} and \autoref{sec:prospects} are obtained directly with this formalism. Experimentally, however, the density matrix must instead be reconstructed from the measured angular distributions of the $t$, $\bar t$ and $Z$ decay products through a quantum state tomography procedure. In this appendix we present this procedure for the full $12 \times 12$ density matrix of the $t\bar t Z$ spin state.

\subsection{Spherical Tensor Representation for Spin-1/2 and Spin-1}
First let us consider the density matrix for a spin-1/2 particle 
\begin{align}
    \rho_{1/2} = \frac 1 2\left(\mathbb{I}_2 + \vec{b}\cdot\vec{\sigma}\right)\,,
\end{align}
where $\vec b = (b_1\,,b_2\,,b_3)$ is the polarisation vector and $\vec \sigma = (\sigma_1\,,\sigma_2\,,\sigma_3)$ are the Pauli matrices. Define the spin-1/2 spherical tensor operators $t_M^L$~\cite{Leader:2001nas,Bernal:2023jba},  
\begin{align}
    t_{\pm 1}^1 = \mp \frac{1}{\sqrt{6}}(\sigma_1\pm i\sigma_2)\,,~\text{and}~t_0^1 = \frac{1}{\sqrt{3}}\sigma_3\,. 
\end{align}
They provide an orthonormal basis and satisfy
\begin{align}
t^L_{-M} = (-1)^M\left(t^L_M\right)^\dagger\,,\,\,\,\, \Tr{t^L_M\left(t^{L'}_{M'}\right)^\dagger} = \frac 2 3 \delta_{LL'}\delta_{MM'}\,.
\end{align}
Explicitly,
\begin{align}
    t_1^1 = -\sqrt{\frac{2}{3}}\begin{pmatrix}
        0 & 1 \\
        0 & 0
    \end{pmatrix}\,,\,\,\,
    t_{-1}^1 = \sqrt{\frac{2}{3}}\begin{pmatrix}
        0 & 0 \\
        1 & 0
    \end{pmatrix}\,,\,\,\,
    t_0^1 = \frac{1}{\sqrt{3}}\begin{pmatrix}
        1 & 0 \\
        0 & -1
    \end{pmatrix}\,.
\end{align}
The spin-1/2 density matrix can then be written as
\begin{align}
    \label{eq:spinhalfdens}
     \rho_{1/2} = \frac 1 2 \left(\mathbb{I}_2 + A_{11}t_1^1 + A_{1-1}t_{-1}^1 + A_{10}t_0^1\right) = \frac 1 2 \left(\mathbb{I}_2 + A_{LM}t_M^L\right)\,,
\end{align}
where the multipole parameters, $A_{LM}$ with $L=1$ and $-1\leq M \leq 1$, satisfy $A_{L-M}=(-1)^MA_{LM}^\ast$ and are related to the components of the polarisation vector by
\begin{align}
    A_{11} = -\sqrt{3}\left(\frac{b_1 - i b_2}{\sqrt{2}}\right)\,,\,\,\,
    A_{1-1} = \sqrt{3}\left(\frac{b_1 + i b_2}{\sqrt{2}}\right)\,,\,\,\,
    A_{10} = \sqrt{3}b_3\,.
\end{align}

The density matrix for a spin-1 particle can be expressed in an analogous way
\begin{align}
    \label{eq:spin1dens}
    \rho_1 = \frac 1 3\left(\mathbb{I}_3+A_{LM}T_M^L\right)\,,
\end{align}
where $L=1,2$ and $-L\leq M\leq L$. The spherical tensor operators, $T^{L}_{M}$, in terms of spin-1 component operators, $J_x$, $J_y$, and $J_z$ are given by~\cite{Aguilar-Saavedra:2022wam}

\begin{align}
    T_{\pm1}^1 &= \mp\frac{\sqrt{3}}{2}\left(J_x\pm i~J_y\right)\,, && T_0^1 = \sqrt{\frac{3}{2}}J_z, \nonumber\\
    T_{\pm 2}^2 &= \frac{2}{\sqrt{3}}\left( T_{\pm 1}^1\right)^2\,, && T_{\pm 1}^2 = \sqrt{\frac{3}{2}}\left( T_{\pm 1}^1 T_0^1 + T_0^1 T_{\pm 1}^1\right)\,, \\
    T_{0}^2 &= \frac{\sqrt{2}}{3}\left( T_1^1 T_{-1}^1 + T_{-1}^1 T_1^1 + 2(T_0^1)^2\right)\nonumber\,.
\end{align}
As in the spin-1/2 case, they form an orthonormal basis satisfying
\begin{align}
T^L_{-M} = (-1)^M\left(T^L_M\right)^\dagger\,,\,\,\,\, \Tr{T^L_M\left(T^{L'}_{M'}\right)^\dagger} = 3 \delta_{LL'}\delta_{MM'}\,.
\end{align}
Explicitly,
\begin{align}
    T_{-1}^1 = \sqrt{\frac{3}{2}}\begin{pmatrix}
        0 & 0 & 0 \\
        1 & 0 & 0 \\
        0 & 1 & 0
    \end{pmatrix}\,, && T_0^1 = \sqrt{\frac{3}{2}}\begin{pmatrix}
        1 & 0 & 0 \\
        0 & 0 & 0 \\
        0 & 0 & -1
    \end{pmatrix}\,, && T_1^1 = \sqrt{\frac{3}{2}}\begin{pmatrix}
        0 & -1 & 0 \\
        0 & 0 & -1 \\
        0 & 0 & 0
    \end{pmatrix}\,, \\
    T_{-2}^2 = \sqrt{3}\begin{pmatrix}
        0 & 0 & 0 \\
        0 & 0 & 0 \\
        1 & 0 & 0 
    \end{pmatrix}\,, && T_{-1}^2 = \sqrt{\frac 3 2}\begin{pmatrix}
        0 & 0 & 0 \\
        1 & 0 & 0 \\
        0 & -1 & 0 
    \end{pmatrix}\,, && T_{0}^2 = \frac{1}{\sqrt{2}}\begin{pmatrix}
        1 & 0 & 0 \\
        0 & -2 & 0 \\
        0 & 0 & 1 
    \end{pmatrix}\,,\\
    T_{1}^2 = \sqrt{\frac 3 2}\begin{pmatrix}
        0 & -1 & 0 \\
        0 & 0 & 1 \\
        0 & 0 & 0 
    \end{pmatrix}\,, && T_{2}^2 = \sqrt{3}\begin{pmatrix}
        0 & 0 & 1 \\
        0 & 0 & 0 \\
        0 & 0 & 0 
    \end{pmatrix}\,.
\end{align}

\subsection{\texorpdfstring{Quantum State Tomography for $t\bar t Z$}{ttZ}}

The $t\bar t Z$ quantum state is represented by a $12\times 12$ density matrix, which can be parametrized in terms of the spherical tensor basis introduced in the previous section. Using~\autoref{eq:spinhalfdens} and~\autoref{eq:spin1dens} the $t \bar t Z$ density matrix reads\footnote{We parametrize the density matrix using the spherical tensor basis, which is conventionally used for the $Z$ boson quantum tomography. An equivalent parametrization in the ``Cartesian'' basis is given in Ref.~\cite{Rahaman:2022dwp}.}
\begin{align}
\label{eq:ttZDensMat}
    \rho &= \frac{1}{12}\Biggl(\mathbb{I}_{12} + A_{L_1 M_1}^{1} t_{M_1}^{L_1}\otimes\mathbb{I}_{2}\otimes\mathbb{I}_{3} + A_{L_2 M_2}^2 \mathbb{I}_{2}\otimes t_{M_2}^{L_2}\otimes\mathbb{I}_{3} 
    + A_{L_3 M_3}^3 \mathbb{I}_{2}\otimes\mathbb{I}_{2}\otimes T_{M_3}^{L_3} \notag\\
    &+ C_{L_1 M_1 L_2 M_2}^{12} t_{M_1}^{L_1}\otimes t_{M_2}^{L_2}\otimes\mathbb{I}_{3} + C_{L_1 M_1 L_3 M_3}^{13} t_{M_1}^{L_1}\otimes\mathbb{I}_{2}\otimes T_{M_3}^{L_3} + C_{L_2 M_2 L_3 M_3}^{23} \mathbb{I}_{2}\otimes t_{M_2}^{L_2}\otimes T_{M_3}^{L_3} \notag \\
    &+ C_{L_1 M_1 L_2 M_2 L_3 M_3}^{123} t_{M_1}^{L_1}\otimes t_{M_2}^{L_2}\otimes T_{M_3}^{L_3}\Biggr)\,,
\end{align}
where $L_{1,2}=1$, $-L_{1,2}\leq M_{1,2}\leq L_{1,2}$, $L_3=1,2$ and $-L_3\leq M_3 \leq L_3$. Here $A_{L_i M_i}^i$ are the polarisations of particle $i$ with $i=1$ for the top quark, $i=2$ for the antitop quark, and $i=3$ for the $Z$ boson, $C_{L_i M_i L_j M_j}^{ij}$ are the two-particle spin correlations between particles $i$ and $j$, and $C_{L_1 M_1 L_2 M_2 L_3 M_3}^{123}$ are the three-particle spin correlations. 

The spin information of the top quarks and $Z$ boson is imprinted in the angular distributions of their decay products. Consider the fully leptonic final state $t(\rightarrow b\ell^+_1\nu_\ell)~\bar{t}(\rightarrow\bar{b}\ell^-_2\bar{\nu}_\ell)~Z(\rightarrow \ell^+_3\ell^-_3)$. The decay matrices for the top and antitop quarks are given by~\cite{Boudjema:2009fz}
\begin{align}
    \Gamma_{1,2}=\begin{pmatrix}
        \frac{1+\beta_{1,2}\cos\theta_{1,2}}{2} & \frac{\beta_{1,2}\sin\theta_{1,2}}{2} e^{i\varphi_{1,2}} \\
        \frac{\beta_{1,2}\sin\theta_{1,2}}{2}e^{-i\varphi_{1,2}} &  \frac{1-\beta_{1,2}\cos\theta_{1,2}}{2}
    \end{pmatrix}\,,
    \label{eq:Gamma_mat_t}
\end{align}
where $(\theta_{1,2},\varphi_{1,2})$ are the polar and azimuthal angles of the charged lepton $\ell_{1,2}$ in their mother rest frame, and $\beta_{1,2}$ are their spin analysing powers. For top quark decays to charged leptons, $\beta_1=+1$ for $\ell^+_1$ and $\beta_2=-1$ for $\ell^-_2$~\cite{Jezabek:1994qs}. For the $Z$ boson decaying into two charged leptons, the decay matrix reads
\begin{equation}
\resizebox{\textwidth}{!}{$
\Gamma_3
=
\frac{1}{4}
\begin{pmatrix}
1+\delta_3+(1-3\delta_3)\cos^2\theta_3 - 2\eta_3\cos\theta_3
&
\frac{1}{\sqrt{2}}\Big[(1-3\delta_3)\sin2\theta_3 - 2\eta_3\sin\theta_3\Big] e^{i\varphi_3}
&
(1-3\delta_3)(1-\cos^2\theta_3) e^{2i\varphi_3}
\\[0.6em]
\frac{1}{\sqrt{2}}\Big[(1-3\delta_3)\sin2\theta_3 - 2\eta_3\sin\theta_3\Big] e^{-i\varphi_3}
&
4\delta_3+2(1-3\delta_3)\,\sin^2\theta_3
&
-\frac{1}{\sqrt{2}}\Big[(1-3\delta_3)\sin2\theta_3 + 2\eta_3\sin\theta_3\Big] e^{i\varphi_3}
\\[0.6em]
(1-3\delta_3)(1-\cos^2\theta_3) e^{-2i\varphi_3}
&
-\frac{1}{\sqrt{2}}\Big[(1-3\delta_3)\sin2\theta_3 + 2\eta_3\sin\theta\Big] e^{-i\varphi_3}
&
1+\delta_3+(1-3\delta_3)\cos^2\theta_3 + 2\eta_3\cos\theta_3
\end{pmatrix}
$}\,,
\label{eq:Gamma_matZ}
\end{equation}
with $(\theta_3,\varphi_3)$ denoting the angles of the negatively charged lepton, $\ell^-_3$, in the $Z$ rest frame. The parameters $\eta_3$ and $\delta_3$ are the spin analysing powers of $\ell_3^-$ analogous to $\beta$ in the top quark decay. Neglecting the charged lepton mass, they are given by~\cite{Goncalves:2025mvl,Goncalves:2026njf}
\begin{align}
\eta_3 = \frac{C_L^2-C_R^2}{C_L^2+C_R^2}\,,\,\,\,\, \delta_3=0\,,
\end{align}
where $C_{L,R}$ are the left and right chiral couplings of the $Z$ boson to charged leptons. 

We define a basis to compute the polar and azimuthal angles of the charged leptons in their mother rest frames. We start by boosting all particles to the $t\bar{t}Z$ rest frame. The procedure to measure the angles of particle A's decay product is as follows: particle A is brought along the z direction by rotations around the z-axis by $\phi$, and around the y-axis by $\theta$. The same rotations are applied to its daughter. The daughter is then boosted to A's rest frame. The $\{\hat{x},\,\hat{y},\,\hat{z}\}$ basis for particle A is then defined as 
\begin{itemize}
    \item $\hat{z}=\hat{p}_{A}$ 
    \item $\hat{x}=\hat{p}_{B}\times\hat{p}_{C}$
    \item $\hat{y}=\hat{z}\times\hat{x}$\,, 
\end{itemize}
with $p_{B}$ and $p_{C}$ in the $t\bar{t}Z$ rest frame and $p_{A}$ after the described rotations. 

The normalized differential distribution for the charged leptons, $\ell_1^+$, $\ell_2^-$, and $\ell_3^-$ is then given by
\begin{align}
    \frac{1}{\sigma} \frac{\dd\sigma}{\dd\Omega_1 \dd\Omega_2 \dd\Omega_3} 
    &= \left( \frac{2}{4\pi} \right)^2\left(\frac{3}{4\pi}\right)
    \Tr\left\{ \rho \, \big(\Gamma_1 \otimes \Gamma_2 \otimes \Gamma_3 \big)^T \right\} \\
    &= \frac{1}{(4\pi)^3} \Bigg[
        1 + A^1_{L_1 M_1} B_{L_1}^1 Y^{M_1}_{L_1}(\Omega_1)
          + A^2_{L_1 M_1}  B_{L_2}^2 Y^{M_2}_{L_2}(\Omega_2) \nonumber \\
    &+ A^3_{L_3 M_3}  B_{L_3}^3 Y^{M_3}_{L_3}(\Omega_3) + C_{L_1 M_1 L_2 M_2}^{12} B_{L_1}^1  B_{L_2}^2 
          Y^{M_1}_{L_1}(\Omega_1) Y^{M_2}_{L_2}(\Omega_2) \nonumber \\
    &+ C_{L_1 M_1 L_3 M_3}^{13} B_{L_1}^1  B_{L_3}^3 
          Y^{M_1}_{L_1}(\Omega_1) Y^{M_3}_{L_3}(\Omega_3)
    + C_{L_2 M_2 L_3 M_3}^{23} B_{L_2}^{2}  B_{L_3}^{3} 
          Y^{M_2}_{L_2}(\Omega_2) Y^{M_3}_{L_3}(\Omega_3) \nonumber \\
    &+ C_{L_1 M_1 L_2 M_2 L_3 M_3}^{123} B_{L_1}^1 B_{L_2}^2 B_{L_3}^3 Y^{M_1}_{L_1}(\Omega_1) Y^{M_2}_{L_2}(\Omega_2) Y^{M_3}_{L_3}(\Omega_3) \nonumber
      \Bigg],
\end{align}
where $L_{1,2}=1$, $-L_{1,2}\leq M_{1,2}\leq L_{1,2}$, $L_3=1,2$ and $-L_3\leq M_3 \leq L_3$, and $B_{1}^{1,2}=\frac{2\sqrt{\pi}}{3}\beta_{1,2}$, $B_1^3=-\sqrt{2\pi}\,\eta_3$ and $B_2^3=\sqrt{2\pi/5}\,(1-3\delta_3)$. The A and C parameters can be computed from the differential distribution using the orthogonality of the spherical harmonics
\begin{align}
    \int\,\frac{1}{\sigma}\frac{\dd\sigma}{\dd\Omega_1\dd\Omega_2\dd\Omega_3}Y_{L_i}^{M_i}(\Omega_i)^\ast\dd\Omega_1\dd\Omega_2\dd\Omega_3 &= \frac{B_{L_i}^i}{4\pi}A_{L_i M_i}^i\,,\,\,\,i=1,2,3 \label{eq:polarisations} \\
    \int\,\frac{1}{\sigma}\frac{\dd\sigma}{\dd\Omega_1\dd\Omega_2\dd\Omega_3}Y_{L_i}^{M_i}(\Omega_i)^\ast Y_{L_j}^{M_j}(\Omega_j)^\ast\dd\Omega_1\dd\Omega_2\dd\Omega_3 &= \frac{B_{L_i}^i}{4\pi}\frac{B_{L_j}^j}{4\pi} C_{L_i M_i L_j M_j}^{ij}\,,~i\neq j\,, \label{eq:twopartcorr} \\
    \int\,\frac{1}{\sigma}\frac{\dd\sigma}{\dd\Omega_1\dd\Omega_2\dd\Omega_3}Y_{L_1}^{M_1}(\Omega_1)^\ast Y_{L_2}^{M_2}(\Omega_2)^\ast Y_{L_3}^{M_3}(\Omega_3)^\ast\dd\Omega_1\dd\Omega_2\dd\Omega_3 &= \frac{B_{L_1}^1}{4\pi}\frac{B_{L_2}^2}{4\pi}\frac{B_{L_3}^3}{4\pi} C_{L_1 M_1 L_2 M_2 L_3 M_3}^{123}\,. \label{eq:threepartcorr}
\end{align}

\section{Entanglement from Polarised Beam Collisions}
\label{app:pol}

In the main text, we evaluated the partially inclusive entanglement maps in the $(m_{t\bar t},\cos\theta_{Z})$ plane for unpolarised beams (see \autoref{fig:1-1_negativity_partial} and \autoref{fig:1-2_negativity_partial}), and we claimed in \autoref{sec:partial} and \autoref{sec:feesibility} that the one-to-one and one-to-other negativities depend only weakly on the choice of beam polarisation.
For completeness, we collect in this appendix the corresponding maps for the two polarised configurations used in \autoref{sec:prospects},
\be
{\rm Pol}+ : ({\cal P}_{e^-},{\cal P}_{e^+})=(+0.8,-0.3)\,,
\qquad
{\rm Pol}- : ({\cal P}_{e^-},{\cal P}_{e^+})=(-0.8,+0.3)\,,
\ee
to back up this claim.
All plots are evaluated on the partially inclusive fictitious state $\rho_{\Sigma}(m_{t\bar t},\cos\theta_{Z})$ in the helicity basis at $\sqrt{s}=1$~TeV.

\begin{figure}[t!]
\centering
\includegraphics[width=0.32\textwidth]{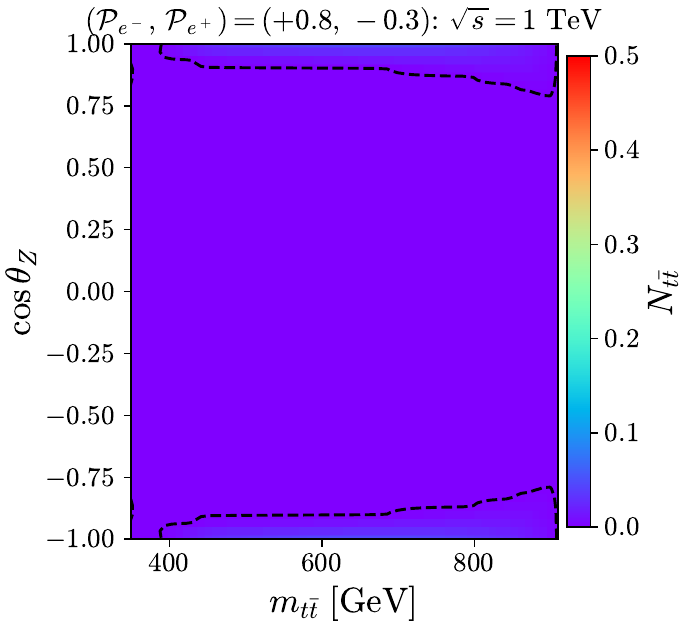}
\includegraphics[width=0.32\textwidth]{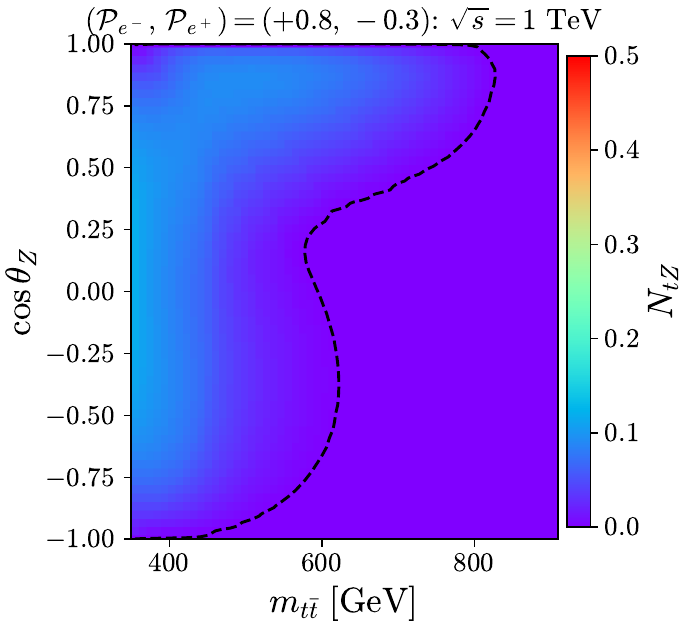}
\includegraphics[width=0.32\textwidth]{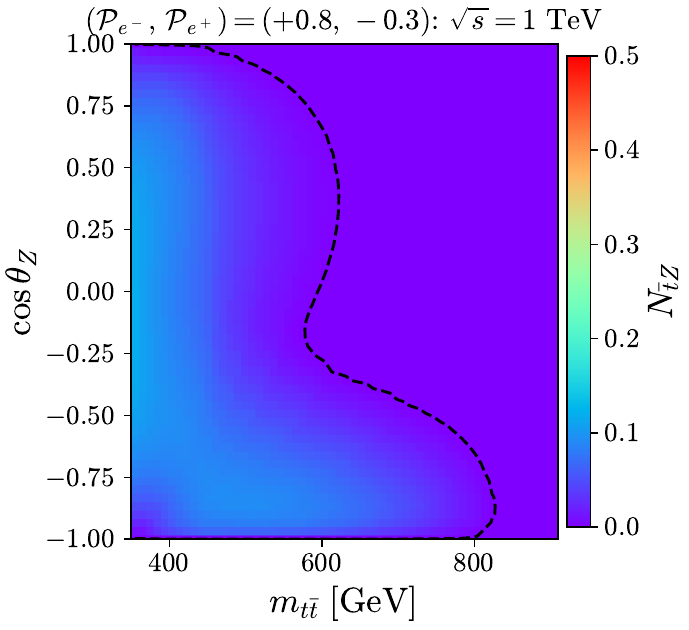}
\includegraphics[width=0.32\textwidth]{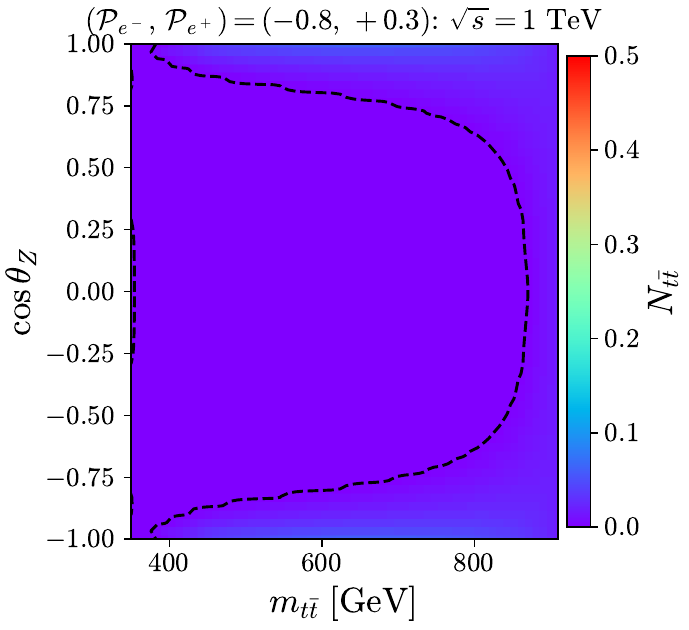}
\includegraphics[width=0.32\textwidth]{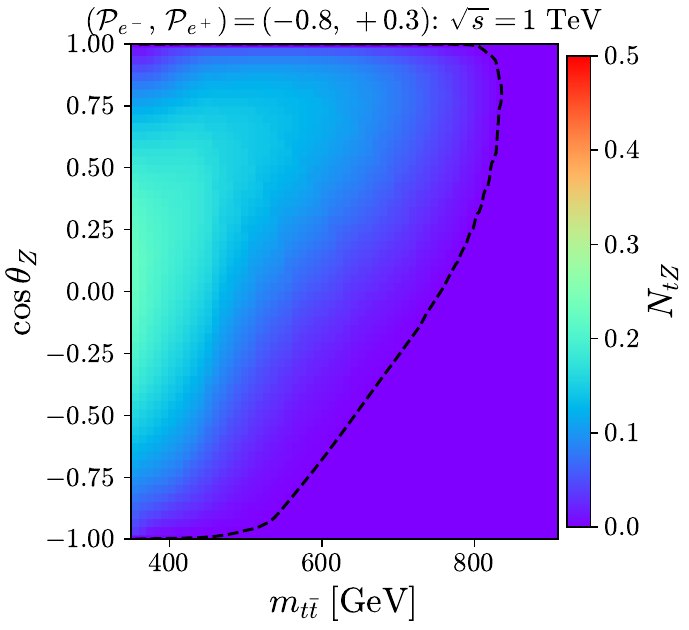}
\includegraphics[width=0.32\textwidth]{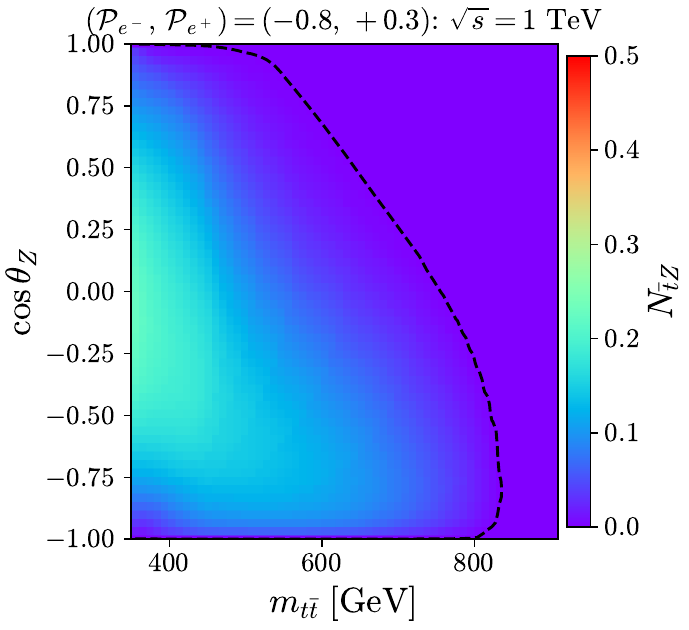}
\caption{\small One-to-one negativities of the partially inclusive $t\bar t Z$ spin state $\rho_{\Sigma}$ at $\sqrt{s}=1$~TeV for polarised beams, in the $(m_{t\bar t},\cos\theta_{Z})$ plane in the helicity basis: $N_{t\bar t}$ (left), $N_{tZ}$ (middle), and $N_{\bar tZ}$ (right).
\emph{Upper row:} Pol$+$ with $({\cal P}_{e^-},{\cal P}_{e^+})=(+0.8,-0.3)$.
\emph{Lower row:} Pol$-$ with $({\cal P}_{e^-},{\cal P}_{e^+})=(-0.8,+0.3)$.
The dashed black contours mark the threshold $N=10^{-3}$, below which each measure is treated as effectively vanishing.
The colour bar saturates at the algebraic upper bound $\frac{1}{2}$.}
\label{fig:1-1_negativity_partial_pol}
\end{figure}

\begin{figure}[t!]
\centering
\includegraphics[width=0.32\textwidth]{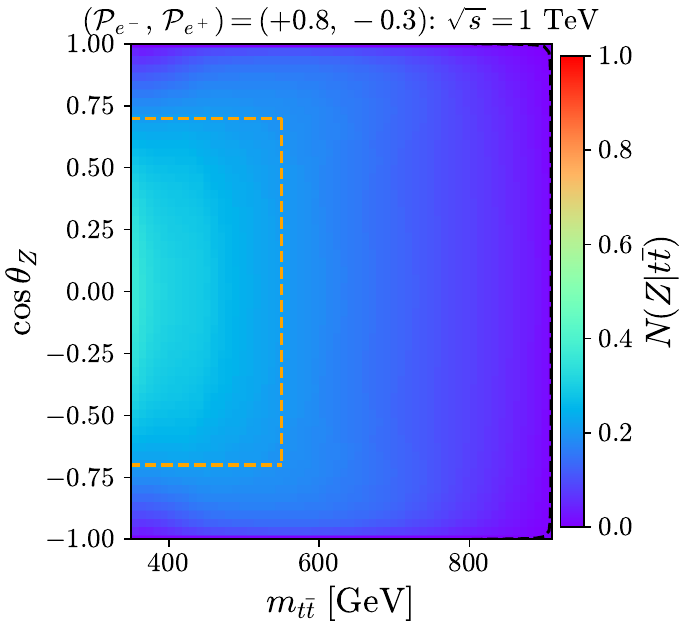}
\includegraphics[width=0.32\textwidth]{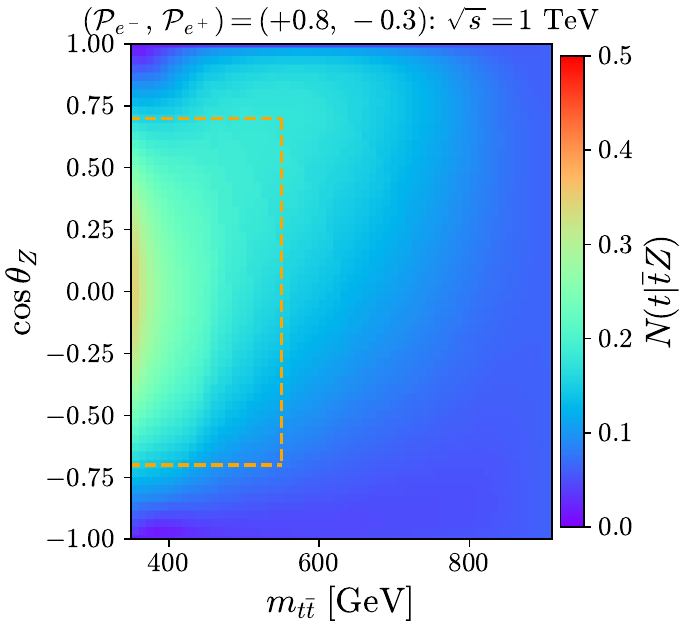}
\includegraphics[width=0.32\textwidth]{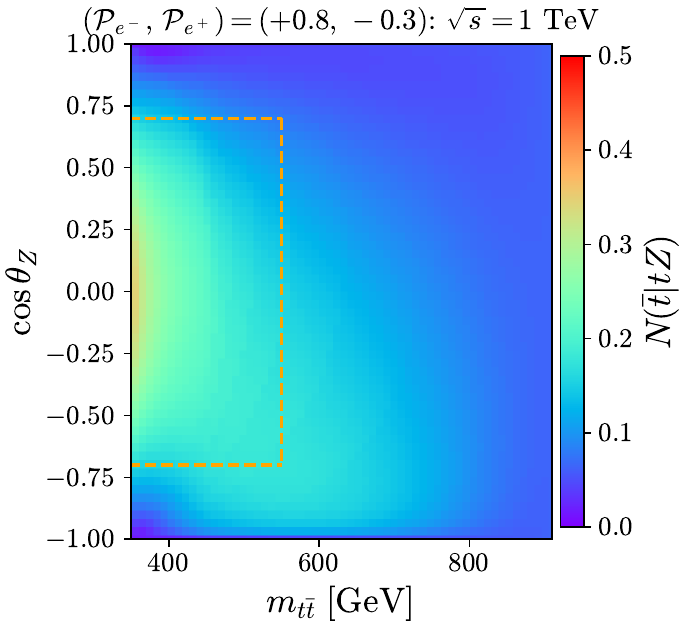}
\includegraphics[width=0.32\textwidth]{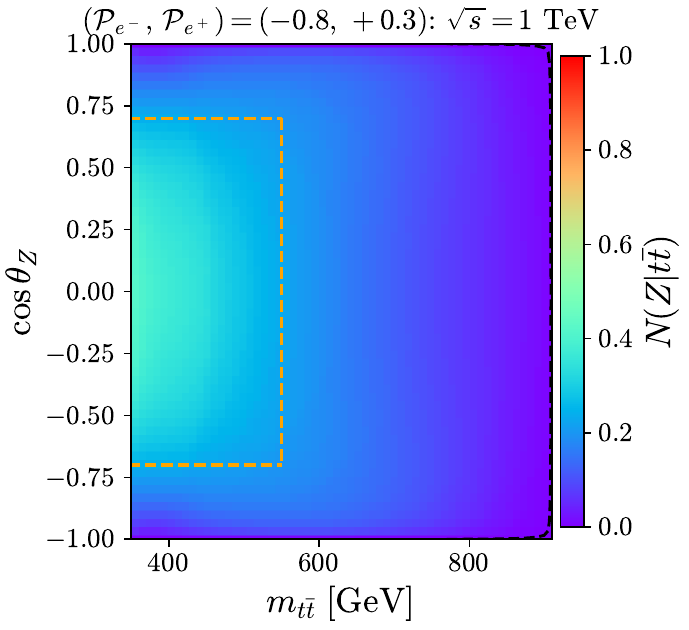}
\includegraphics[width=0.32\textwidth]{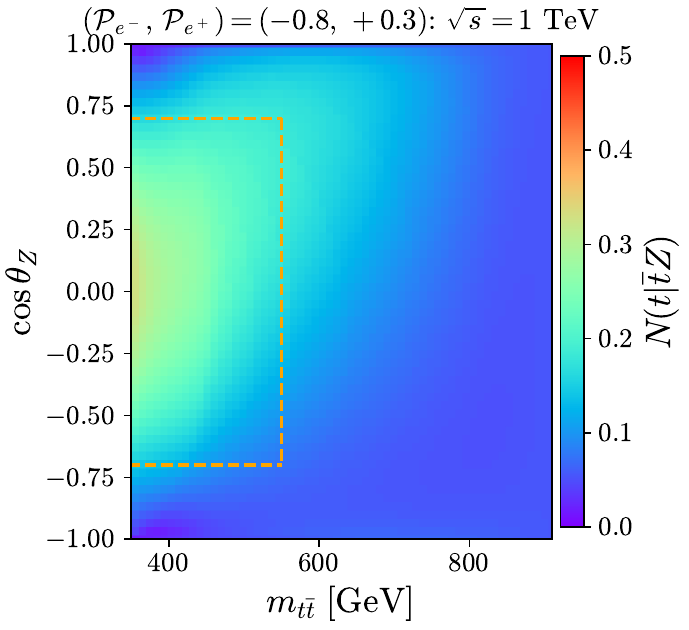}
\includegraphics[width=0.32\textwidth]{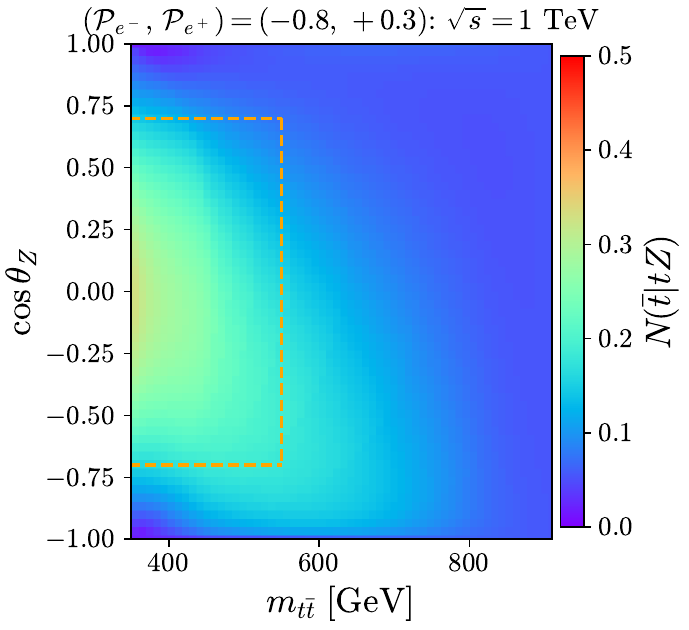}
\caption{\small One-to-other negativities of the partially inclusive $t\bar t Z$ spin state $\rho_{\Sigma}$ at $\sqrt{s}=1$~TeV for polarised beams, in the $(m_{t\bar t},\cos\theta_{Z})$ plane in the helicity basis: $N(Z|t\bar t)$ (left), $N(t|\bar tZ)$ (middle), and $N(\bar t|tZ)$ (right).
\emph{Upper row:} Pol$+$ with $({\cal P}_{e^-},{\cal P}_{e^+})=(+0.8,-0.3)$.
\emph{Lower row:} Pol$-$ with $({\cal P}_{e^-},{\cal P}_{e^+})=(-0.8,+0.3)$.
The colour bar saturates at the algebraic upper bound, $1$ for $N(Z|t\bar t)$ and $\frac{1}{2}$ for $N(t|\bar tZ)$ and $N(\bar t|tZ)$.}
\label{fig:1-2_negativity_partial_pol}
\end{figure}

The one-to-one negativities $N_{t\bar t}$, $N_{tZ}$ and $N_{\bar tZ}$ are shown in \autoref{fig:1-1_negativity_partial_pol}, and the one-to-other negativities $N(Z|t\bar t)$, $N(t|\bar tZ)$ and $N(\bar t|tZ)$ in \autoref{fig:1-2_negativity_partial_pol}.
In each figure the upper row corresponds to Pol$+$ and the lower row to Pol$-$.
A direct comparison with the unpolarised maps in \autoref{fig:1-1_negativity_partial} and \autoref{fig:1-2_negativity_partial} shows that the position, shape and peak heights of the entanglement-positive regions are essentially unchanged: the polarisation acts mainly on the production rate, not on the spin-correlation structure of the partially inclusive state.
This justifies the choice of fiducial regions $\Sigma_{1|2}$ and $\Sigma_{\rm GME}$ from the unpolarised maps in \autoref{sec:partial}, and supports the conclusion in \autoref{sec:feesibility} that the polarisation enters the feasibility study almost exclusively through the fiducial cross section.

\end{sloppypar}

\newpage

\bibliographystyle{utphys}
\bibliography{reference}


\end{document}